\documentclass[iop]{emulateapj}
\usepackage{graphicx}
\usepackage{apjfonts}
\usepackage{amsmath}
\usepackage{color}
\usepackage[bookmarks,breaklinks]{hyperref}
   \hypersetup{
     colorlinks,
     citecolor=blue,
     linkcolor=blue,
    }  
\usepackage{natbib}
\newcommand{\radboud}{1}
\newcommand{\cfa}{2}
\newcommand{\mpifr}{3}
\newcommand{\astron}{4}

\begin{document}
\title{Quantifying Intrinsic Variability of Sagittarius A* using Closure Phase Measurements of the Event Horizon Telescope }
\shorttitle{Quantifying Intrinsic Variability of Sgr A* with Closure Phases}
\author{Freek Roelofs\altaffilmark{\radboud}, 
		Michael D. Johnson\altaffilmark{\cfa}, 
		Hotaka Shiokawa\altaffilmark{\cfa},
		Sheperd S.\ Doeleman\altaffilmark{\cfa},
  \&   Heino Falcke\altaffilmark{\radboud,\mpifr,\astron}
}
\altaffiltext{\radboud}{Department of Astrophysics, Institute for Mathematics, Astrophysics and Particle Physics, Radboud University, PO Box 9010, 6500 GL Nijmegen, The Netherlands}
\altaffiltext{\cfa}{Harvard-Smithsonian Center for Astrophysics, 60
  Garden Street, Cambridge, MA 02138, USA}
\altaffiltext{\mpifr}{Max-Planck-Institut f\"{u}r Radioastronomie, Auf
  dem H\"{u}gel 69, D-53121 Bonn, Germany}
\altaffiltext{\astron}{ASTRON, The Netherlands Institute for Radio
  Astronomy, Postbus 2, NL-7990 AA Dwingeloo, The Netherlands}

\shortauthors{Roelofs et al.}
\email{f.roelofs@astro.ru.nl}

\begin{abstract}
General relativistic magnetohydrodynamic (GRMHD) simulations of accretion disks and jets associated with supermassive black holes show variability on a wide range of timescales. On timescales comparable to or longer than the gravitational timescale $t_G=GM/c^3$, variation may be dominated by orbital dynamics of the inhomogeneous accretion flow. Turbulent evolution within the accretion disk is expected on timescales comparable to the orbital period, typically an order of magnitude larger than $t_G$. For Sgr A*, $t_G$ is much shorter than the typical duration of a VLBI experiment, enabling us to study this variability within a single observation. Closure phases, the sum of interferometric visibility phases on a triangle of baselines, are particularly useful for studying this variability. In addition to a changing source structure, variations in observed closure phase can also be due to interstellar scattering, thermal noise, and the changing geometry of projected baselines over time due to Earth rotation. We present a metric that is able to distinguish the latter two from intrinsic or scattering variability. This metric is validated using synthetic observations of GRMHD simulations of Sgr A*. When applied to existing multi-epoch EHT data of Sgr A*, this metric shows that the data are most consistent with source models containing intrinsic variability from source dynamics, interstellar scattering, or a combination of those. The effects of black hole inclination, orientation, spin, and morphology (disk or jet) on the expected closure phase variability are also discussed.
\end{abstract}


\section{Introduction}
\label{sec:introduction}
\subsection{The Event Horizon Telescope}
The Event Horizon Telescope (EHT) is a very long baseline interferometry (VLBI) array observing supermassive black holes at the centers of galaxies on event horizon scales at mm-wavelengths \citep{Doeleman2009}. The prime target of the EHT is Sagittarius A* (Sgr A*), the radio source associated with the $4\cdot10^6M_{\odot}$ \citep{Ghez2008, Gillessen2009, Chatzopoulos2015} black hole at the Galactic Center. Due to its proximity, the apparent size of the event horizon as predicted by general relativity ($\sim 50$ $\mu$as) is the largest of all black hole candidates \citep{Johannsen2012}. EHT observations over several epochs have confirmed the existence of structures on the scale of the event horizon in Sgr A* \citep{Doeleman2008, Fish2011}. This makes Sgr A* the most promising candidate for imaging the black hole ``shadow'' imprinted on the surrounding accretion flow due to gravitational lensing in the curved spacetime \citep{Bardeen1973, Falcke2000, Takahashi2004}. Spatially resolving this shadow and the structure of the surrounding emission will allow for tests of general relativity \citep{Bambi2009, Johannsen2010, Falcke2013, Psaltis2015, Goddi2016} and give information about the nature of the accretion flow, which could be dominated by an accretion disk or relativistic jet \citep{Falcke2000b, Yuan2003, Fish2009, Moscibrodzka2009, Dexter2010, Chan2015, Gold2017}.

\subsection{Time-variable emission}
Light curves of Sgr A* have been analyzed over many years and wavelengths to study the processes of total emission variability \citep[e.g.][]{Yusef-Zadeh2006, Marrone2008, Porquet2008, Do2009, Meyer2009, Dexter2014}, and variations on Schwarzschild radius scales have been observed with EHT observations \citep{Fish2011}. Emission from Sgr A* may be expected to exhibit variations due to several mechanisms near the event horizon. First, the rotating accretion flow can imprint fluctuations with orbital timescales on the horizon-scale emission. The period of the prograde innermost stable circular orbit (ISCO) ranges from about 4 minutes for a maximally spinning black hole to half an hour for a non-spinning black hole with the mass of Sgr A* \citep{Bardeen1972}. Furthermore, the magnetorotational instability \citep[MRI;][]{Balbus1991, Balbus1996} causes stochastic turbulence in the disk that may cause variations on time scales comparable to the ISCO period. In addition to the dynamics of the source itself, interstellar scattering introduces spurious refractive substructure that is expected to be variable on timescales of about a day at 1.3 mm \citep{Johnson2015}. The measured interferometric quantities are also variable due to the nature of the observations. Since a typical VLBI observation takes several hours, the projected lengths and orientations of the baselines as seen from the source change as the Earth rotates. Different Fourier components of the source brightness distribution are thus measured at different times, causing the complex visibilities measured on each baseline to change on timescales of a few hours. Finally, instrumental noise causes these quantities to fluctuate with each measurement.

A source that is time variable over the course of an observation poses challenges for reconstructing an image, though the shadow of Sgr A* may still be recovered by averaging, normalizing, and smoothing the measured visibilities \citep{Lu2016}. A variable source such as Sgr A* further provides opportunities for studying dynamics in the accretion flow through direct use of interferometric quantities without the need for imaging. \citet{Johnson2015cov} showed that the angular velocity of the accretion flow emission pattern may be estimated by analyzing the lagged covariance between visibility amplitudes on baselines with approximately equal length and slightly different orientation. \citet{Kim2016} developed a Bayesian framework for fitting EHT visibility amplitudes to GRMHD models that include source variability. Closure phase is another example of an interferometric quantity of which variability may be studied, and will be the main focus of this paper.

\subsection{Closure phase}
Closure phase is the sum of interferometric visibility phases on a closed triangle of baselines \citep{Jennison1958, Rogers1974}. It is a robust interferometric observable as phase corruptions introduced by the atmosphere at each station cancel as the phases are added. These phase corruptions are severe at mm-wavelengths and particularly difficult to deal with in VLBI observations as the atmospheres at the different sites are uncorrelated due to the large distances between them. Most mm-VLBI phase information is therefore obtained from closure phases. A closure phase value deviating from 0 or 180 degrees indicates that the source is not point-symmetric \citep{Monnier2007}. For a point-symmetric source, there is a central point such that every source element has an identical counterpart at the same distance and opposite direction from that point. Except for this rule, closure phases are generally difficult to interpret intuitively as they are the sum of three Fourier phases of the image and strongly dependent on the baseline triangle on which they are measured. However, by comparing observed closure phases to those simulated from a source model, one may still use them to draw conclusions on the source morphology in situations where imaging is not possible due to, e.g., poor $uv$-coverage.

\citet{Ortiz2016} measured non-zero closure phases of Sgr A* at 3.5 mm, and showed that the asymmetry can be explained by refractive scattering alone.
At the same wavelength, nonzero closure phases of Sgr A* have been measured by \citet{Brinkerink2016}, which could be explained by interstellar scattering or intrinsic source structure. \cite{Fish2016} measured a nonzero median closure phase for EHT observations at 1.3 mm spanning four years, indicating persistent asymmetric structure at much smaller scales. More data at 1.3 mm are required to pursue the detailed nature of the asymmetry.

Closure phases and their behavior as a function of time have been simulated by several groups. \citet{Doeleman2009cl} showed that closure phases from models of orbiting inhomogeneities, or ``hot-spots'', in the accretion flow exhibited periodicities that could be used to time orbital dynamics near the black hole. \citet{Fish2009} showed that polarimetric quantities could also be used to track structural variation. \citet{Dexter2010} computed closure phases from general relativistic magnetohydrodynamics (GRMHD) movies, which included variability due to orbital dynamics, turbulence, and Earth rotation. They also investigated dependence on sky orientation. \citet{Dexter2013} did the same for a tilted accretion disk. They both found that the closure phase is strongly variable for those sky orientations where the visibility amplitudes on one of the baselines are small. \citet{Broderick2016} used observed 1.3 mm closure phases to constrain the orientation of radiatively inefficient accretion flow models of Sgr A*. \citet{Encinas2016} computed closure phases for disk and jet dominated GRMHD models and studied their variability due to Earth rotation. \citet{Medeiros2016} generated closure phases from GRMHD movies, and investigated differences in the closure phase variability for different source models and baselines, identifying regions in the $uv$-plane where the (closure) phase variability is strong. They did not address contributions form thermal noise and interstellar scattering. Our focus is on developing a metric that makes a distinction between image variability (due to the source itself and interstellar scattering), and observational variability (due to thermal noise and Earth rotation).

\subsection{Outline}
In this work, we develop a metric that gives a measure of the closure phase variability in a given data set that is not due to thermal noise or Earth rotation. This metric can be used to test for the presence of image (source and/or scattering) variability, quantify the amount of image variability, and compare the amount of image variability to and between different source simulations. The metric is introduced and validated using simulated closure phases from general relativistic magnetohydrodynamics (GRMHD) simulations of Sgr A*. The metric is then used to test for image variability in the closure phases measured on the California-Hawaii-Arizona triangle of EHT baselines during multiple epochs in 2009-2013 by \citet{Fish2016}. Finally, the metric outcomes as a function of several GRMHD parameters are compared.

The details of the GRMHD simulations are discussed in Section \ref{sec:simulations}. Section \ref{sec:Nmetric} introduces the metric $\mathcal{Q}$ to quantify intrinsic closure phase variability. Section \ref{sec:ehtdata} describes how this metric may be used to compare the closure phase simulations directly to EHT data. The results of this procedure applied to several source models and the EHT closure phase measurements of Sgr A* by \citet{Fish2016} are presented in Section \ref{sec:results}. Our conclusions are summarized and suggestions for future directions are given in Section \ref{sec:summary}.

\section{Simulating Sgr A* closure phases}
\label{sec:simulations}
We generated synthetic observations from existing GRMHD simulations in order to validate our metric and test for intrinsic variability in the 2009-2013 EHT data from \citet{Fish2011}. The movies were generated for a range of GRMHD parameters in order to investigate their effect on closure phase variability. The GRMHD simulations are introduced in Section \ref{sec:grmhd}. Section \ref{sec:clnoise} describes how we calculated closure phases from GRMHD movies, and Section \ref{sec:behavior} shows how they behave for different baseline triangles and source types (static or variable).

\subsection{GRMHD simulations}
\label{sec:grmhd}
Using Faraday rotation measurements, the mass accretion rate of Sgr A* has been limited to a maximum of only $\sim10^{-7} M_{\odot}$/yr \citep{Marrone2007}. For a system with such a low accretion rate, the plasma density of the accretion disk is low and cooling is inefficient as there are few particle interactions. The formation of a thin disk (with a thickness $\lesssim 20$\% of its radius) is prevented by the pressure of the hot gas. The viscous energy of the disk is then advected into the black hole instead of radiated away \citep{Narayan1998}. Such thick disks have been modeled as radiatively ineffecient accretion flows \citep[RIAF;][]{Yuan2003}. 

Advances in computing power over recent decades have allowed for complex numerical simulations of such accretion flows, pioneered by \citet{Gammie2003}. These simulations evolve a magnetized plasma according the equations of general relativistic magnetohydrodynamics (GRMHD). GRMHD simulations have succesfully recovered observational parameters of black holes such as Sgr A* and M87, and are widely used as a theoretical basis for interpreting EHT observations \citep[e.g.][]{Moscibrodzka2009, Dexter2010, Dexter2012, Drappeau2013, Chan2015, Moscibrodzka2016, Gold2017}.

In this work, we used the GRMHD simulation $\texttt{b0-high}$ from \citet{Shiokawa2013} as a basis. The initial configuration of this simulation was a RIAF type torus-shaped thick disk (inner radius at 12$r_{\mathrm{G}}$, pressure maximum at 24$r_{\mathrm{G}}$) aligned to the spin axis of a black hole with spin $a_*=0.9375$ (0.94 hereafter). The same configuration was used for a black hole with spin $a_*=0$ as well. The initial magnetic field configuration was poloidal. The adiabatic index $\Gamma$ was set to 13/9, corresponding to a collisionless plasma of relativistic electrons and nonrelativistic protons \citep{Shapiro1973}. The \texttt{HARM3D} code \citep{Noble2009} was run from these initial conditions until the MRI turbulence saturated at $t=8000t_{\mathrm{G}}$, after which the simulation was run for another $6500t_{\mathrm{G}}$ where the results were recorded with a time resolution of 0.5$t_{\mathrm{G}}$.

For each time step, images were generated from the physical quantities calculated by the GRMHD prescription using a general relativistic ray-tracing code \citep{Noble2007}. This code integrates the emissivity and absorption of the synchrotron-emitting plasma along the geodesics running from the pixels of a camera far away from the source to the simulation domain. One difference from the results in \citet{Shiokawa2013} is that the fast light approximation was not used for our ray-traced images, which means that the fluid evolved as the photons propagated outwards. Photons were distributed along each geodesic running from the camera to the vicinity of the disk, and propagated along the geodesic at each time step while the radiative transfer function was integrated for each of them using the current GRMHD frame \citep{Shiokawa2017}. This generally makes the resulting light curves and closure phase time series somewhat smoother, but does not have a large effect on our results. The black hole mass was set to $4.5\cdot10^6M_{\odot}$, consistent with mass measurements from stellar dynamics \citep{Ghez2008, Gillessen2009} and resulting in a gravitational time scale $t_{\mathrm{G}}=GM/c^3$ of 22 s. For the $a_*=0.94$ case, the ISCO radius is $2.04r_{\mathrm{G}}$, and the ISCO period is 24.25$t_{\mathrm{G}}=9$ min. For $a_*=0$, the ISCO radius is $6r_{\mathrm{G}}$, and the ISCO period is 34 min. The eventual image depends on three additional free parameters. The accretion rate $\dot{M}$ is a global scaling factor that determines the total flux. It was set such that the average total flux was compatible with the observed $3.68 \pm 0.66$ Jy at 231.9 GHz for Sgr A* \citep{Bower2015}. The proton-to-electron temperature ratio $T_p/T_e$ was set to 3 in accordance with the best-fit model for Sgr A* from \citet{Moscibrodzka2009}. The final free parameter is the inclination angle $i$ between the black hole spin axis and the line of sight. Although the best-fit model from \citet{Moscibrodzka2009} only fits the observational data well for high inclinations, a broad range of inclinations was considered in order to study the closure phases behavior as a function of inclination. Images were generated for inclinations of 2, 18, 30, 45, 60, and 85 degrees. 

In the radiative transfer model described above, the emission originates solely in the accretion disk. However, by using a different prescription for $T_p/T_e$, one can construct a simulation where the emission also contains contributions from a relativistic jet. Although a relativistic jet has not been observed directly for Sgr A*, some evidence for the presence of a jet exists. 

For example, Sgr A* shows a flat to inverted spectrum: the flux density increases towards higher frequencies up to the sub-mm bump at $\sim10^{12}$ Hz \citep{Falcke2013}. Flat or inverted spectra are often observed in sources that contain a jet, and can be explained by the dominance of different jet components at different frequencies due to optical depth effects. A larger part of the radio core becomes optically thick as the frequency decreases, so that the flux is dominated by emission further down the jet \citep[e.g.][]{Blandford1979, Falcke1995, Hada2011}. Another implication of the optical
depth effect (i.e. the changing peak radio frequency as a function of position along the jet) in
combination with a relativistic outflow is the expectation of time lags between the light curves
at different radio frequencies as the emitting material propagates down the jet. Time lags
between light curve maxima at different frequencies have indeed been measured by \citet{Brinkerink2015}. The measured time lags indicate a mildly relativistic outflow, with a lower velocity estimate of $0.5c$.

\citet{Moscibrodzka2013} were able to improve the \citet{Moscibrodzka2009} GRMHD model fit to the spectrum of Sgr A* by increasing the electron temperature in the jet region rather than imposing a uniform $T_p/T_e$ across the simulation domain. This causes the observed synchrotron emission to contain contributions from the jet sheath. The inner parts of the jet do not produce much radiation because of the low plasma density there. For our simulations, we used the more physical prescription
\begin{equation}
\frac{T_p}{T_e}=R_{\mathrm{high}}\frac{\beta^2}{1+\beta^2}+R_{\mathrm{low}}\frac{1}{1+\beta^2}
\end{equation}
from \citet{Moscibrodzka2016}. $\beta=P_{\mathrm{gas}}/P_{\mathrm{mag}}$ is the ratio of the gas and magnetic pressure. $\beta$ will generally be high in the disk region and low in the jet region. $R_{\mathrm{high}}$ and $R_{\mathrm{low}}$ describe the electron-proton coupling in the disk (high) and jet (low) regions, and were set to 20 and 1, respectively, so that the electrons in the jet region are effectively heated. This simulation will be referred to as the ``jet model'', and the uniform $T_p/T_e$ case will be referred to as the ``disk model''.

\begin{figure*}
\begin{center}
\resizebox{0.16\hsize}{!}{\includegraphics{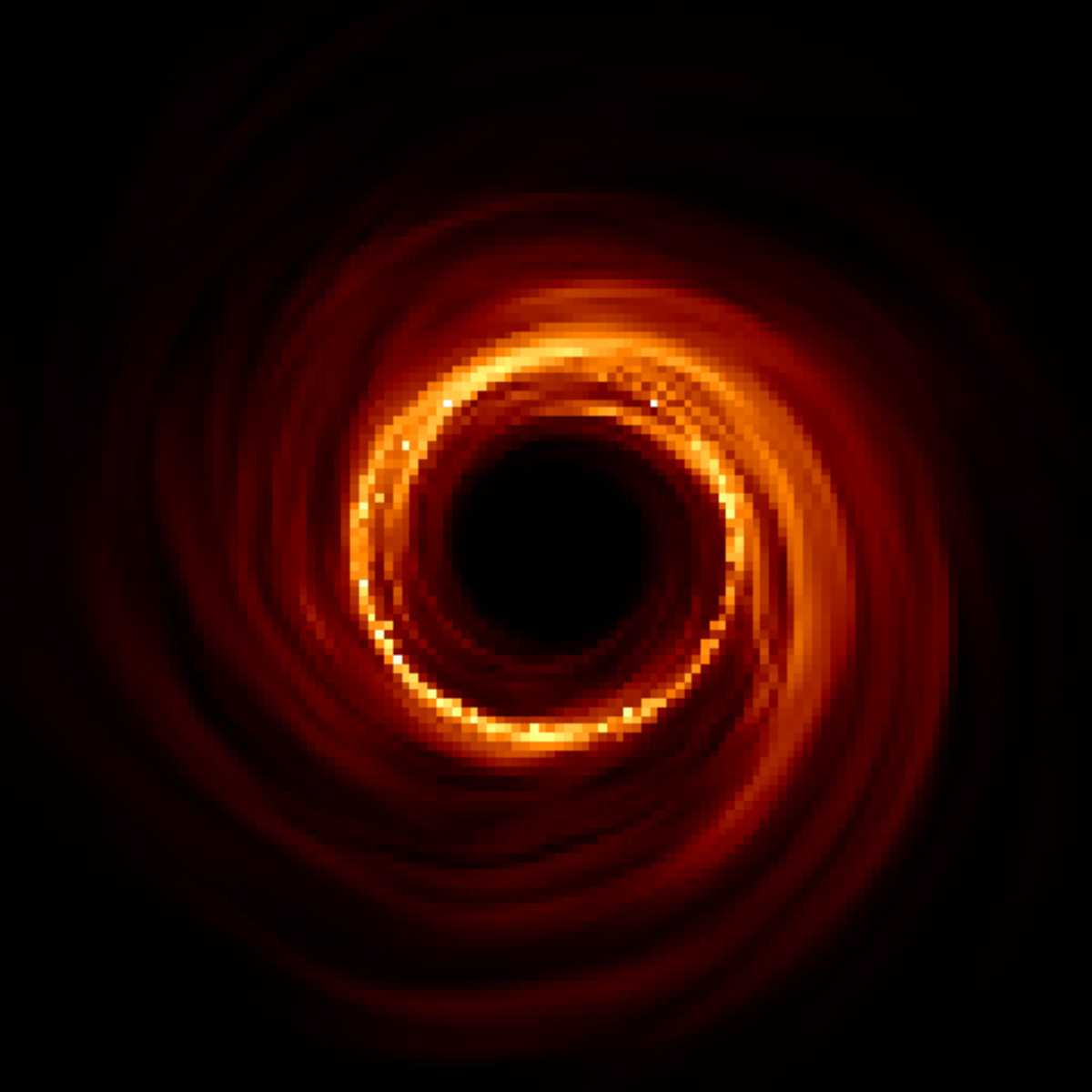}}
\resizebox{0.16\hsize}{!}{\includegraphics{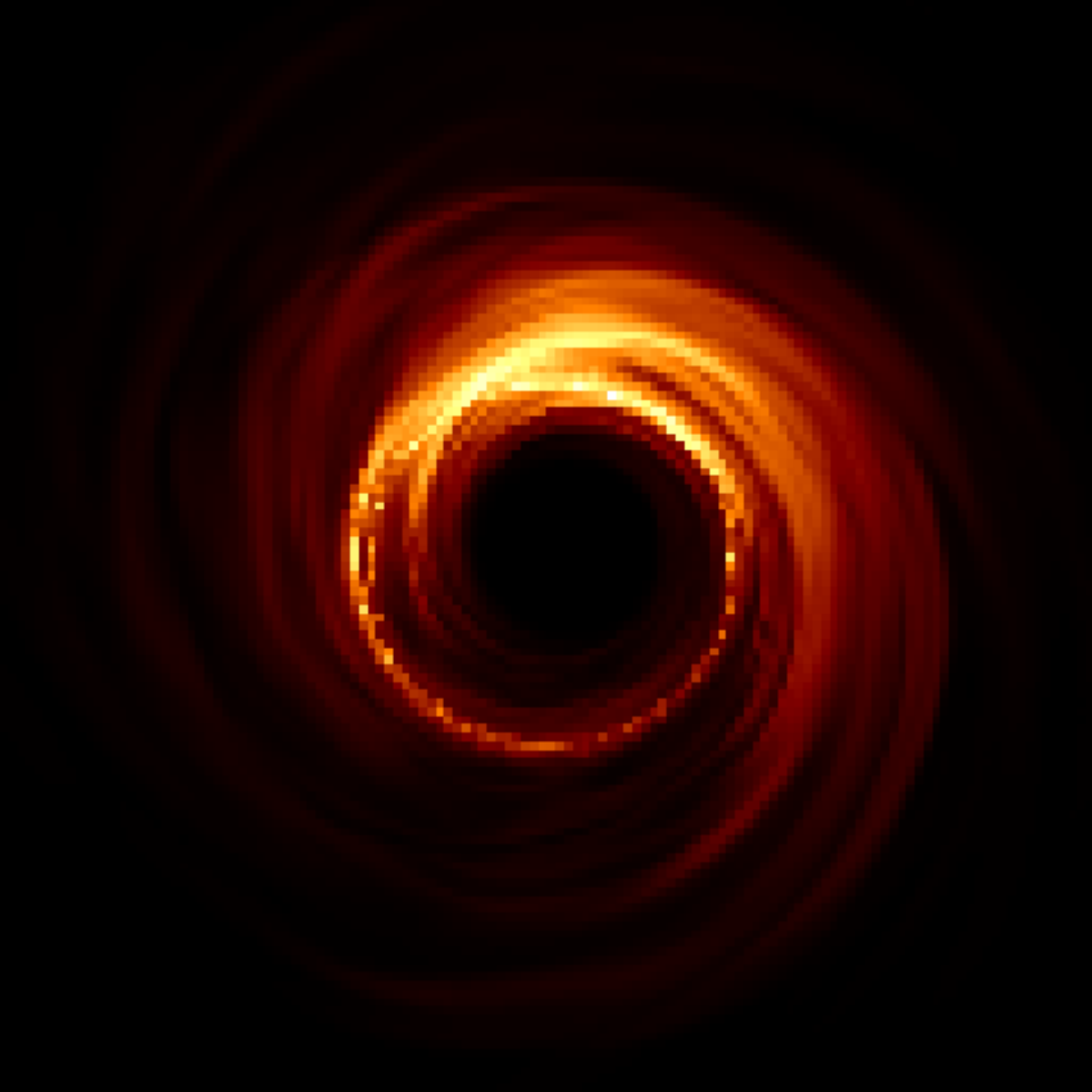}}
\resizebox{0.16\hsize}{!}{\includegraphics{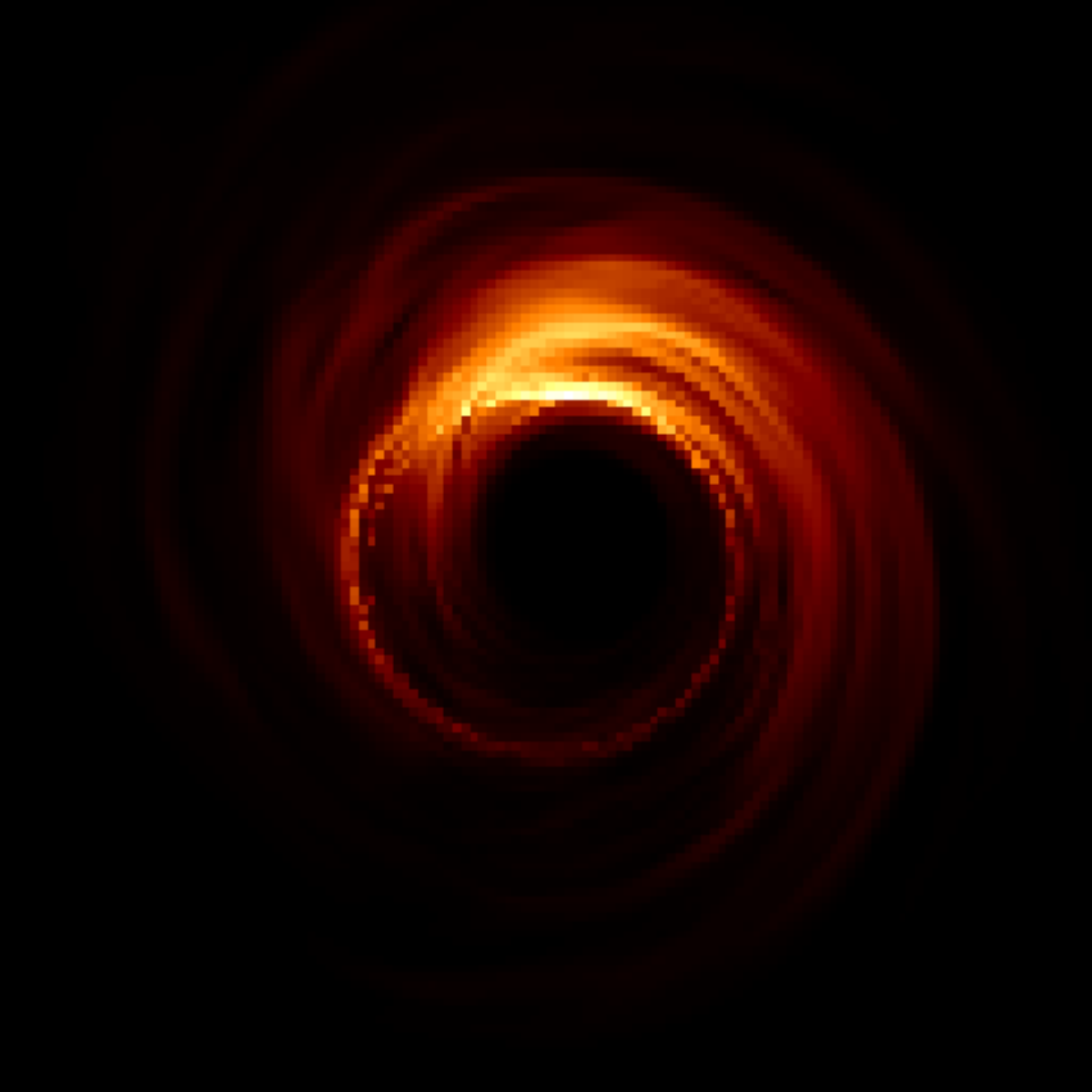}}
\resizebox{0.16\hsize}{!}{\includegraphics{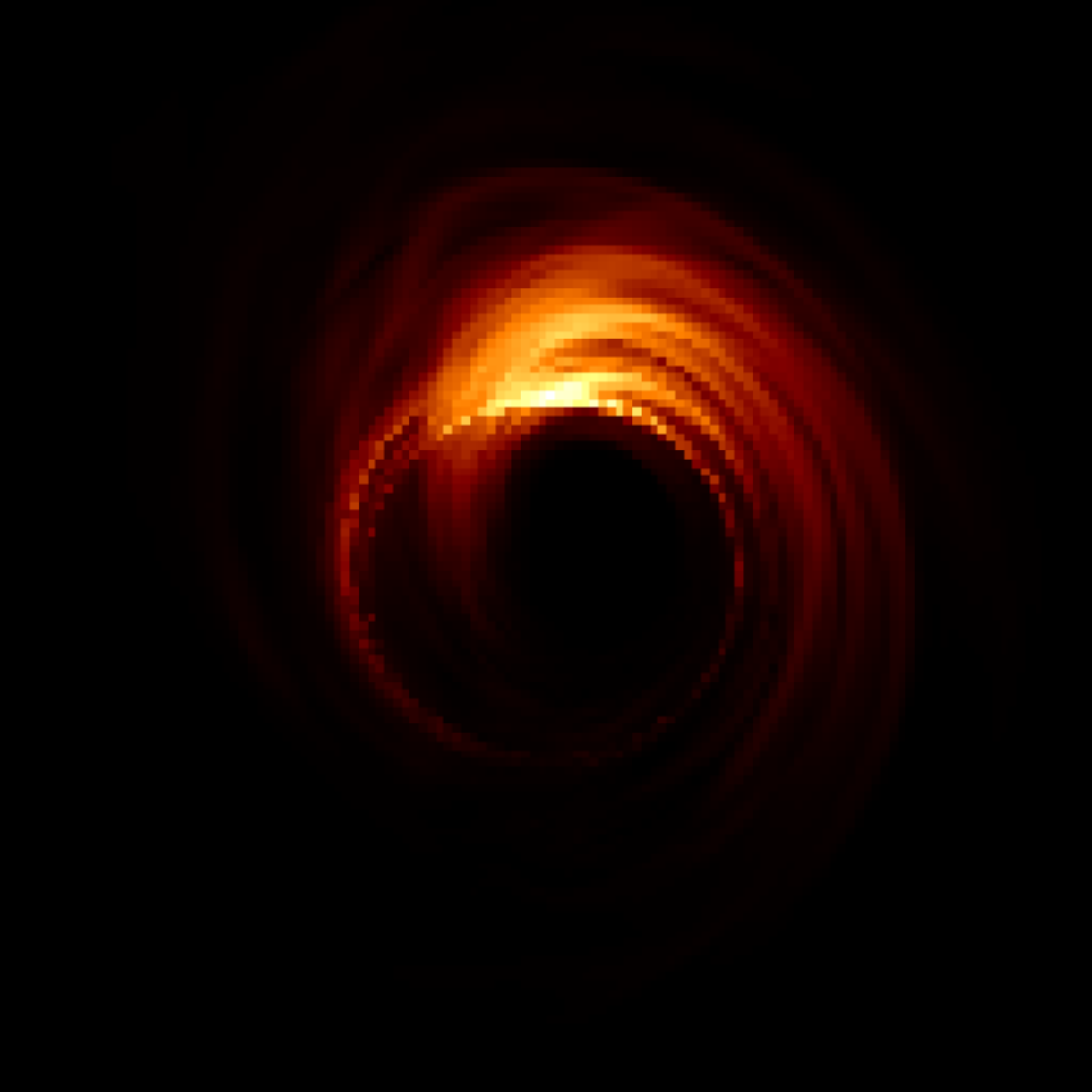}}
\resizebox{0.16\hsize}{!}{\includegraphics{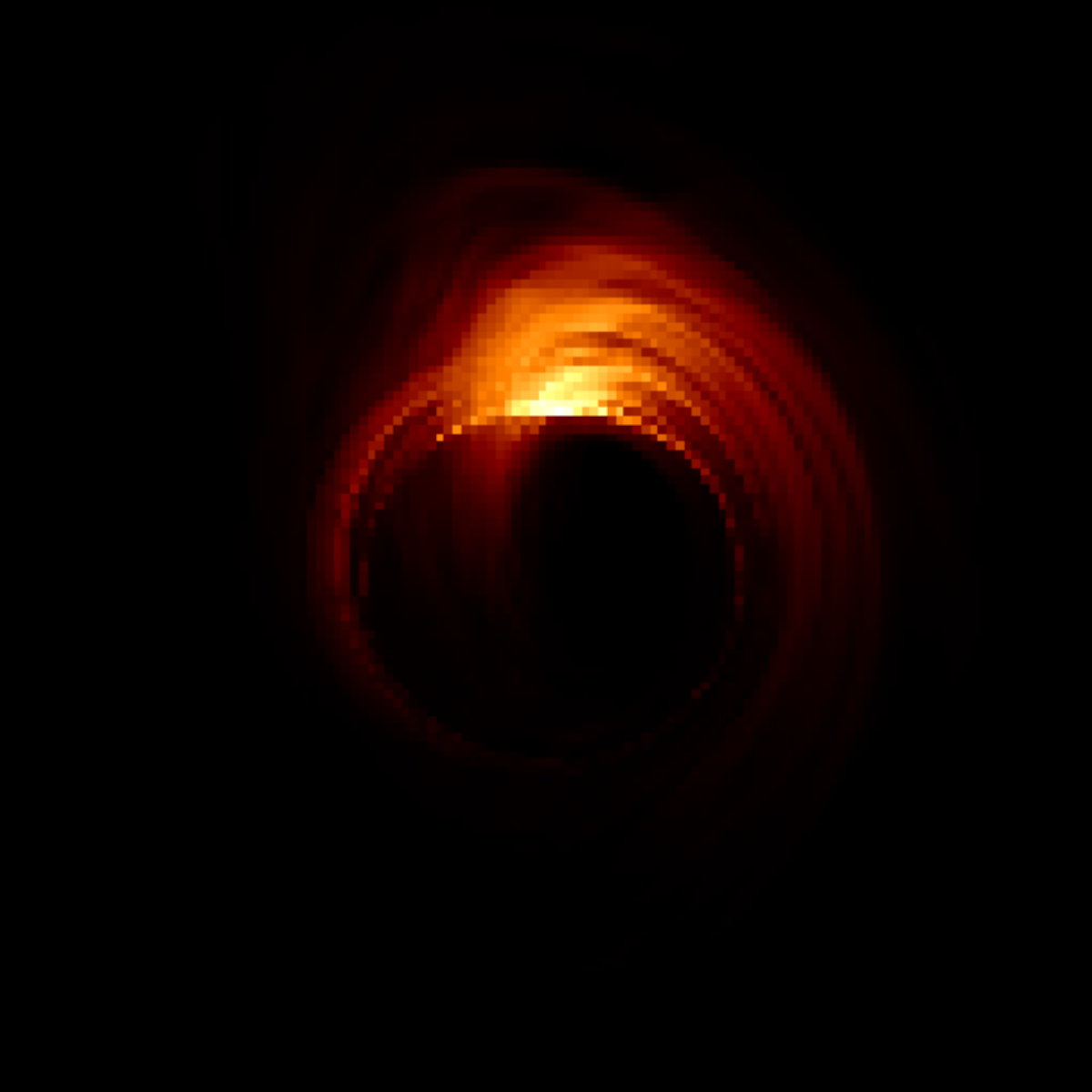}}
\resizebox{0.16\hsize}{!}{\includegraphics{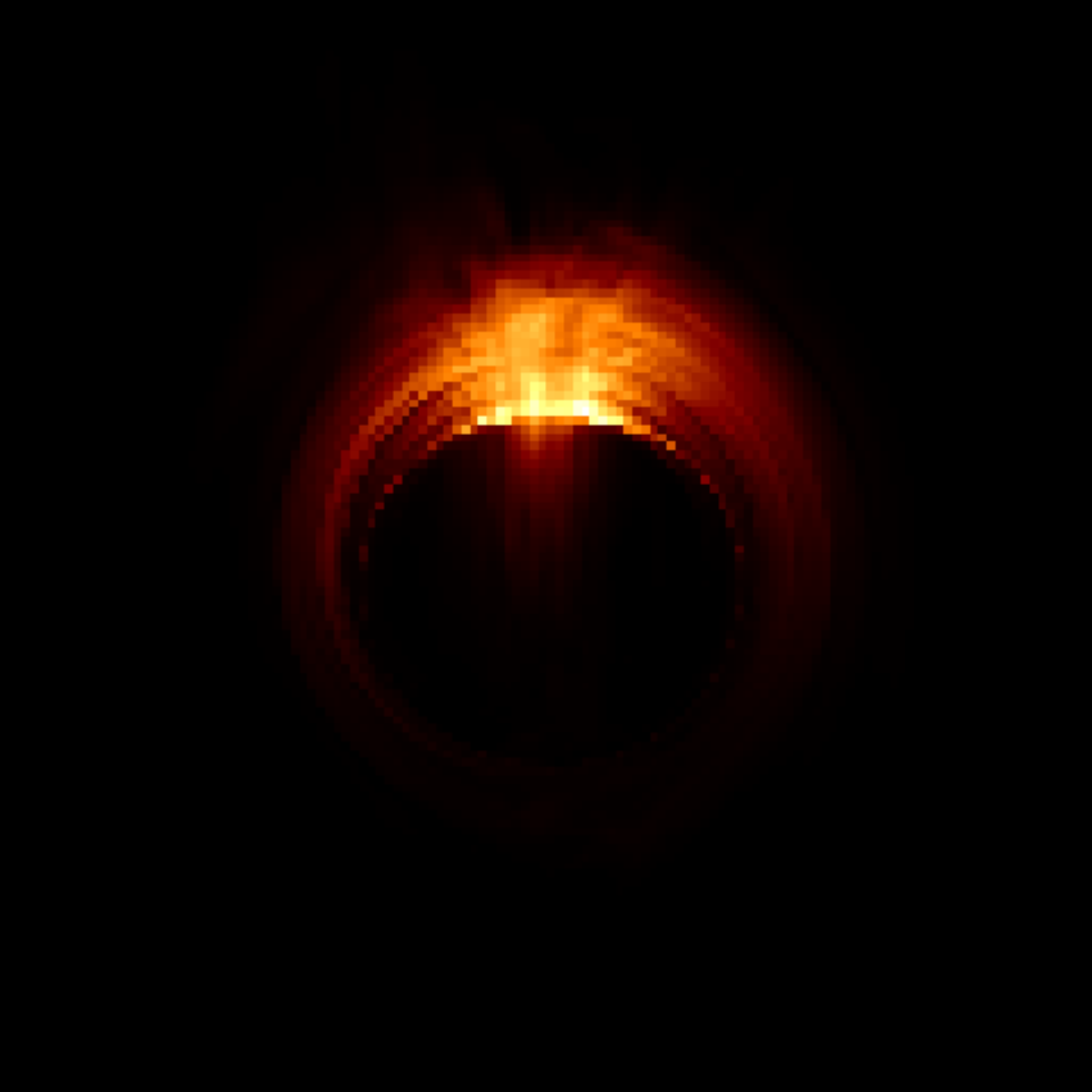}} 
\resizebox{0.16\hsize}{!}{\includegraphics{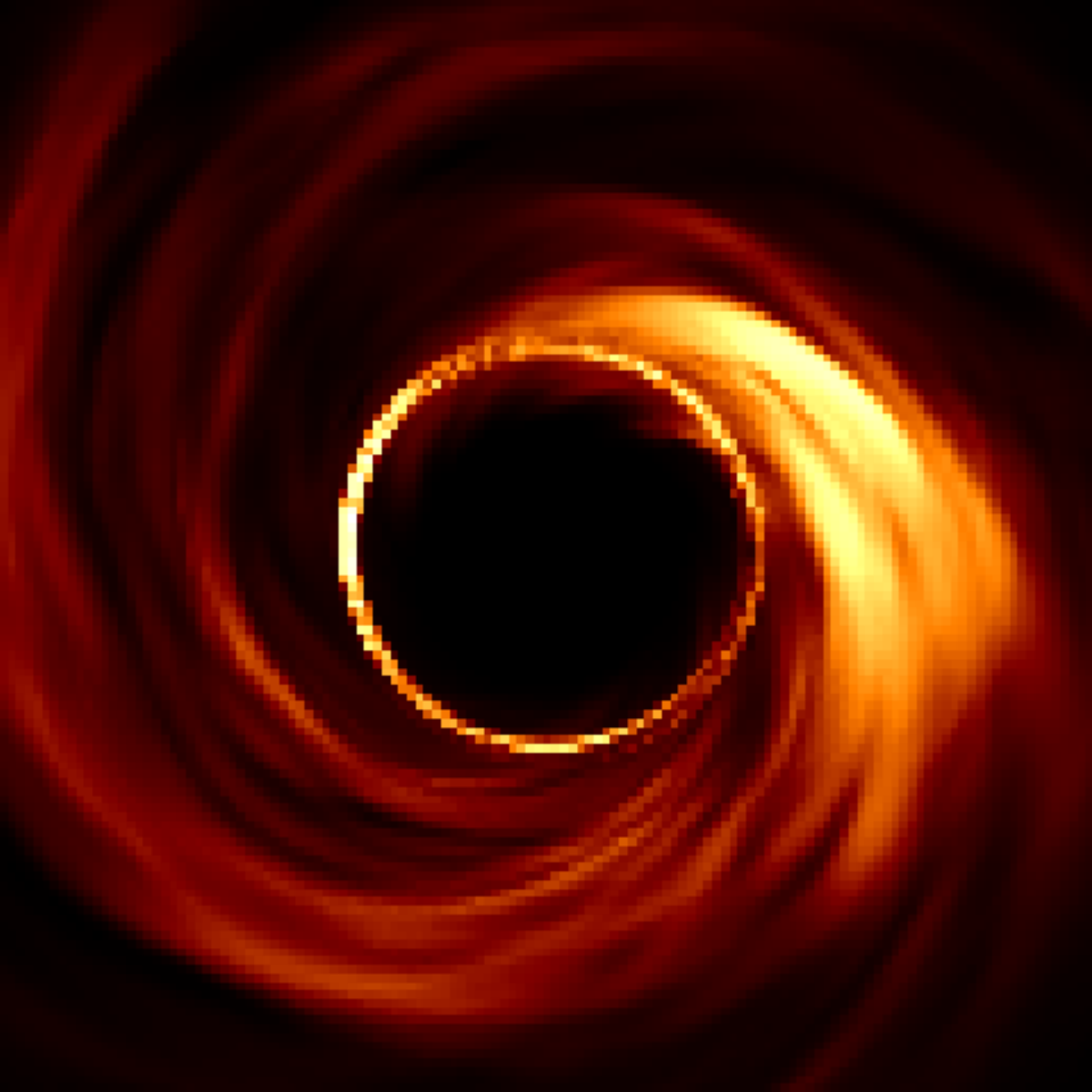}}
\resizebox{0.16\hsize}{!}{\includegraphics{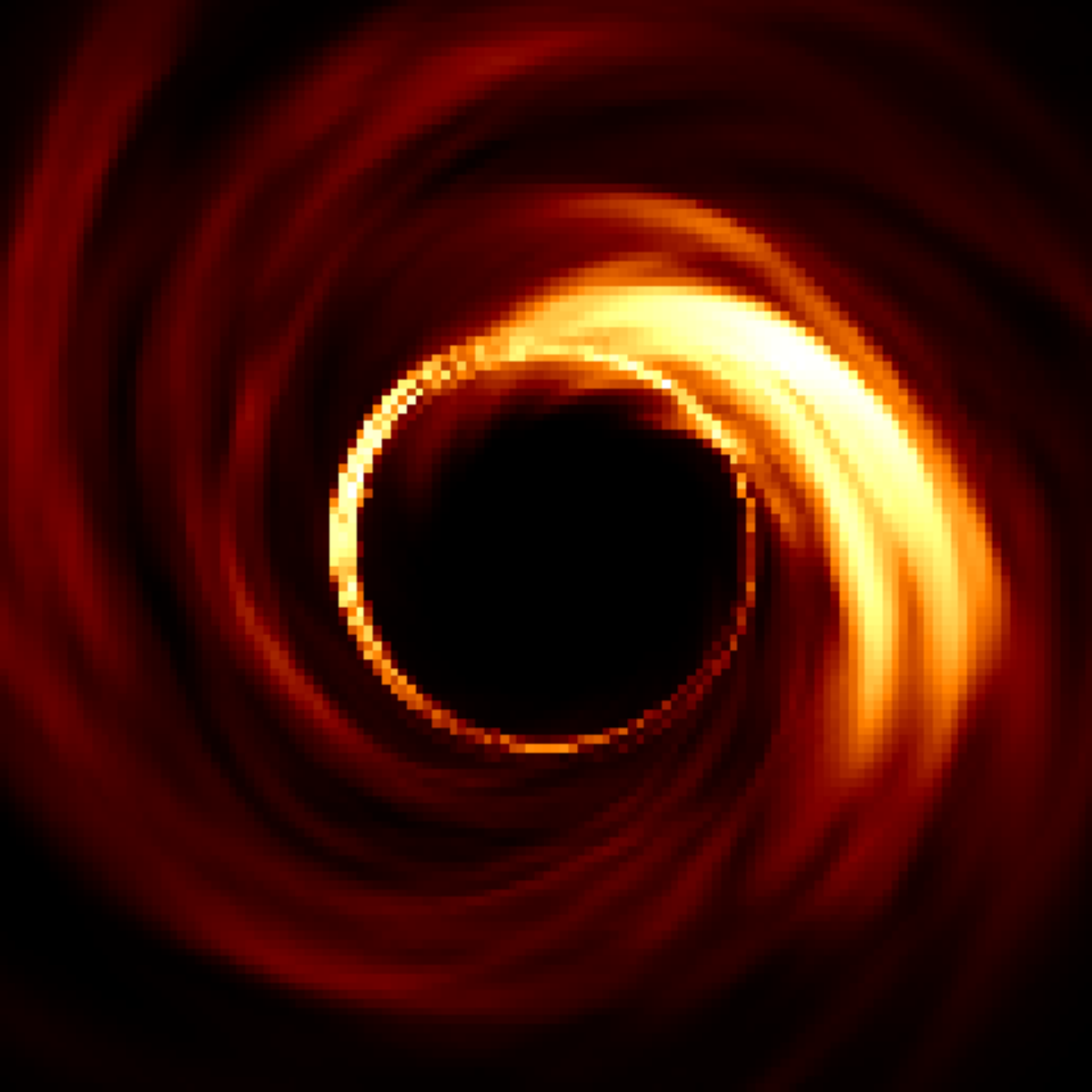}}
\resizebox{0.16\hsize}{!}{\includegraphics{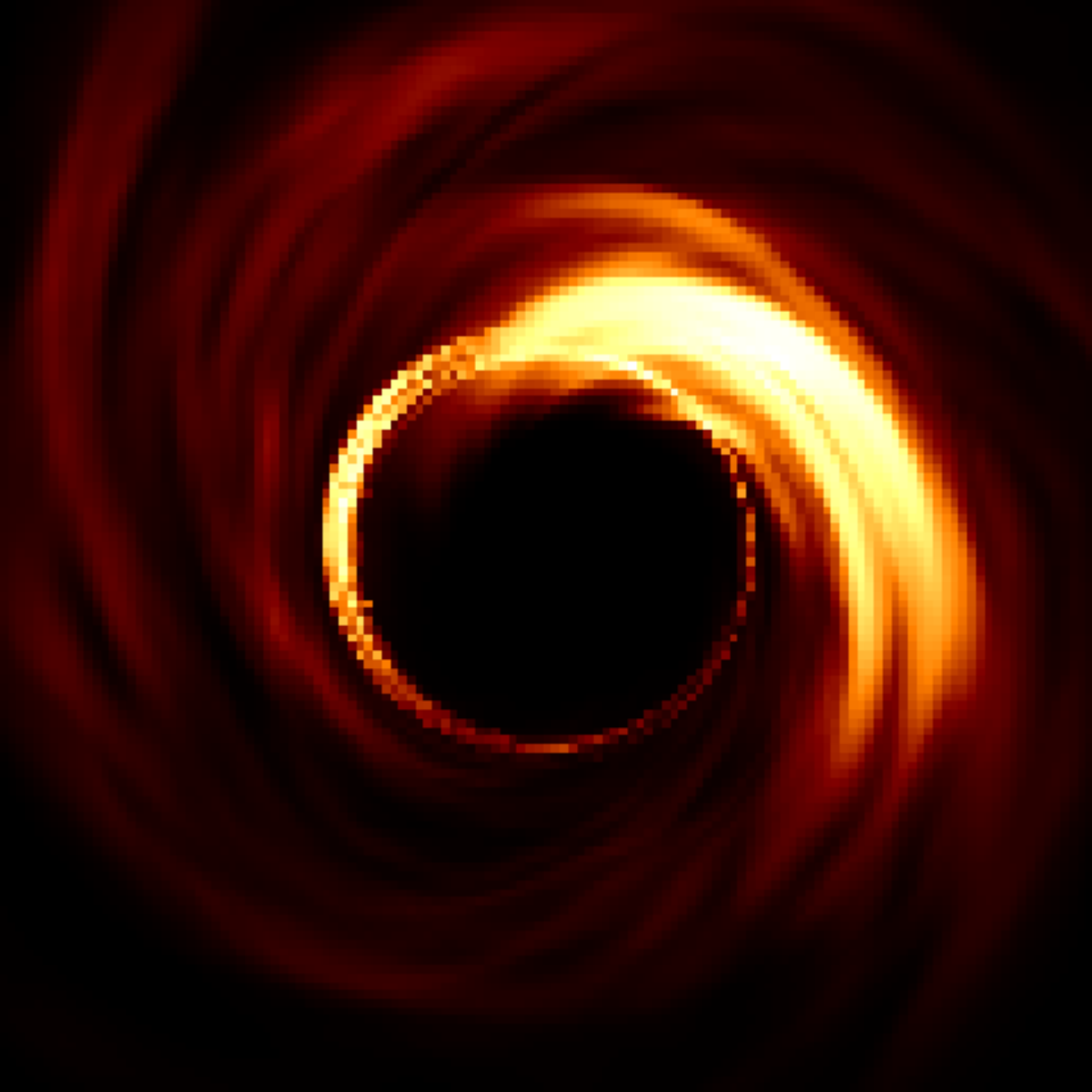}}
\resizebox{0.16\hsize}{!}{\includegraphics{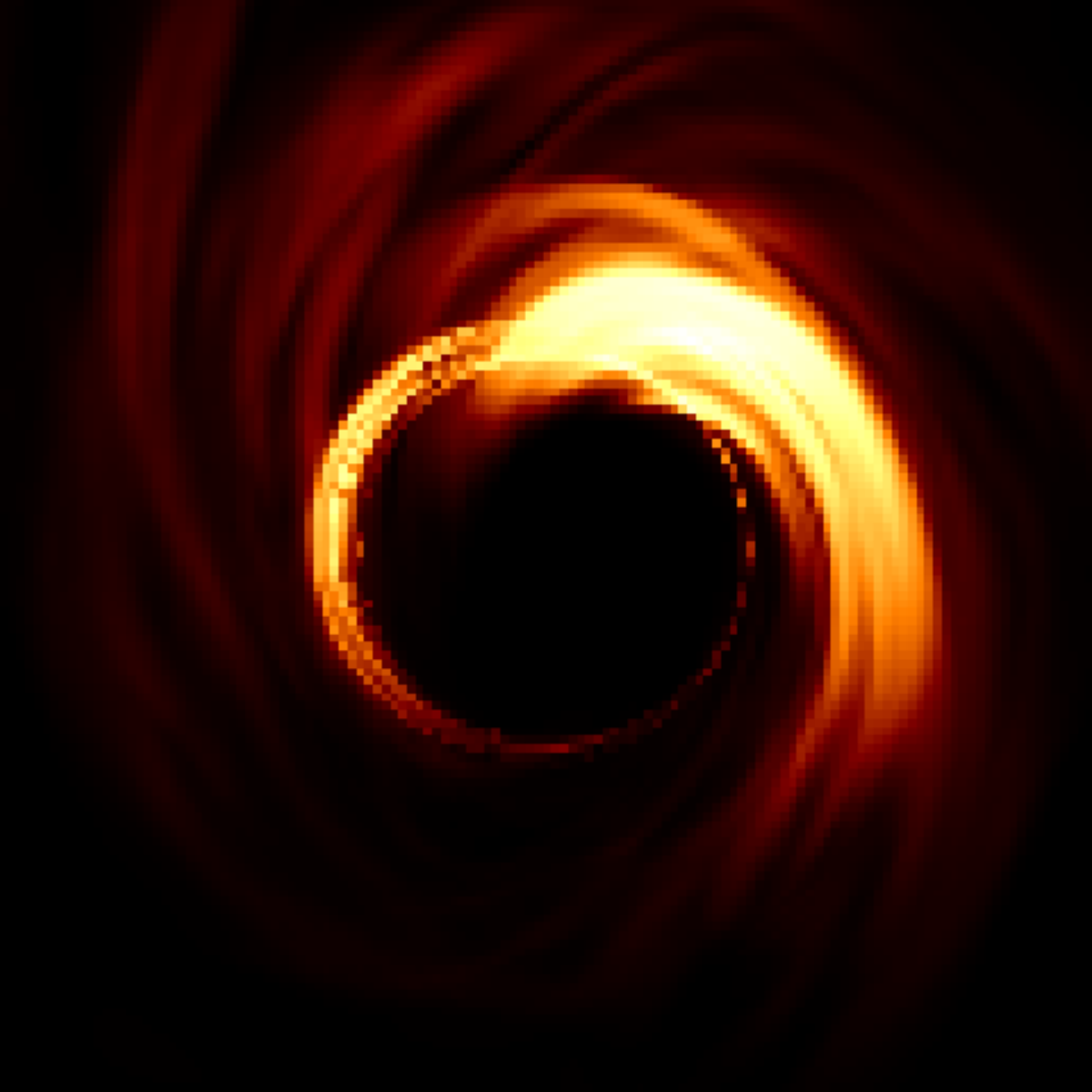}}
\resizebox{0.16\hsize}{!}{\includegraphics{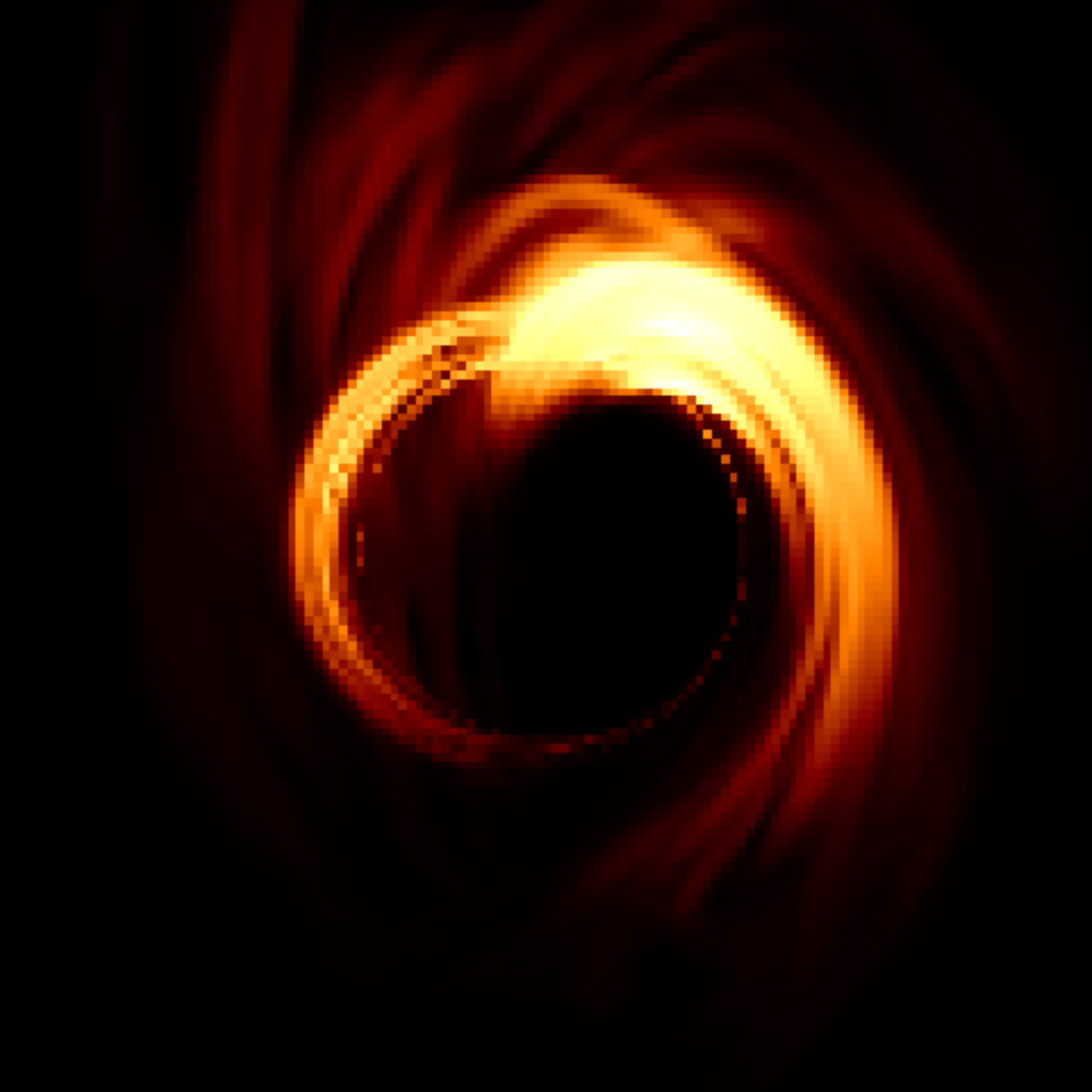}}
\resizebox{0.16\hsize}{!}{\includegraphics{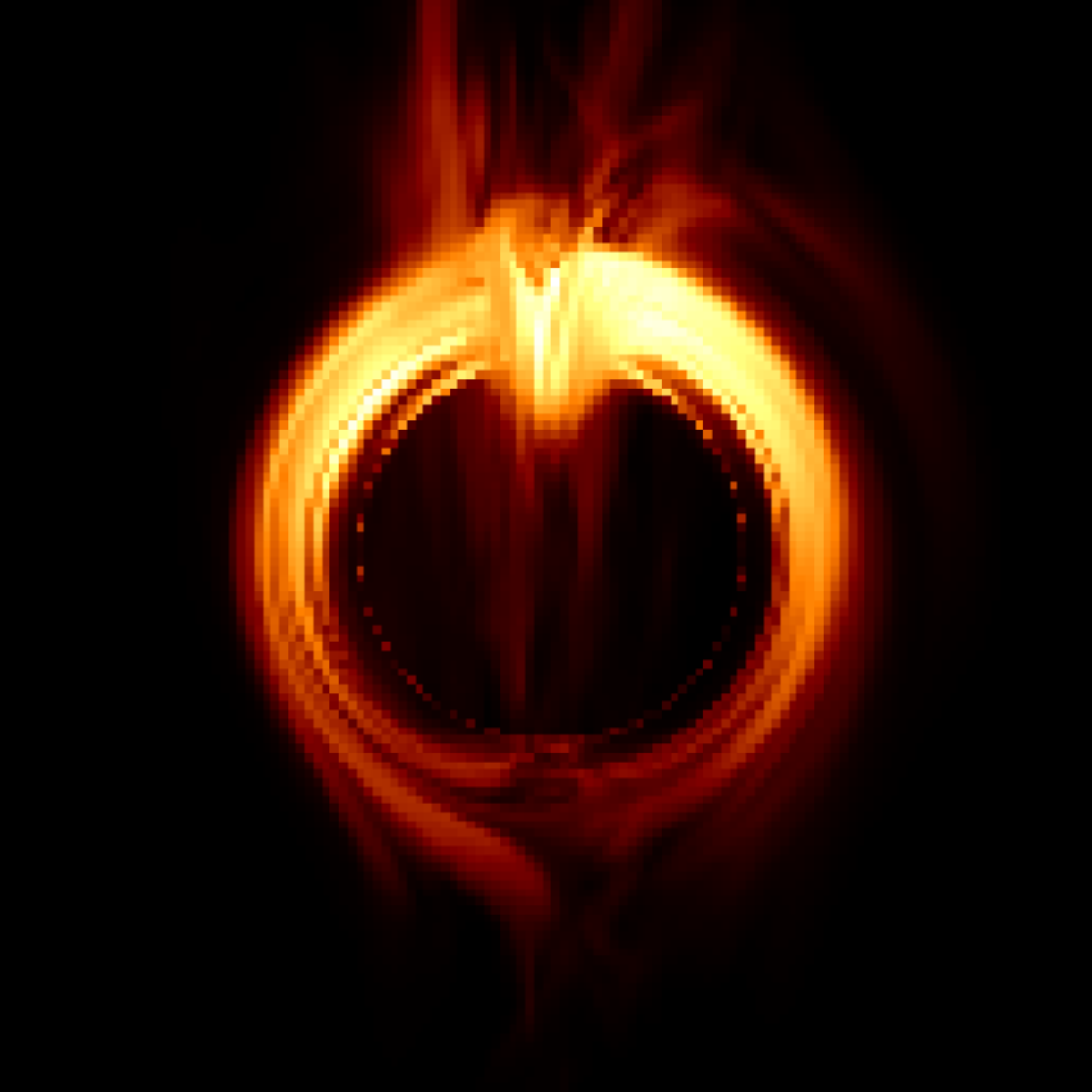}} 
\resizebox{0.16\hsize}{!}{\includegraphics{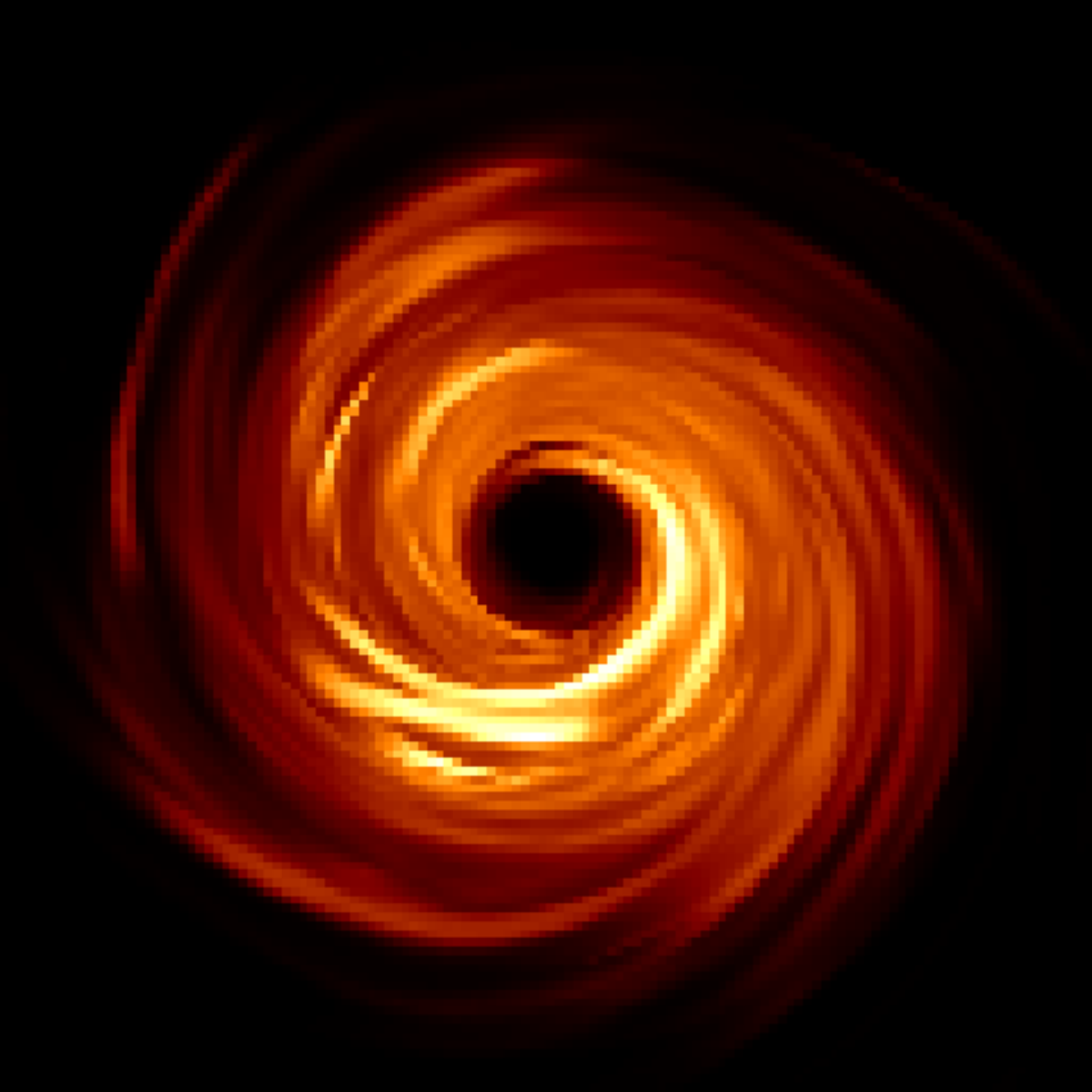}}
\resizebox{0.16\hsize}{!}{\includegraphics{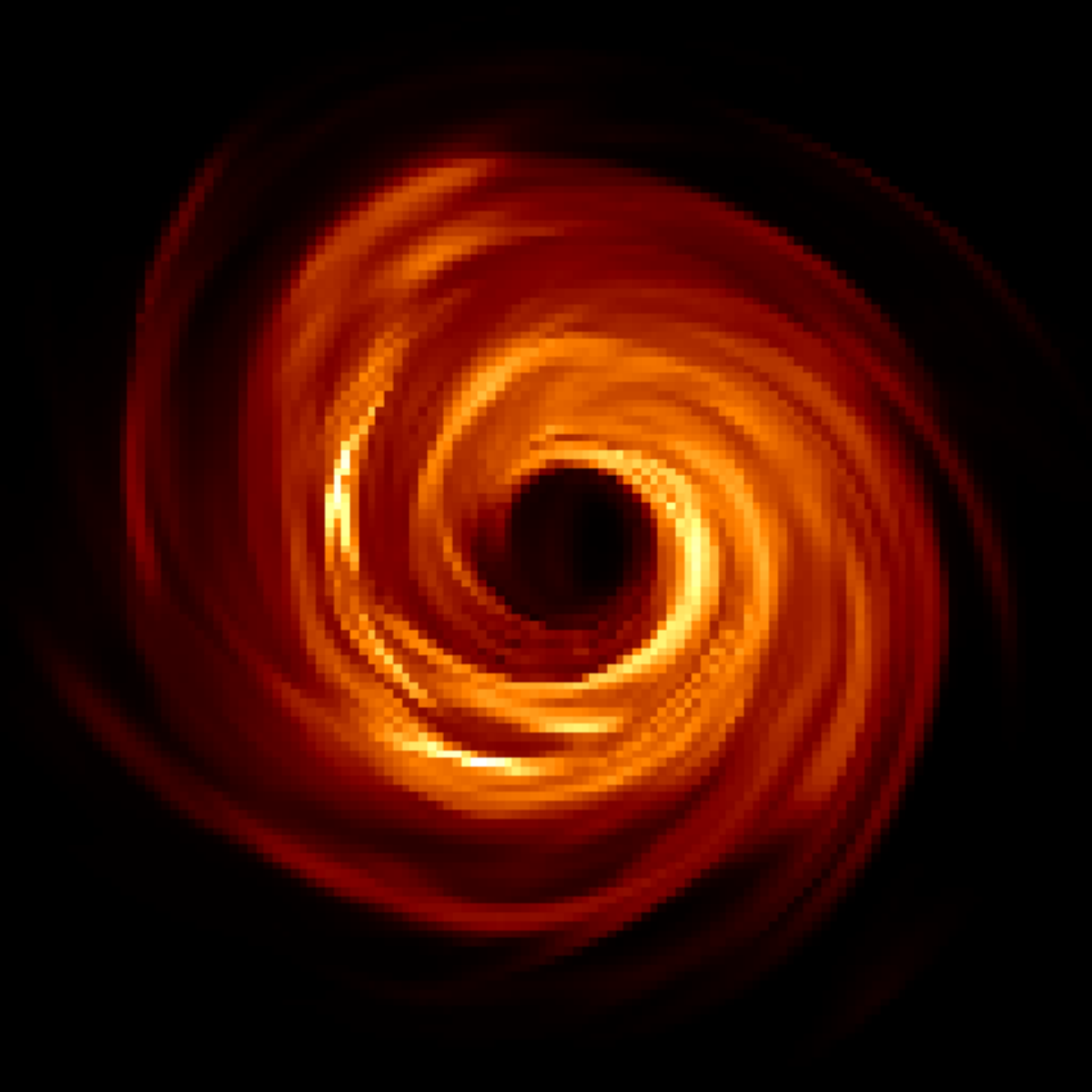}}
\resizebox{0.16\hsize}{!}{\includegraphics{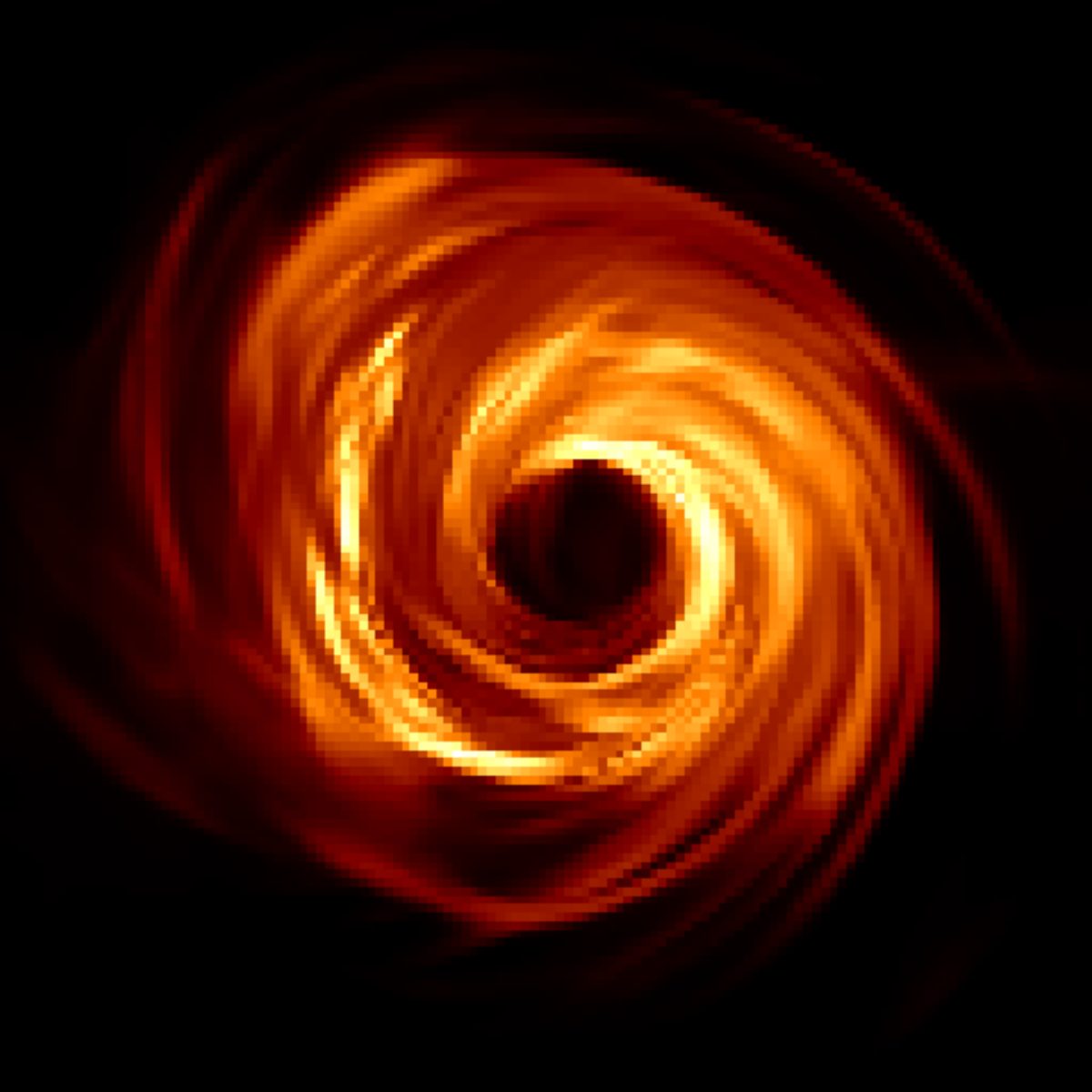}}
\resizebox{0.16\hsize}{!}{\includegraphics{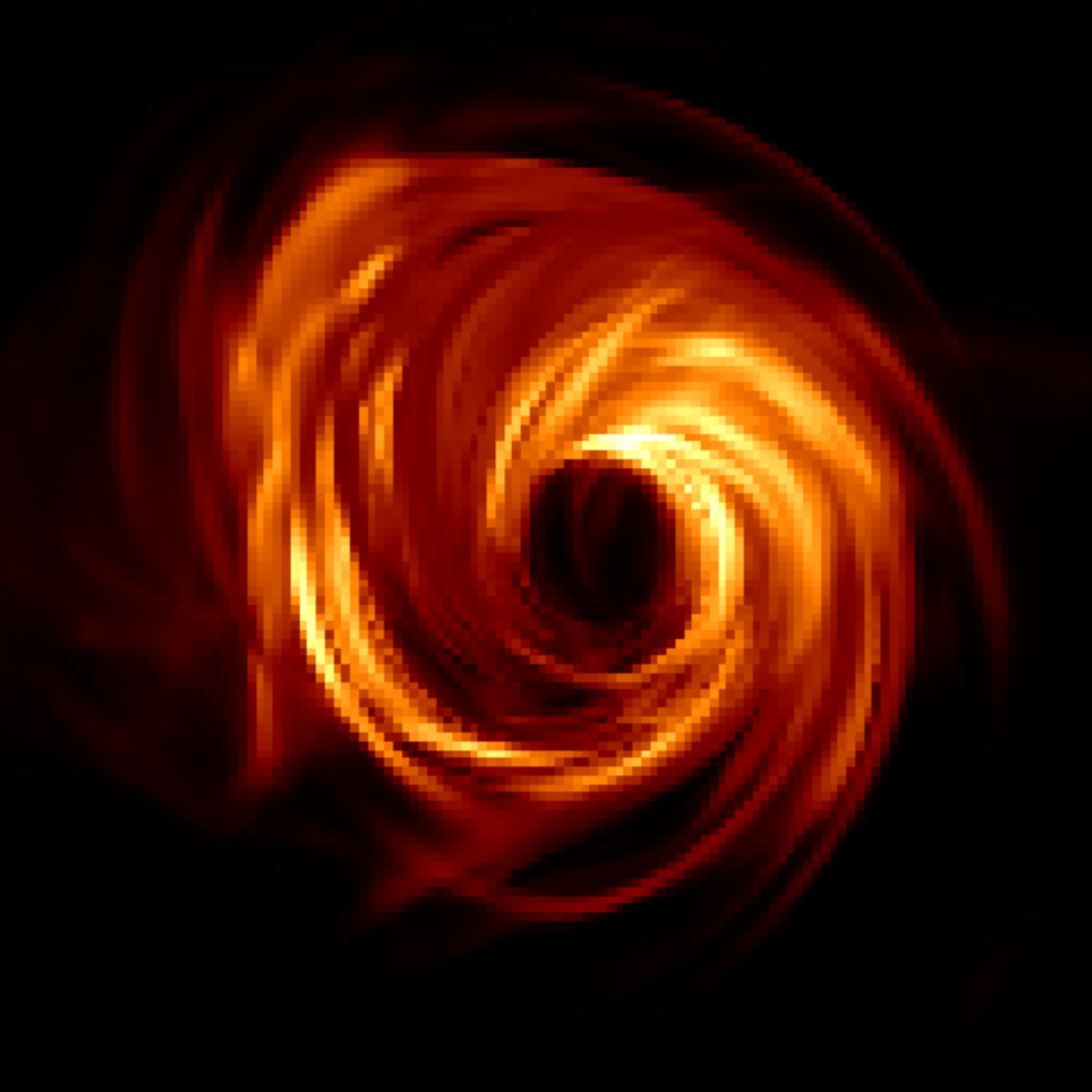}}
\resizebox{0.16\hsize}{!}{\includegraphics{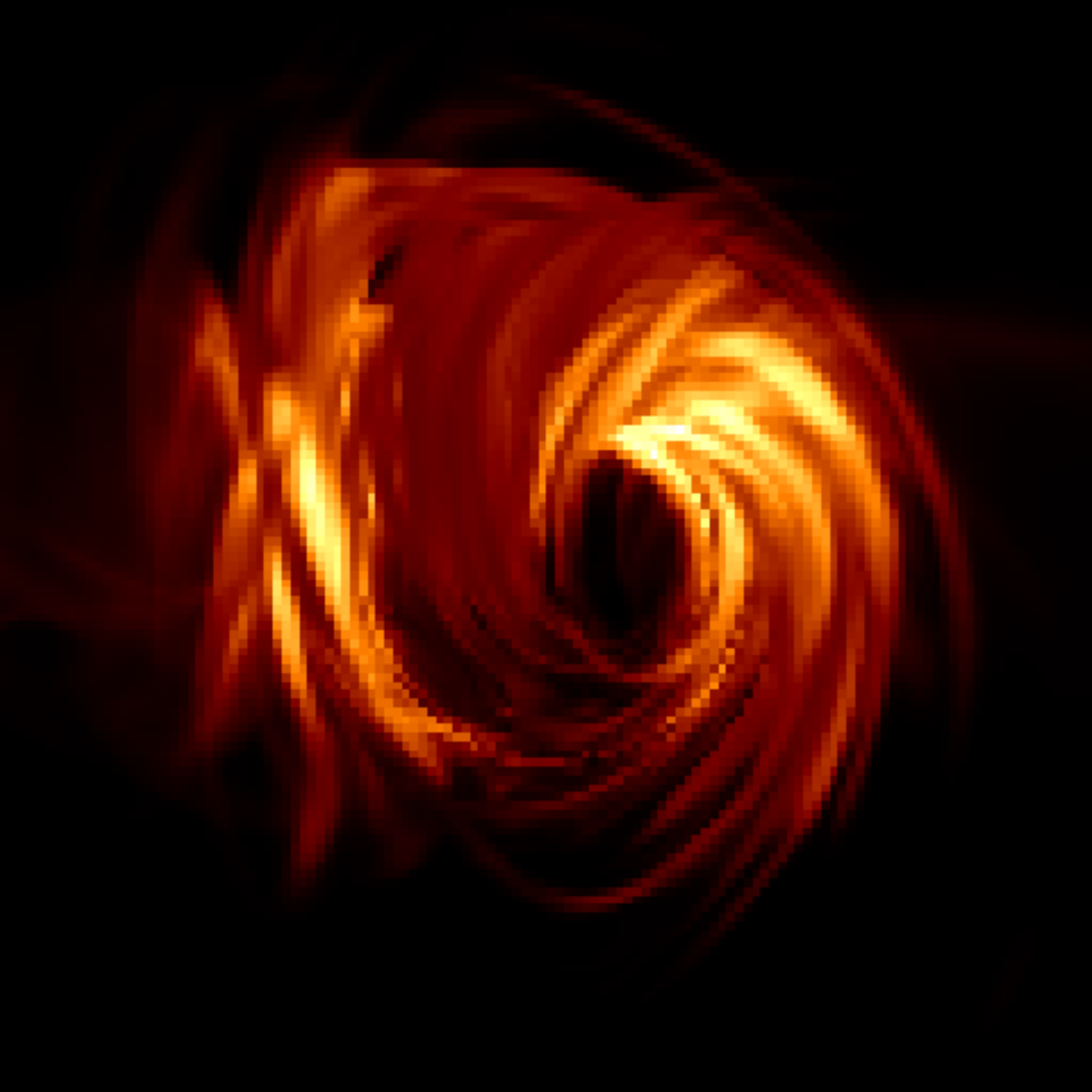}}
\resizebox{0.16\hsize}{!}{\includegraphics{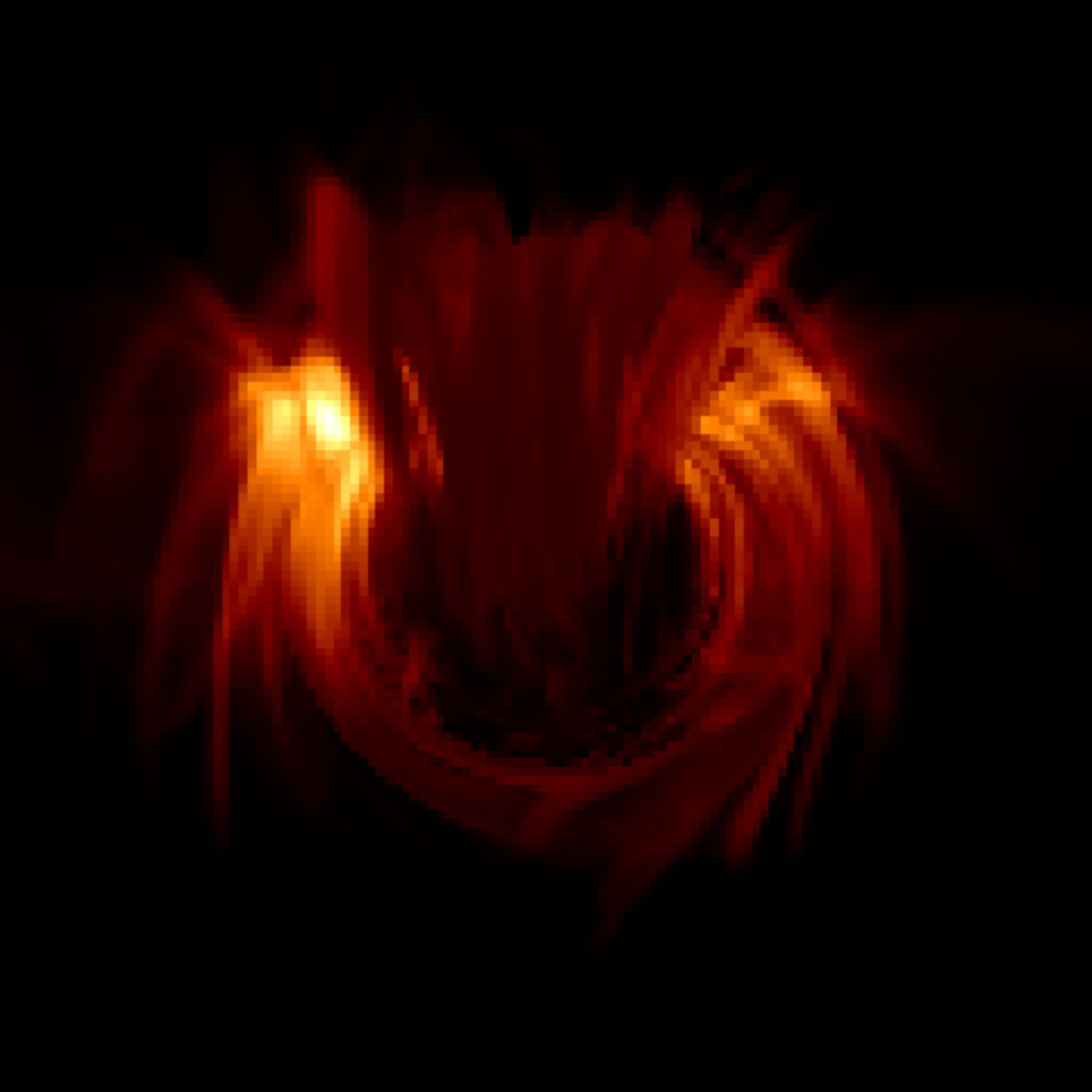}} 

\end{center}
\caption{Middle frames of the GRMHD movies of Sgr A* used to generate closure phases. Columns from left to right are inclinations 2, 18, 30, 45, 60, and 85 degrees, and rows from top to bottom are the $a_*=0.94$ disk model, the $a_*=0$ disk model, and the $a_*=0.94$ jet model.}
\label{fig:movies}
\end{figure*}

The middle frames of the three different simulations at all inclinations are shown in Figure \ref{fig:movies}. At low inclinations, the disk and jet model appear qualitatively similar, though in the jet model about half of the emission originates from the jet region. For the jet model, an elongated structure becomes apparent towards high inclinations, caused mainly by the optically thick disk covering the central part of the crescent. The bright spots are dominated by disk emission in this case.

In the following sections, we will introduce our analysis techniques using examples from the $a_*=0.94$ disk model. The results for this model will be compared to the other models in Section \ref{sec:comp}.

\subsection{Generating closure phases with noise}
\label{sec:clnoise}
EHT observations of the movies were simulated using the \texttt{eht-imaging} software\footnote{\url{https://github.com/achael/eht-imaging}} \citep{Chael2016}. The simulated array included the Submillimeter Array (SMA) on Hawaii, the Submillimeter Telescope (SMT) in Arizona, the Large Millimeter Telescope (LMT) in Mexico, the Atacama Large Millimeter Array (ALMA) in Chile, the South Pole Telescope (SPT), the Plateau de Bure Interferometer (PDB) in France, and the Pico Veleta Observatory (PV) in Spain. A Fourier transform was performed on each movie frame in order to obtain the visibilities as a function of $uv$-coordinates, depending on the antenna locations and Earth orientation at the time of observation. Thermal noise was added to the complex visibilities as jitter characterized by a circular Gaussian with standard deviation \citep{Thompson2001}
\begin{equation}
\sigma=\frac{1}{0.88}\sqrt{\frac{\mathrm{SEFD}_1\times\mathrm{SEFD}_2}{2\Delta\nu t_{\mathrm{int}}}}.
\end{equation}
Here, $\Delta\nu$ is the observing bandwidth, which was set to 4 GHz. $t_{\text{int}}$ is the integration time, which was set to the frame duration of 11.085 s. $\mathrm{SEFD}_1$ and $\mathrm{SEFD}_2$ are the system equivalent flux densities of the antennas on a particular baseline. The factor 1/0.88 results from losses due to 2-bit quantization of the signal. In the limit of high signal-to-noise, the resulting error in the visibility amplitudes $|\mathcal{V}|$ is $\sigma$, and the error in radians on the visibility phase is $\sigma/|\mathcal{V}|$. Closure phases were calculated as the phase of the bispectrum, which is the product of three complex visibilities $\mathcal{V}_{ij}=|\mathcal{V}_{ij}|e^{i\phi_{ij}}$ on a closed triangle of baselines:
\begin{equation}
\mathcal{B}=\mathcal{V}_{12}\mathcal{V}_{23}\mathcal{V}_{31}=|\mathcal{V}_{12}| |\mathcal{V}_{23}| |\mathcal{V}_{31}|e^{i\left(\phi_{12}+\phi_{23}+\phi_{31}\right)}=|\mathcal{B}|e^{i\phi_c}.
\end{equation}
Here, $|\mathcal{B}|$ is the triple amplitude and $\phi_c$ is the closure phase. The noise on the bispectrum is Gaussian in the limit of high signal-to-noise (SNR) \citep[][see Sec. \ref{sec:ehttrack} for a discussion on the validity regime of the high-SNR approximation]{Thompson2001}, and its standard deviation is calculated as
\begin{equation}
\sigma_{\mathcal{B}}\approx|\mathcal{B}|\sqrt{\frac{\sigma_{12}^2}{|\mathcal{V}_{12}|^2} + \frac{\sigma_{23}^2}{|\mathcal{V}_{23}|^2} + \frac{\sigma_{31}^2}{|\mathcal{V}_{31}|^2}}.
\label{eq:berr}
\end{equation}
The error in radians on the closure phase is then \citep{Chael2016}
\begin{equation}
\label{eq:cerr}
\sigma_c\approx\frac{\sigma_{\mathcal{B}}}{|\mathcal{B}|}.
\end{equation}

\subsection{Closure phase behavior}
\label{sec:behavior}
\begin{figure*}
\begin{center}
\resizebox{0.9\hsize}{!}{\includegraphics{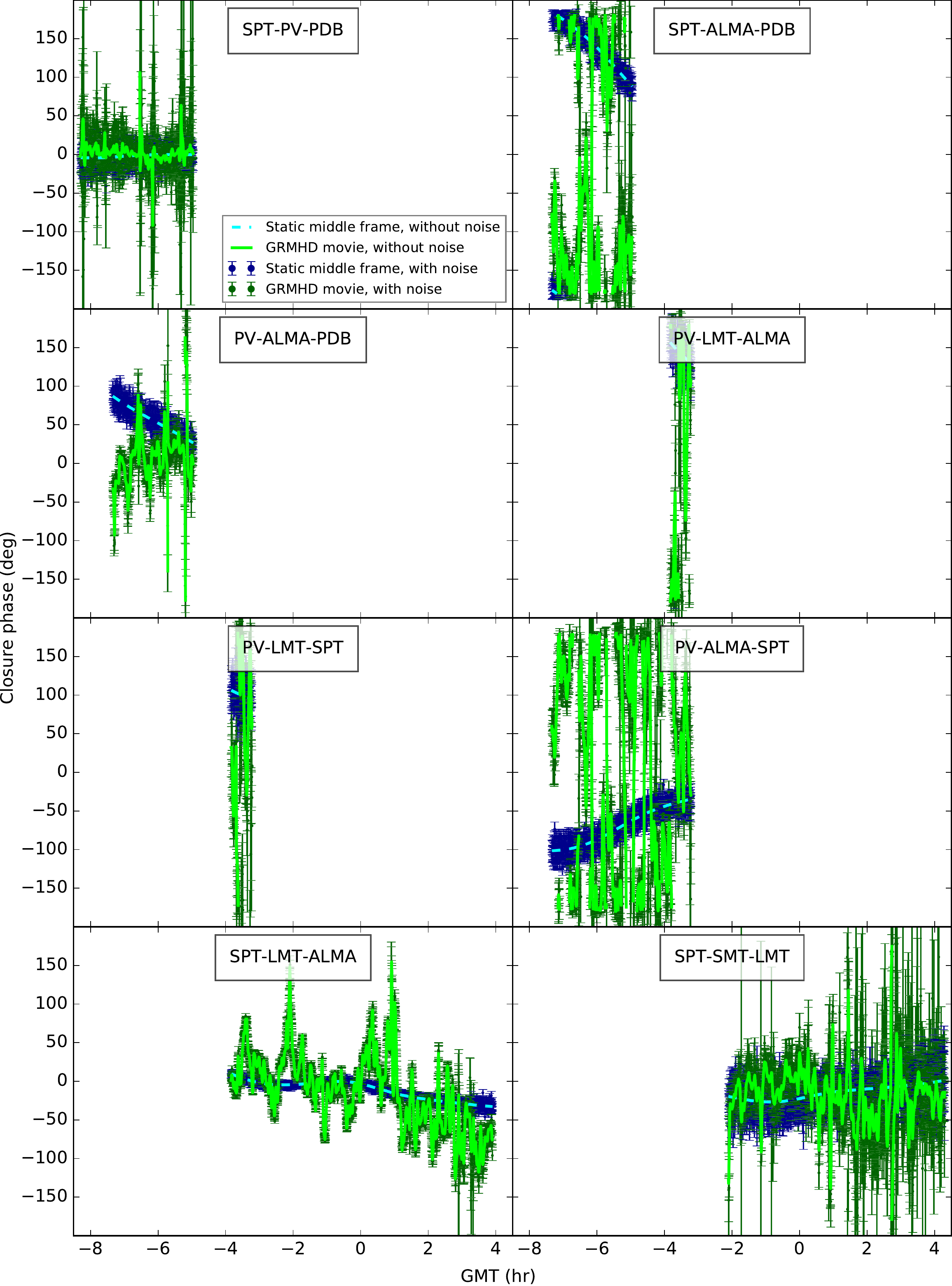}}
\end{center}
\caption{Closure phase as a function of time at the first half of the triangles with mutual visibility of Sgr A* for the $a_*=0.94$ disk movie (dark green with noise ($t_{\mathrm{int}}=11$ s, $\Delta\nu=4$ GHz), light green solid line without noise) and middle frame of the movie observed as a static source (dark blue with noise, light blue dashed line without noise) at an inclination of 2 degrees.}
\label{fig:cphase_time_1}
\end{figure*}

\begin{figure*}
\begin{center}
\resizebox{0.9\hsize}{!}{\includegraphics{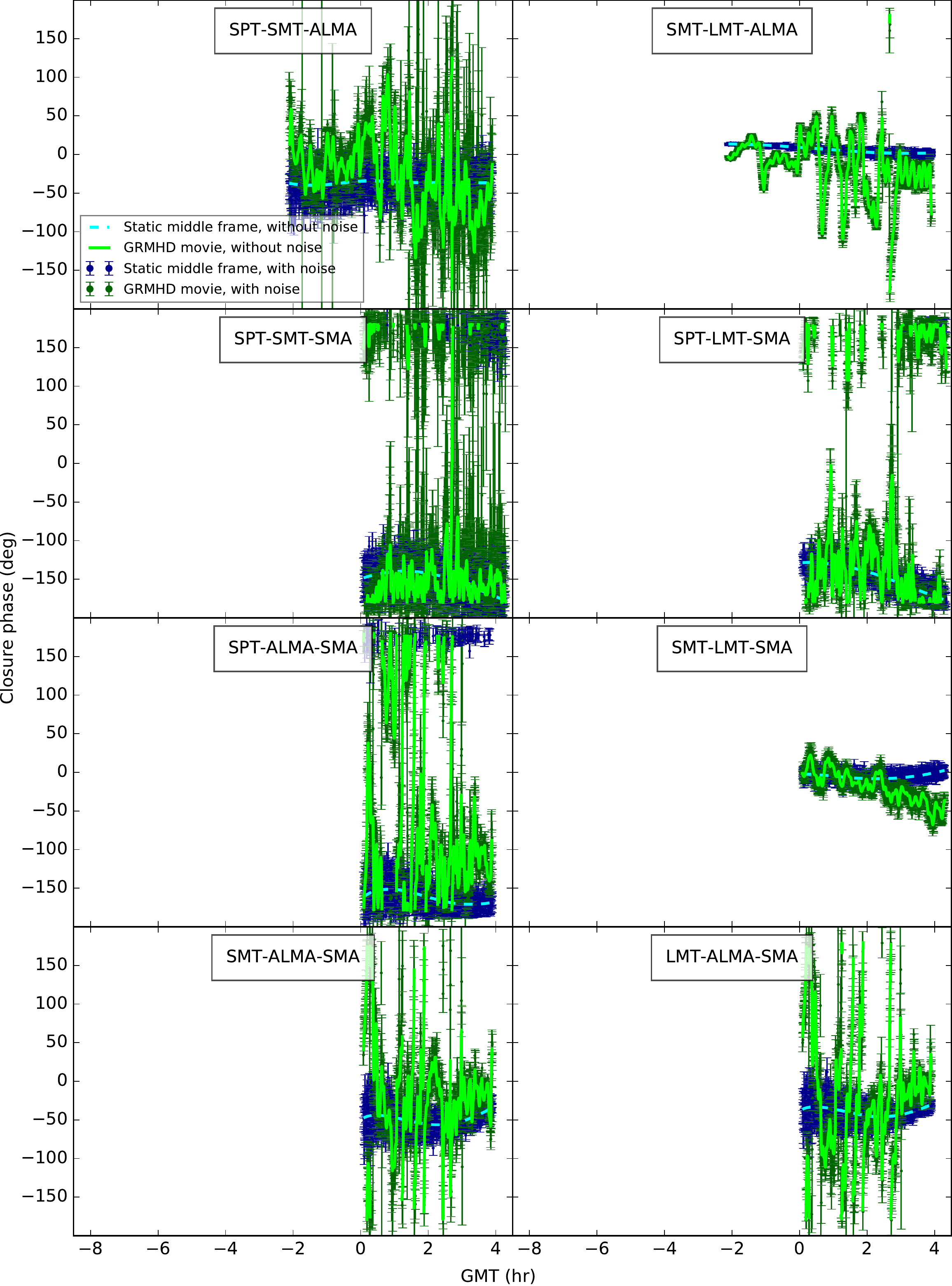}}
\end{center}
\caption{Same as Figure \ref{fig:cphase_time_1}, but for the second half of the triangles with mutual visibility of Sgr A*.}
\label{fig:cphase_time_2}
\end{figure*}

Figures \ref{fig:cphase_time_1} and \ref{fig:cphase_time_2} show closure phases from the $a_*=0.94$ disk movie at $2^{\circ}$ inclination as a function of time for all triangles with mutual visibility of Sgr A*. Since the light blue points represent the observation of the middle frame as a static source, all time variability in these data is caused by baseline evolution due to Earth rotation. For the dark blue points, there is additional variability from thermal noise. The light green points also contain variability from baseline evolution, but show fluctuations caused by source variability on top of that. Finally, the dark green points exhibit variability caused by the source, Earth rotation, and thermal noise. 

These plots already show that the source variability is much more rapid than variability caused by Earth rotation. The fluctuations are also larger. Similarly, \citet{Gold2017} have shown that intrinsic variability dominates over variability from Earth rotation for simulated polarimetric observations with the EHT for a variety of GRMHD simulations. The simulated closure phase behavior is strongly dependent on the triangle on which it is measured due to the noise properties of the antennas and the different Fourier components measured by the different baselines. In general, thermal noise will dominate the observed variability more strongly for longer baselines. These baselines probe the small scale source structure and thus measure small visibility amplitudes. Because these visibilities are closer to the origin of the complex plane, a given Gaussian noise fluctuation here will generally result in a larger phase fluctuation. For this array, the triangles SPT-LMT-ALMA and SMT-LMT-ALMA in particular seem to be very good candidates for measuring the closure phase variations caused by intrinsic source variability, due to the high sensitivity of interferometric baselines to the LMT and ALMA sites.

\begin{figure*}
\begin{center}
\resizebox{\hsize}{!}{\includegraphics{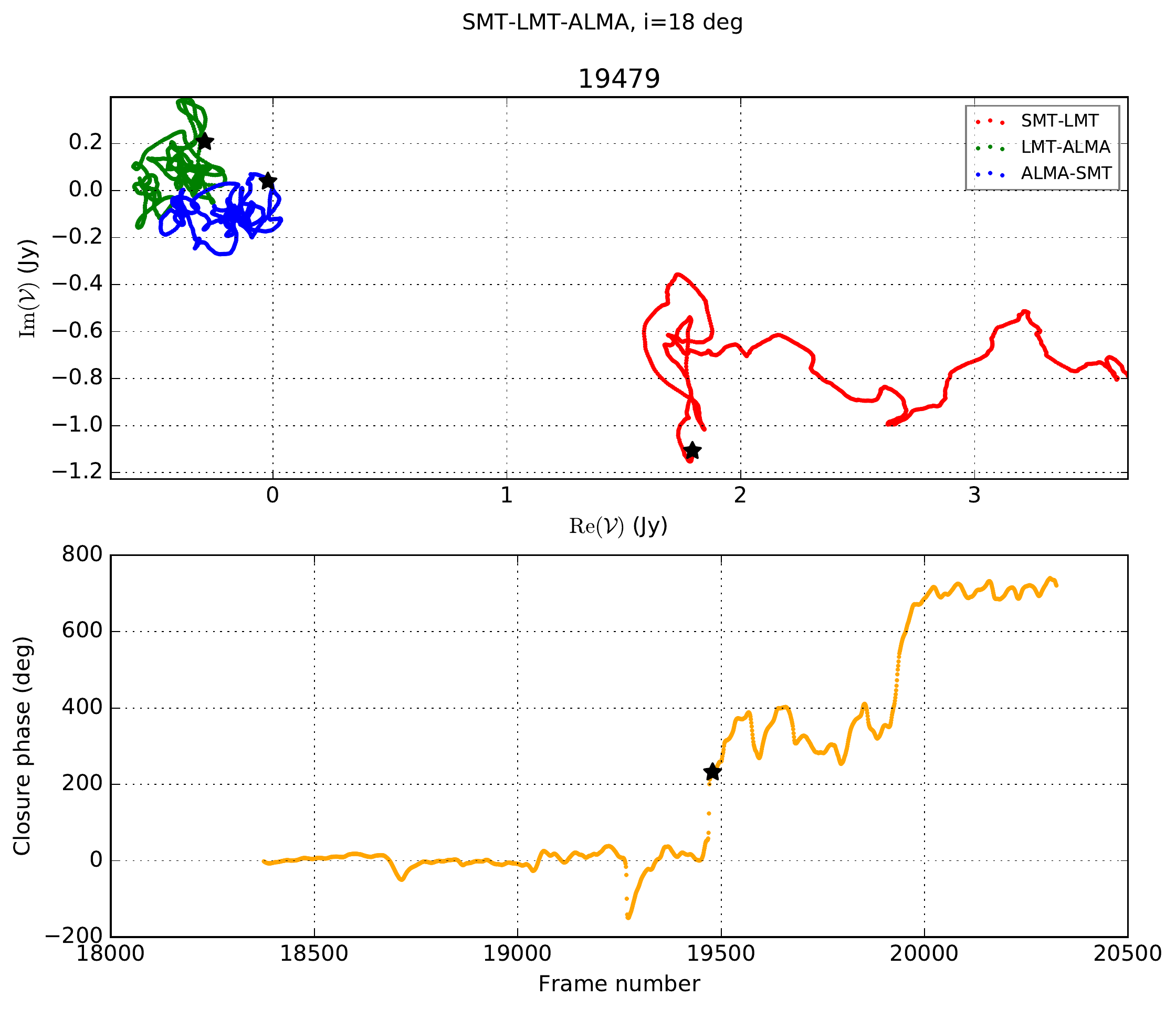}}
\end{center}
\caption{Movie frame of the complex visibilities on individual baselines (top) and the resulting unwrapped closure phase as a function of GRMHD frame number (bottom) at the SMT-LMT-ALMA triangle for the $a_*=0.94$ disk model with inclination 18 degrees. Earth rotation was included, but thermal noise was not. The black stars indicate the complex visibilities and closure phase of the current frame (19479). The different colors in the complex visibility plane indicate the complex visibilities of the previous frames at the different baselines. The frame spacing is 11 seconds. The closure phase thus swings by more than 100 degrees within half a minute around frame 19250. The rapid variations around frame 19250 and 19450 are caused by a close passage of the origin by the complex visibility at the ALMA-SMT baseline (blue). After the passage, the closure phase returns to approximately the same value (modulo 360 degrees). This figure is available online as an animation.}
\label{fig:vismov}
\end{figure*}

An interesting feature of the simulated closure phase tracks is the occurrence of wraps and jumps, as illustrated with an example in Figure \ref{fig:vismov}. These are manifestations of the same phenomenon: the visibility on one of the baselines (in this case ALMA-SMT) closely passing the origin of the complex plane due to some source fluctuation. A $\sim180$ degree closure phase flip, such as the ones around frame numbers 19250 and 19450, may occur if the visibility passes the origin so closely that the sampling time is too long to track its movement. As the visibility fluctuates within a certain area of the complex plane (the source structure does not change too drastically), the closure phase value returns to approximately the same value (modulo 360 degrees) after passing the origin. 

\section{Framework for studying intrinsic closure phase variability}
\label{sec:Nmetric}

In this section, a framework is set up to quantify intrinsic closure phase variability in a measured closure phase track containing variations from image (source and scattering) and observational (thermal noise and baseline evolution) variability. We construct a metric $\mathcal{Q}$ that represents the fraction of variability in a given closure phase track that is not due to thermal noise. Variability due to Earth rotation can be taken out by detrending the data set on long time scales. The constructed metric is based on circular statistics. Section \ref{sec:cstat} introduces some of the techniques involved in circular statistics, and in particular describes a way to estimate the spread on a circular data set based on \citet[][]{Mardia2000}. Section \ref{sec:q} introduces the metric $\mathcal{Q}$.

\subsection{Circular statistics}
\label{sec:cstat}
Closure phases are an example of circular data, which can be described as points on the unit circle characterized by an angle $\theta$. The periodic nature of circular data demands a different statistical analysis approach than for linear data. For example, naively calculating a linear average of circular data will not give a representative result. Angle measurements at e.g. $150^{\circ}$ and $-170^{\circ}$ would average to $-10^{\circ}$ instead of the desired $170^{\circ}$, as $+180^{\circ}$ and $-180^{\circ}$ are equivalent phases. 

\begin{figure}[h!]
\begin{center}
\resizebox{0.98\hsize}{!}{\includegraphics{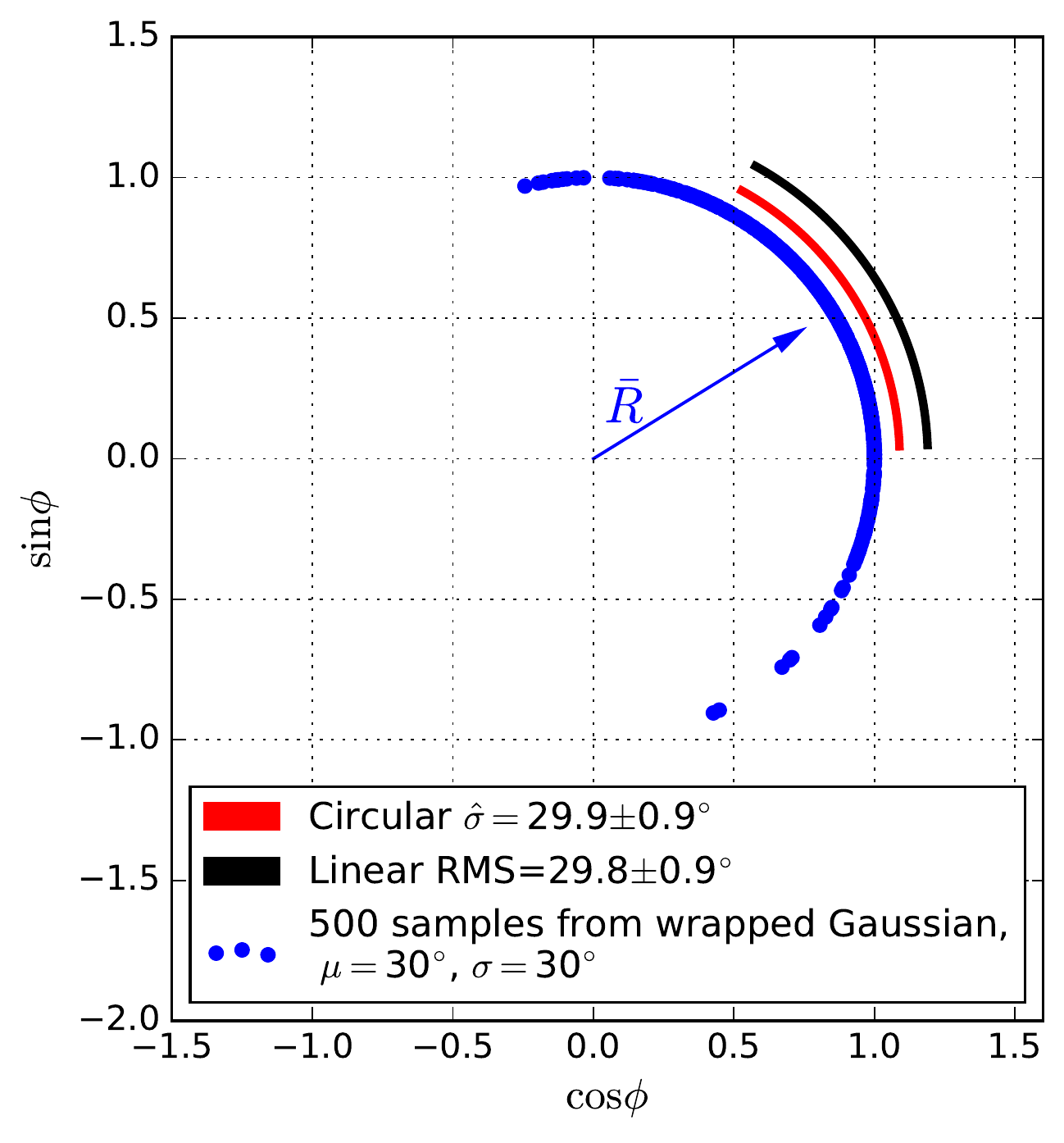}}
\resizebox{0.98\hsize}{!}{\includegraphics{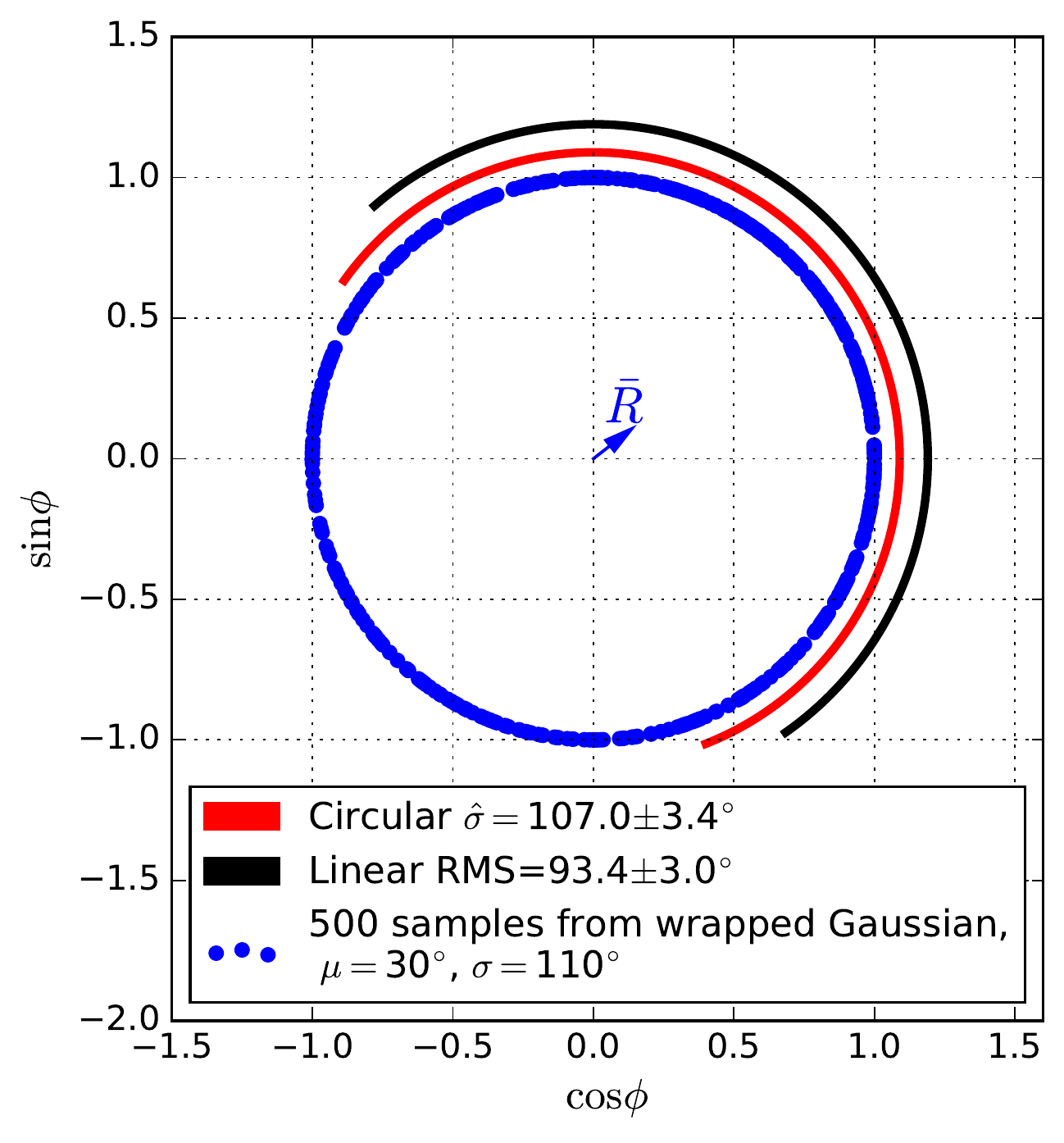}}
\end{center}
\caption{Sines and cosines of 500 phases sampled from a wrapped normal distribution with mean $\mu=30^{\circ}$ and standard deviations $\sigma=30^{\circ}$ (upper panel) and $\sigma=110^{\circ}$ (lower panel. The vectors point to the averages of the sines and cosines, their angles with respect to the point (1,0) defining the average phase. The length $\bar{R}$ of this vector is a measure for how concentrated the points are. The circular standard deviation can be estimated as $\hat{\sigma}=\sqrt{-2\ln\bar{R}}$ (Eq. \ref{eq:sigma}), and is indicated with a red arc. The black arc indicates the linear root-mean-square of the sampled phases. The estimates of the linear RMS and $\hat{\sigma}$ have an accuracy limited by the number of data points. The errors have been calculated with Equation \ref{eq:sigmaerr}.}
\label{fig:circavg}
\end{figure}

In circular statistics, the correct quantity to use is the mean direction $\bar{\theta}$, which is the direction of the center of mass of the unit vectors. In Cartesian coordinates, the center of mass lies at 
\begin{equation}
\mathbf{\bar{x}} = (\bar{C},\bar{S}) = \left(\frac{1}{n}\sum_{j=1}^n \cos\theta_j, \frac{1}{n}\sum_{j=1}^n \sin\theta_j\right) = \bar{R}(\cos\bar{\theta}, \sin\bar{\theta}),
\end{equation}
where $n$ is the number of measurements, and 
\begin{equation}
\label{eq:Rbar}
\bar{R}=\left|\left<e^{i\theta_j}\right>\right|=\sqrt{\bar{C}^2+\bar{S}^2}=\sqrt{\left(\frac{1}{n}\sum_{j=1}^n \cos\theta_j\right)^2 + \left(\frac{1}{n}\sum_{j=1}^n \sin\theta_j\right)^2}
\end{equation}
is the mean resultant length. The mean direction is then
\begin{equation}
\bar{\theta} = \mathrm{atan2}\left(\bar{C}, \bar{S}\right).
\end{equation}
This process of averaging sines and cosines is illustrated with an example of phases sampled from a wrapped normal distribution (see also Eqs. \ref{eq:gauss} and \ref{eq:gausswrap}) in Figure \ref{fig:circavg}. The mean resultant length $\bar{R}$ is a measure for the concentration of the points: a uniform angular distribution\footnote{Or, in fact, any distribution where for each angle $\theta$ there is an angle $\theta+\pi$.} would give $\bar{R}\rightarrow 0$ as $n\rightarrow\infty$, while $\bar{R}=1$ would indicate that every measurement gave the same value. $V=1-\bar{R}$ is therefore known as the circular variance.

For investigating closure phase variability, it would be useful to have a more intuitive measure for the spread of the circular data. In linear statistics, one could assume that the data is drawn from a normal distribution with mean $\mu$ and compute the variance to estimate its standard deviation $\sigma$. In order to do this with circular data, the normal distribution
\begin{equation}
\label{eq:gauss}
f(\theta; \mu, \sigma) = \frac{1}{\sigma\sqrt{2\pi}}\exp\left(-\frac{(\theta-\mu)^2}{2\sigma^2}\right),
\end{equation}
must be wrapped around the unit circle. To do this, $f(\theta; \mu, \sigma)$ can be split into segments of length $2\pi$, which are then summed to obtain the wrapped normal distribution
\begin{equation}
\label{eq:gausswrap}
f_w(0\leq\theta<2\pi; \mu, \sigma)=\sum_{k=-\infty}^{\infty}f(\theta+2\pi k; \mu, \sigma).
\end{equation}
In order to learn how to estimate the standard deviation $\sigma$ from the data, it is useful to look at the sample moments $\bar{C} = \frac{1}{n}\sum_{j=1}^n \cos\theta_j$ and $\bar{S} = \frac{1}{n}\sum_{j=1}^n \sin\theta_j$. Taken together, they define the first trigonometric moment about the zero direction
\begin{equation}
m_1=\bar{C}+i\bar{S}=\bar{R}e^{i\bar{\theta}}.
\end{equation}
This can be extended to define the $p$th trigonometric moment about the zero direction as 
\begin{equation}
m_p=\frac{1}{n}\sum_{j=1}^n \cos (p\theta_j)+\frac{i}{n}\sum_{j=1}^n \sin (p\theta_j)=\bar{R}_pe^{i\bar{\theta}_p},
\end{equation}
where $\bar{R}_p$ and $\bar{\theta}_p$ are the mean resultant length and direction of the $p\theta_j$.

These trigonometric moments are strongly related to the Fourier transform of the underlying population distribution. The Fourier transform $\phi(t)$ of a probability density function $f(x)$ of a random variable $x$ in linear statistics is equal to the expectation value $\mathrm{E}$ of the quantity $e^{itx}$ according to $f(x)$:
\begin{equation}
\phi(t)=\mathrm{E}[e^{itx}]=\int_{-\infty}^{\infty}f(x)e^{itx}dx.
\end{equation}
$\phi(t)$ is also known as the characteristic function of $f(x)$ 

In circular statistics, $t$ must be confined to integer values. For a non-integer $t$, the expectation values $\mathrm{E}[e^{it\theta}]$ and $\mathrm{E}[e^{it(\theta+2\pi)}]$ are generally not equal, which they should be as the phases are equivalent. Thus, for a circular distribution the characteristic function becomes the sequence of complex numbers
\begin{equation}
\Phi_p=\mathrm{E}[e^{ip\theta}], \quad p=0,\pm1, \pm2, \ldots.
\end{equation}
For $p\geq0$, this can also be written as
\begin{equation}
\Phi_p=\mathrm{E}[\cos (p\theta)]+i\mathrm{E}[\sin (p\theta)]=\rho_pe^{i\mu_p}, \quad \rho_p\geq0. 
\end{equation}
Note that the $\Phi_p$ are the population versions of the moments $m_p$. Thus, by calculating the $\Phi_p$ of the wrapped normal distribution in terms of $\mu$ and $\sigma$, we can equate them to the measured moments $m_p$ to obtain an estimate for them. This way of estimating distribution parameters from measured data is known as the method of moments \citep[e.g.][]{Bowman2004}.

The $\Phi_p$ for a wrapped distribution are equal to the Fourier transform of the unwrapped distribution $\phi(t)$ evaluated at integer values $p$:
\begin{equation}
\begin{split}
\Phi_p&=\mathrm{E}[e^{ip\theta}]=\int_0^{2\pi} f_w(\theta)e^{ip\theta}d\theta=\int_0^{2\pi}\sum_{k=-\infty}^\infty f(\theta+2\pi k)e^{ip\theta}d\theta\\&=\sum_{k=-\infty}^\infty\int_{2\pi k}^{2\pi(k+1)}f(\theta)e^{ip\theta}d\theta=\int_{-\infty}^{\infty}f(\theta)e^{ip\theta}d\theta=\phi(p).
\end{split}
\end{equation}
The Fourier transform of the normal distribution is
\begin{equation}
\phi(t)=e^{i\mu t-t^2\sigma^2/2},
\end{equation}
so that for the wrapped normal distribution
\begin{equation}
\Phi_p=e^{i\mu p-p^2\sigma^2/2}, \quad p=0,\pm1, \pm2, \ldots,
\end{equation}
and thus
\begin{equation}
\Phi_1=\mathrm{E}[e^{i\theta}] = e^{i\mu-\sigma^2/2}=e^{-\sigma^2/2}e^{i\mu}.
\end{equation}
Equating this to the population version $m_1=\bar{R}e^{i\bar{\theta}}$ gives estimates of $\mu$ and $\sigma$: 
\begin{equation}
\label{eq:sigma}
\hat{\mu}=\bar{\theta}, \quad \hat{\sigma}=\sqrt{-2\ln\bar{R}},
\end{equation}
where the circumflex indicates that $\hat{\mu}$ and $\hat{\sigma}$ are estimates of $\mu$ and $\sigma$, respectively.

Figure \ref{fig:circavg} shows that when one takes a sample from a wrapped normal distribution, $\hat{\sigma}$ provides a better estimate of the standard deviation than the conventional linear RMS. This effect becomes more important for large standard deviations since the tail of the Gaussian that extends beyond $\pm180^{\circ}$ then contains a large part of its total area, and the number of data points at a given angle contains contributions from multiple parts of the original Gaussian that has been wrapped around the unit circle. The linear RMS underestimates the standard deviation because the domain is restricted to [$-180^{\circ}$, $180^{\circ}$], and it does not account for contributions from wrappings. For $\sigma=110^{\circ}$, the true value is well beyond the error of the linear RMS estimate, but within the error of the circular standard deviation $\hat{\sigma}$.

\subsection{$\mathcal{Q}-metric$}
\label{sec:q}
Measured closure phases fluctuate from thermal noise in addition to changes from an evolving image or baseline. We now introduce a useful metric to characterize the amount of intrinsic variability in a measured time series of $n$ closure phases $\theta_j$ with measurement uncertainties $\epsilon_j$. This metric compares the measured total variation in the closure phase track $\hat{\sigma}^2$ to the variation expected from thermal noise $\tilde{\epsilon}^2$. 
\subsubsection{Definition}
The $\mathcal{Q}$-metric is defined as
\begin{equation}
\label{eq:Nmetric}
\begin{split}
\mathcal{Q}&\equiv\frac{\hat{\sigma}^2-\tilde{\epsilon}^2}{\hat{\sigma}^2}, \text{ where}\\  \tilde{\epsilon} &\equiv \left \langle \hat{\sigma}_{\epsilon}\right \rangle, \quad \hat{\sigma}_\epsilon\equiv\sqrt{-2\mathrm{ln}\bar{R}_{\epsilon}}, \quad \bar{R}_{\epsilon}\equiv\left| \frac{1}{n} \sum_{j=1}^{n} e^{i \delta \theta_j} \right|,
\end{split}
\end{equation}
where $\delta\theta_j$ is a zero-mean Gaussian random variable with standard deviation $\epsilon_j$. $\tilde{\epsilon}$ is the expected spread on the closure phases due to thermal noise. In practice, it can be calculated from the measurement uncertainties $\epsilon_j$ using a Monte Carlo approach. This entails generating $N$ ($N=10^5$ for this work) thermal noise realizations and calculating the spread $\hat{\sigma}_{\epsilon}$ for each of these. A thermal noise realization is generated by taking a sample data point from each of the $n$ zero-mean Gaussians with standard deviation equal to the error bars $\epsilon_j$ of each of the $n$ measured closure phases. For each realization of $n$ scattered data points, the spread $\hat{\sigma}_{\epsilon}$ is calculated analogously to the spread $\hat{\sigma}$ of the closure phases themselves. $\tilde{\epsilon}$ is then the average of the $N$ $\hat{\sigma}_{\epsilon}$. 

The metric $\mathcal{Q}$ thus subtracts the spread of the measurements expected from thermal noise $\tilde{\epsilon}^2$ from the total observed spread $\hat{\sigma}^2$. If all variations in the measurements are due to thermal noise, $\mathcal{Q}$ is expected to be zero, whereas if the fluctuations are dominated by intrinsic variability, $\mathcal{Q}$ may increase up to a value of 1 in the noiseless limit. $\mathcal{Q}$ is the fraction of the closure phase spread that cannot be accounted for by thermal noise.

Although we have constructed this metric using Gaussian errors, as are expected for measurements with SNR > 1, it can also be applied for weaker measurements and non-Gaussian errors. In this case, $\tilde{\epsilon}$ must simply be defined according to the appropriate distribution function for the sampled closure phases, so that $\mathcal{Q}$ would still be zero if all variability was due to the measurement uncertainties only. This is important because there is not a single correct prescription for the treatment of closure phase statistics, as they depend on the details of how the closure phases are formed. Non-closing errors due to e.g. bandpass variations may be present in closure phase data, leading to an underestimate of $\tilde{\epsilon}$. If these errors can be estimated, which is on-gong work \citep[e.g.][]{Blackburn2017}, they could be included in the error distribution used to determine $\tilde{\epsilon}$. In this work, we assume that non-closing errors are negligible.

\subsubsection{Error estimation}
\label{sec:error}
For an error analysis of $\mathcal{Q}$, consider a closure phase data set with thermal jitter $\tilde{\epsilon}$ and intrinsic circular standard deviation $\sigma_S$, so that
\begin{equation}
\sigma^2=\sigma_S^2+\tilde{\epsilon}^2.
\end{equation}
In this case,
\begin{equation}
\langle\mathcal{Q}\rangle=\frac{1}{1+\left(\frac{\tilde{\epsilon}}{\sigma_S}\right)^2}=\frac{1}{1+S^{-2}},\quad\text{where } S\equiv\frac{\sigma_S}{\tilde{\epsilon}}.
\end{equation}
The data variance $\sigma^2$ can only be estimated within a standard deviation limited by the number of samples $n$, so that its uncertainty
\begin{equation}
\label{eq:sigmaerr}
\delta\hat{\sigma}^2=\sqrt{\frac{2}{n}}\hat{\sigma}^2=\sqrt{\frac{2}{n}}\left(\sigma_S^2+\tilde{\epsilon}^2\right).
\end{equation}
$\tilde{\epsilon}$ can be estimated to high precision with our Monte Carlo approach. With $N\gg n$, the error on $\tilde{\epsilon}$ will be negligible compared to the error on $\hat{\sigma}$, assuming that the error bars indeed correctly represent the thermal noise. The error on $\mathcal{Q}$ can then be calculated from standard error propagation as:
\begin{equation}
\begin{split}
\label{eq:nerr}
\delta\mathcal{Q}&= \left\vert\frac{\partial\mathcal{Q}}{\partial\hat{\sigma}^2}\right\vert\delta\hat{\sigma}^2=\sqrt{\frac{2}{n}}\frac{\tilde{\epsilon}^2}{\hat{\sigma}^4}\left(\sigma_S^2+\tilde{\epsilon}^2\right)=\sqrt{\frac{2}{n}}\frac{1}{1+S^2}\\&=\sqrt{\frac{2}{n}}(1-\mathcal{Q}).
\end{split}
\end{equation}
This expression shows that there is a natural tendency for $\delta\mathcal{Q}$ to decrease with increasing $n$ and $\mathcal{Q}$. 

Note that this simple expression for $\delta\mathcal{Q}$ is only strictly valid if both $\hat{\sigma}^2$ and $\tilde{\epsilon}^2$ do not change across the closure phase track. If the time series is divided into $s$ segments, with a single segment consisting of $n_i$ measurements characterized by intrinsic scatter $\hat{\sigma}_{S, i}$ and noise scatter $\bar{\epsilon_i}$, the total $\hat{\sigma}^2$ and $\tilde{\epsilon}^2$ generalize to 
\begin{equation}
\label{eq:sigmagen}
\hat{\sigma}^2=\frac{1}{n}\sum_{i=1}^sn_i\left(\sigma_{S, i}^2+\tilde{\epsilon}_i^2\right),
\end{equation}
and
\begin{equation}
\tilde{\epsilon}^2=\frac{1}{n}\sum_{i=1}^sn_i\tilde{\epsilon}_i^2,
\end{equation}
so that
\begin{equation}
\label{eq:dqgen}
\delta\mathcal{Q}=\frac{\sqrt{2}}{n}\frac{\tilde{\epsilon}^2}{\hat{\sigma}^4}\sqrt{\sum_{i=1}^sn_i\left(\sigma_{S,i}^2+\tilde{\epsilon}_i^2\right)^2}.
\end{equation}

Alternatively, $\mathcal{Q}$ and its error may be estimated with bootstrapping \citep{Efron1979}. The calculation of $\mathcal{Q}$ outlined above relies on the assumption that the data are relatively homogeneous (the closure phase fluctuates around an average value). However, it may occur that the distribution of the closure phases and their errors may not fluctuate around an average value, which can happen when it wraps around a lot on large triangles or when the time on-source is very short. In this case, the bootstrapping approach may be a good alternative for calculating $\mathcal{Q}$ and its error.

Bootstrapping uses a Monte Carlo approach to estimate the accuracy of sample estimates from the empirical data distribution. The metric $\mathcal{Q}$ is calculated from a set of measured closure phases. In principle, this set of measurements can produce only one value of $\mathcal{Q}$. However, in order to get an idea of the error on $\mathcal{Q}$, one would like to have multiple sets of measurements for which $\mathcal{Q}$ can be calculated. Bootstrapping entails generating new data sets from the same population by randomly resampling data points from the empirical distribution. Since a sample of $n$ measurements is drawn from an empirical distribution constructed of $n$ measurements, the distribution is sampled with replacement, which means that an individual data point may be sampled multiple times. If $n$ is sufficiently large, this approach ensures that the probability of getting two identical resampled data sets is very small. For each resampled data set, the statistic of interest ($\mathcal{Q}$) can be calculated, and its error can be estimated from the distribution of the bootstrapped values.
 
\subsubsection{Detrending for baseline evolution}
The metric $\mathcal{Q}$ does not account for closure phase variations due to Earth rotation. As Earth rotation introduces a slow variation of the observed closure phase (Figs. \ref{fig:cphase_time_1} and \ref{fig:cphase_time_2}), its influence on the metric $\mathcal{Q}$ may be mitigated by detrending the data. There are several methods of doing this. The time series may be divided in shorter segments of a certain size, so that the influence of Earth rotation across a segment is small, whereas the rapid intrinsic fluctuations are still large. $\mathcal{Q}$ and $\delta\mathcal{Q}$ can then be calculated using Equations \ref{eq:sigmagen}-\ref{eq:dqgen}.

In addition, the data within a segment may be detrended by subtracting a linear fit, or taking the closure phase differences at a certain time lag. The latter method was found to do the best job at bringing $\mathcal{Q}$ to zero for closure phases simulated from a static source model, while hardly affecting $\mathcal{Q}$ for a variable source model. The differences $\Delta(\theta_i,\theta_{i+l})$ at lag $l$ can be calculated as
\begin{equation}
\Delta(\theta_j,\theta_{j+l})=\mathrm{min}(\theta_{j+l}-\theta_j, \theta_{j+l}-(\theta_j+360), \theta_{j+l}-(\theta_j-360)).
\end{equation}
$l$ should be set to a time scale long enough to see a change in the closure phase value due to baseline evolution through thermal noise, and to preserve the rapid variability that is due to the source. $l$ should not be set too large as the number of data points is reduced by $l$, and a large lag would not be able to tease out the trend across the track. In the case of the $a_*=0.94$ disk model, a lag of 25 data points or 277 seconds, which is about half an ISCO period, was found to work best. Interstellar scattering is expected to become important on time scales much longer than the detrending window.

\begin{figure}
\begin{center}
\resizebox{\hsize}{!}{\includegraphics{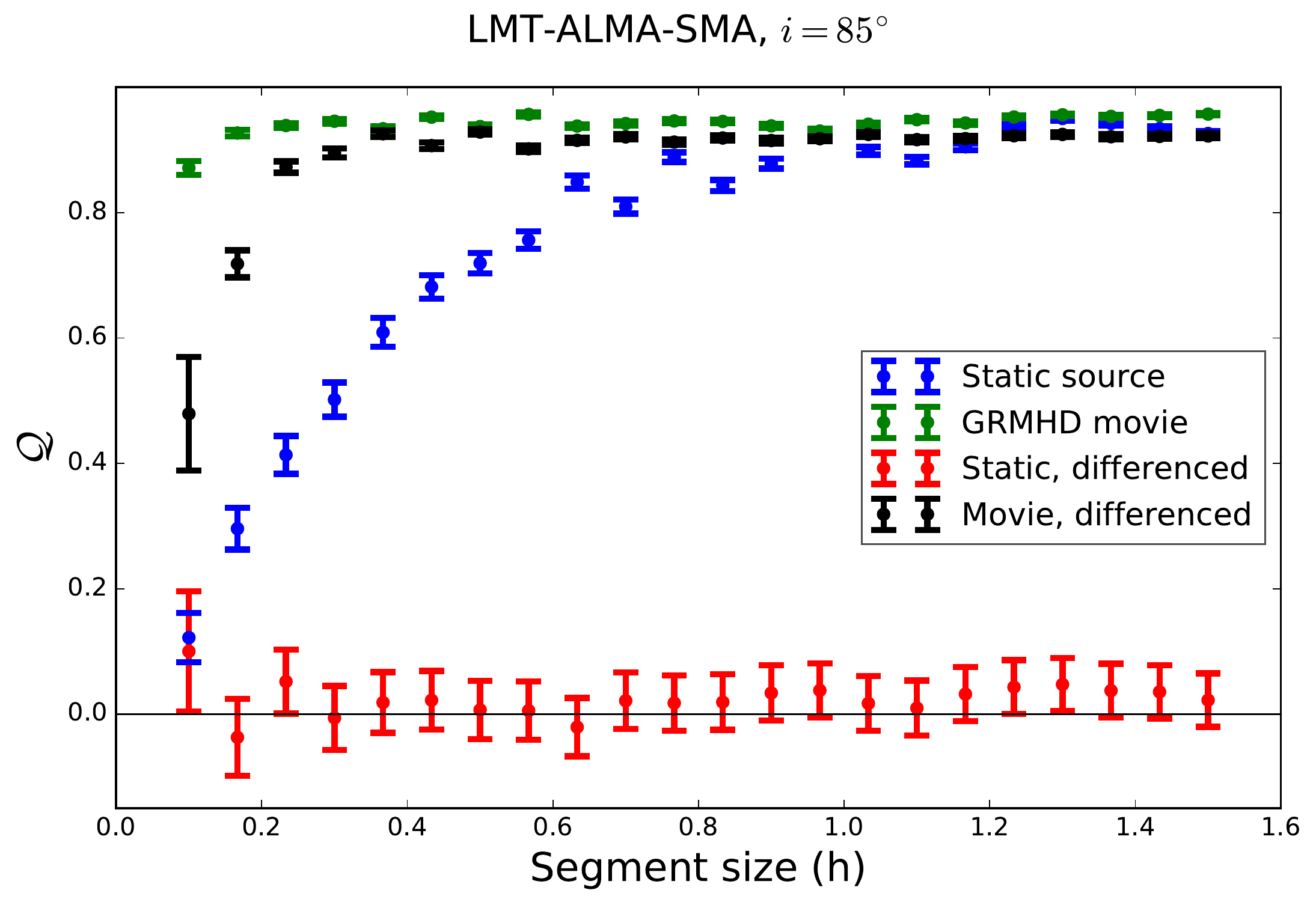}}
\end{center}
\caption{Metric values for the $a_*=0.94$ GRMHD movie at inclination 85 degrees (green) and middle frame of the movie observed as a static source (blue) as a function of segment size at the triangle LMT-ALMA-SMA, with thermal noise characterized by $t_{\mathrm{int}}=11$ s and $\Delta\nu=4$ GHz. The black (movie) and red (static) points are the metric values resulting from differencing the segmented time series at a lag of 25 data points (277 seconds) before calculating $\mathcal{Q}$. The error bars were calculated with Equation \ref{eq:dqgen}.}
\label{fig:metric_seg}
\end{figure}

Figure \ref{fig:metric_seg} shows the metric values as a function of segment size for a variable and static source model, with and without application of differencing. For the non-detrended data of the static source observation (blue), $\mathcal{Q}$ increases as a function of segment size. This could be expected since $\hat{\sigma}$ will be larger for a larger segment if the closure phase is strictly increasing or decreasing throughout the segment, which is often the case due to the slow baseline evolution. For the movie observation (green), $\mathcal{Q}$ usually plateaus at a segment size shorter than $\sim 0.5$ hr, indicating that most of the closure phase variability occurs within $\sim 0.5$ hr time scales. Taking closure phase differences does not affect $\mathcal{Q}$ much for the movie observation (black) as it is dominated by rapid fluctuations, but the values for the static observation (red) are drawn to zero, even for large segment sizes. This makes the metric a robust way of detecting and characterizing intrinsic source variability.

\begin{figure}
\begin{center}
\resizebox{\hsize}{!}{\includegraphics{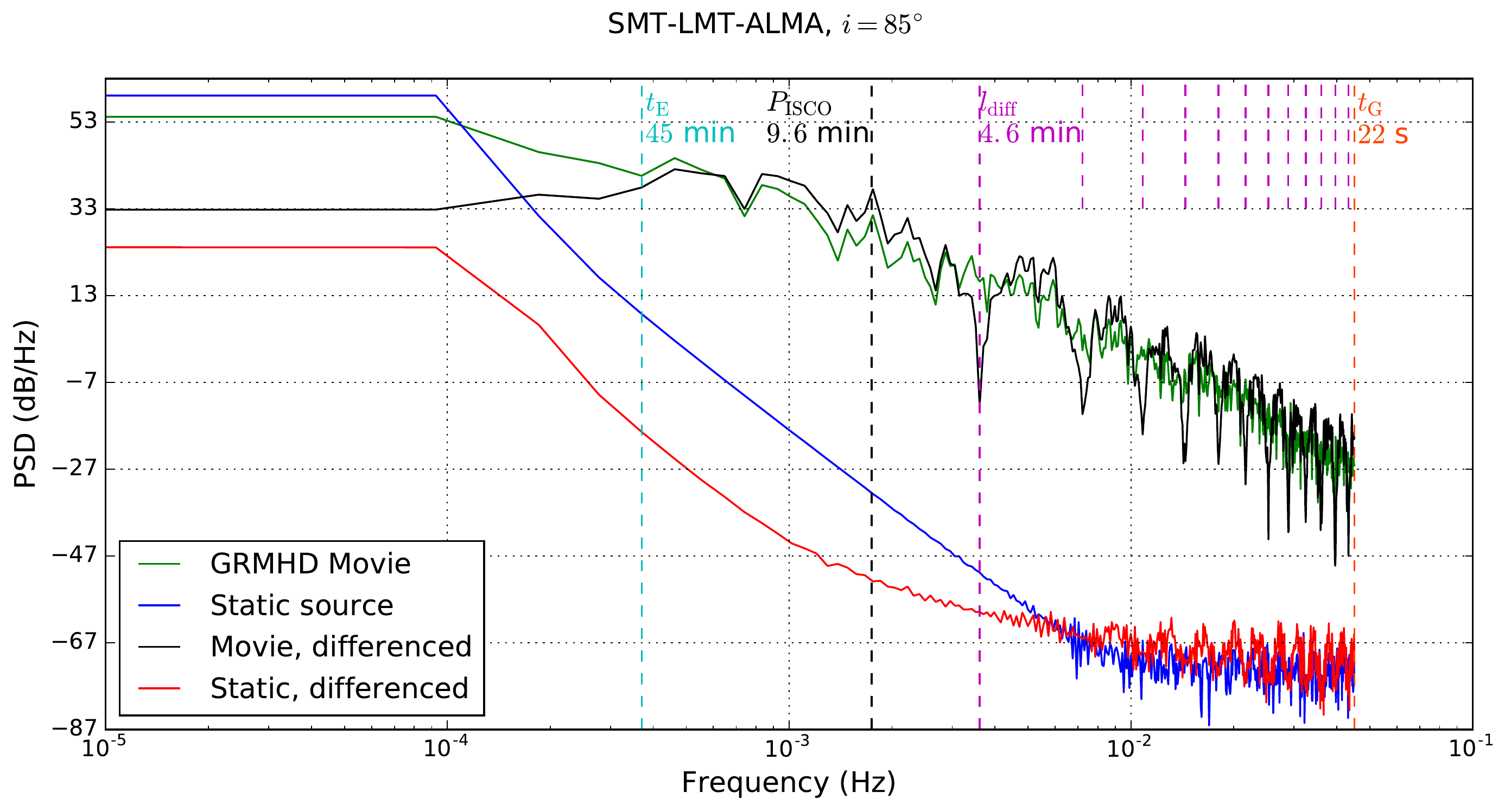}}
\resizebox{\hsize}{!}{\includegraphics{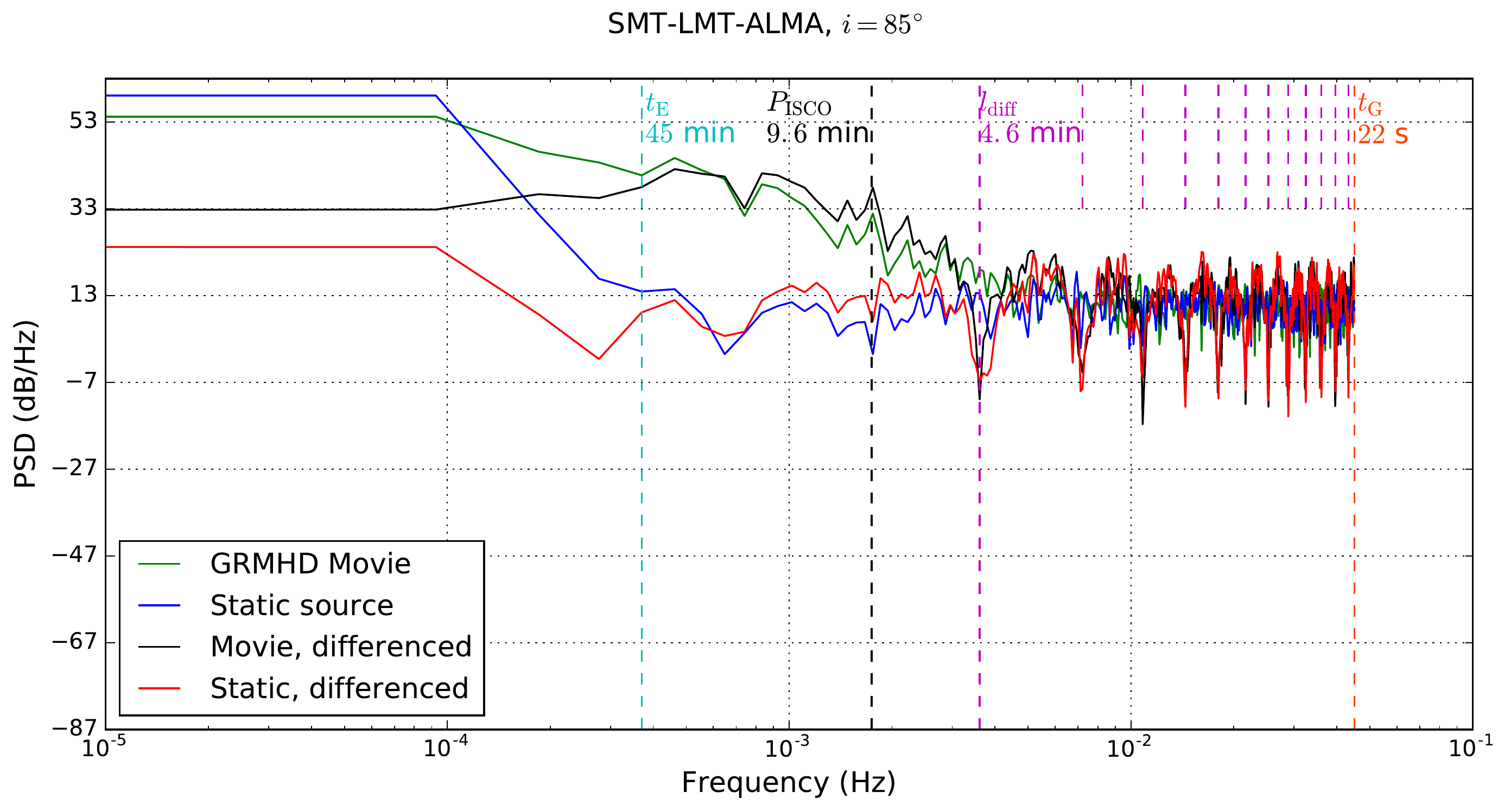}}
\end{center}
\caption{Power spectrum of the closure phase track with (lower panel) and without (upper panel) thermal noise on the SMT-LMT-ALMA triangle. The different lines represent the power spectra for the GRMHD movie and middle frame for an inclination of 85 degrees, with and without detrending the data using the method of differencing the time series at a lag of 25 data points (4.6 minutes). Indicated with vertical lines are the time scales $t_{\mathrm{G}}=GM/c^3$, the detrending lag $l_{\mathrm{diff}}$ and multiples of the corresponding frequency, the period of the innermost stable cirular orbit $P_{\mathrm{ISCO}}$, and the time scale $t_{\mathrm{E}}$ from where the detrended and non-detrended closure phase time series from the GRMHD movie contain approximately equal power again.}
\label{fig:pspec}
\end{figure}

Figure \ref{fig:pspec} shows an example of a power spectrum of the closure phase time series in different cases: the closure phases from the GRMHD movie (green), the differenced closure phases from the GRMHD movie (black), the closure phases from the middle frame of the GRMHD movie (blue), and the differenced closure phases from the static source (red), all with (lower panel) and without (upper panel) thermal noise. Thermal noise introduces a noise floor in the power spectrum at $\sim 10$ dB/Hz, so that no strong signal is present at time scales shorter than $\sim$a few minutes. From these plots, it is clear that detrending using the method of differencing effectively removes power on long time scales (low frequencies). Variability on short time scales (high frequencies), if present, is preserved up to the frequencies where the noise level is reached. The frequency where the detrended and non-detrended GRMHD movie closure phase tracks contain approximately equal power again is $\sim4\cdot10^{-4}$ Hz, corresponding to a timescale of $\sim45$ minutes, which is about 5 periods of the innermost stable circular orbit (ISCO). Some (but not all) of the variability information on larger time scales is thus lost when detrending the data. The dips in the power spectra of the detrended time series are at the frequency corresponding to the differencing time lag of 277 seconds, and multiples of that frequency.

Another metric that could be applied to closure phase measurements in the same way is 
\begin{equation}
\label{eq:qtilde}
\mathcal{\tilde{Q}}\equiv\hat{\sigma}^2-\tilde{\epsilon}^2.
\end{equation}
This metric is useful as its square root gives the spread on the closure phases in degrees that cannot be explained by the thermal noise. It has not been defined as the square root here since for a static source $\tilde{\epsilon}$ may be larger than $\hat{\sigma}$. Our main interest here is in detecting variability rather than measuring the excess variability in degrees, as the latter will be interesting only for high-SNR data where variability can be detected with high confidence. We will therefore focus on the normalized metric $\mathcal{Q}$.

\section{Applying the $\mathcal{Q}$-metric to existing EHT data}
\label{sec:ehtdata}
Before we calculate the $\mathcal{Q}$-metric for existing EHT data and compare the results with simulations, we describe how to process the measured (Sec. \ref{sec:ehttrack}) and simulated (Sec. \ref{sec:simproc}) closure phases in order to make them directly comparable. Section \ref{sec:scat} describes how to add variability introduced by interstellar scattering to the simulations.

\subsection{Generating a single closure phase track from multi-epoch EHT data}
\label{sec:ehttrack}
Closure phase measurements of Sgr A* have been made on the California-Hawaii-Arizona triangle in 2009, 2011, 2012 and 2013, as reported by \citet{Fish2016}. At CARMA, a phased array was used as well as a single comparison dish. At SMA, the phased array and JCMT were operating separately. The data were taken on two types of triangles. Trivial triangles contain two antennas at the same site. The closure phase on these triangles is expected to be zero, since the phases on the long baselines from the same-site antennas to the third antenna cancel, and the structure is not resolved by the short baseline between the same-site antennas, which thus measures zero phase. Triangles consisting of antennas on three different sites are non-trivial triangles. Closure phases on these triangles contain information about the source structure. 

The data from \citet{Fish2016} contain multiple non-trivial triangles because of different antennas used at each site during multiple epochs. In addition, data were taken on two frequency bands centered at 229.089 and 229.601 GHz with a bandwidth of 480 MHz. In order to nevertheless obtain a single closure phase time series, data points taken at the same time, but different triangles and/or frequency bands were circularly averaged after weighing them with the squared bispectral signal-to-noise ratio. Errors were assumed to be Gaussian and the error in radians was calculated as 1/SNR following \citet{Rogers1995}.

\begin{figure}[h!]
\begin{center}
\resizebox{\hsize}{!}{\includegraphics{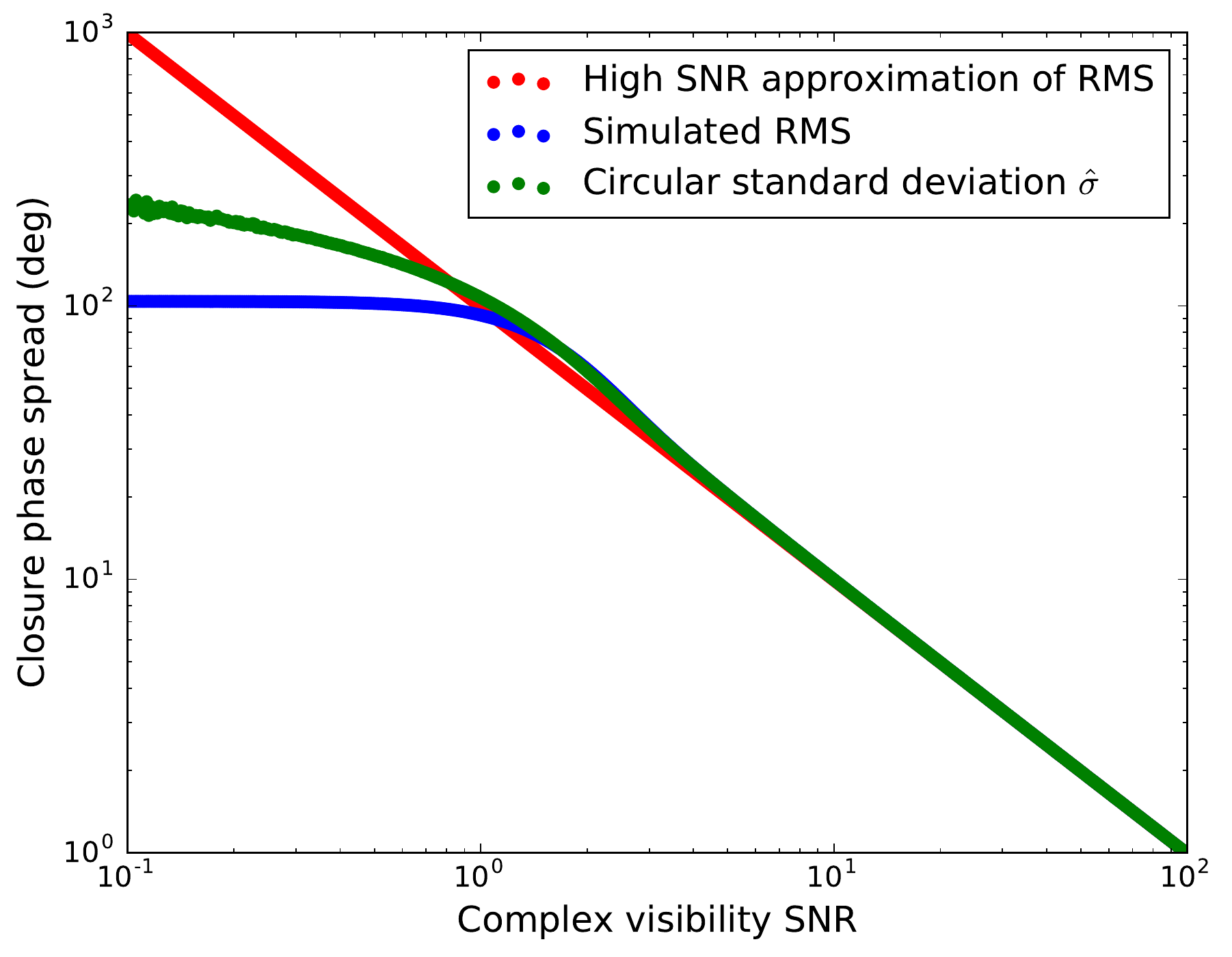}}
\end{center}
\caption{Simulated (blue) and approximate (red) closure phase RMS as a function of the complex visibility signal-to-noise, which was assumed to be equal on all baselines. The simulated RMS on the closure phases plateaus to a value of 60$\sqrt{3}$ as the phase randomizes. The circular standard deviation $\hat{\sigma}$ (green), which does not plateau to a fixed value, is also shown.}
\label{fig:cphase_noise}
\end{figure}

The validity regime of the high-SNR approximation for the closure phase uncertainties is shown in Figure \ref{fig:cphase_noise}. In this calculation, the SNR of the complex visibilities on three individual baselines was kept constant, and $10^7$ different realizations of the thermal noise were generated. The closure phase was calculated for each realization. The actual spread introduced by thermal noise can be calculated as the root-mean-square of the closure phases (blue) or the circular standard deviation $\hat{\sigma}$ (green, Eq. \ref{eq:sigma}), which can be compared to the high-SNR uncertainty approximation (red) in Equations \ref{eq:berr} and \ref{eq:cerr}. For equal SNR on all baselines, this is \citep{Rogers1995}
\begin{equation}
\sigma_c\approx\frac{\sqrt{3}}{\mathrm{SNR}}.
\end{equation}
For closure phase measurements with an estimated uncertainty smaller than 20 degrees, the high-SNR approximation estimates that uncertainty to within $\sim 2\%$ accuracy. As most of the uncertainties of the closure phases measured by \cite{Fish2016} are smaller than this, the high-SNR approximation is indeed valid for these data. 

Figure \ref{fig:cphase_noise} also shows that the linear closure phase RMS (blue) and circular standard deviation $\hat{\sigma}$ (green) behave differently as the SNR decreases. The linear RMS plateaus to a value of 60$\sqrt{3}$, corresponding to a fully randomized phase \citep{Johnson2016}. On the other hand, the circular standard deviation $\hat{\sigma}$ keeps increasing towards lower SNR. This is because as the noise increases the vector average of the closure phases approaches the origin of the unit circle (Fig. \ref{fig:circavg}). In the limit of pure noise, the distance to the origin $\bar{R}\rightarrow 0$ (Eq. \ref{eq:Rbar}). This means that the circular standard deviation goes to infinity (Eq. \ref{eq:sigma}) and does not plateau as the noise increases. The circular standard deviation $\hat{\sigma}$ is in good agreement with the conventional RMS down to an SNR of $\sim1$. Note that the SNR values in Figure \ref{fig:cphase_noise} refer to the expected mean SNR and not to the actual SNR of the data points. A full Monte Carlo approach is needed for a proper analysis of the closure phase statistics in the low SNR regime \citep{Blackburn2017}.

\begin{figure}
\begin{center}
\resizebox{\hsize}{!}{\includegraphics{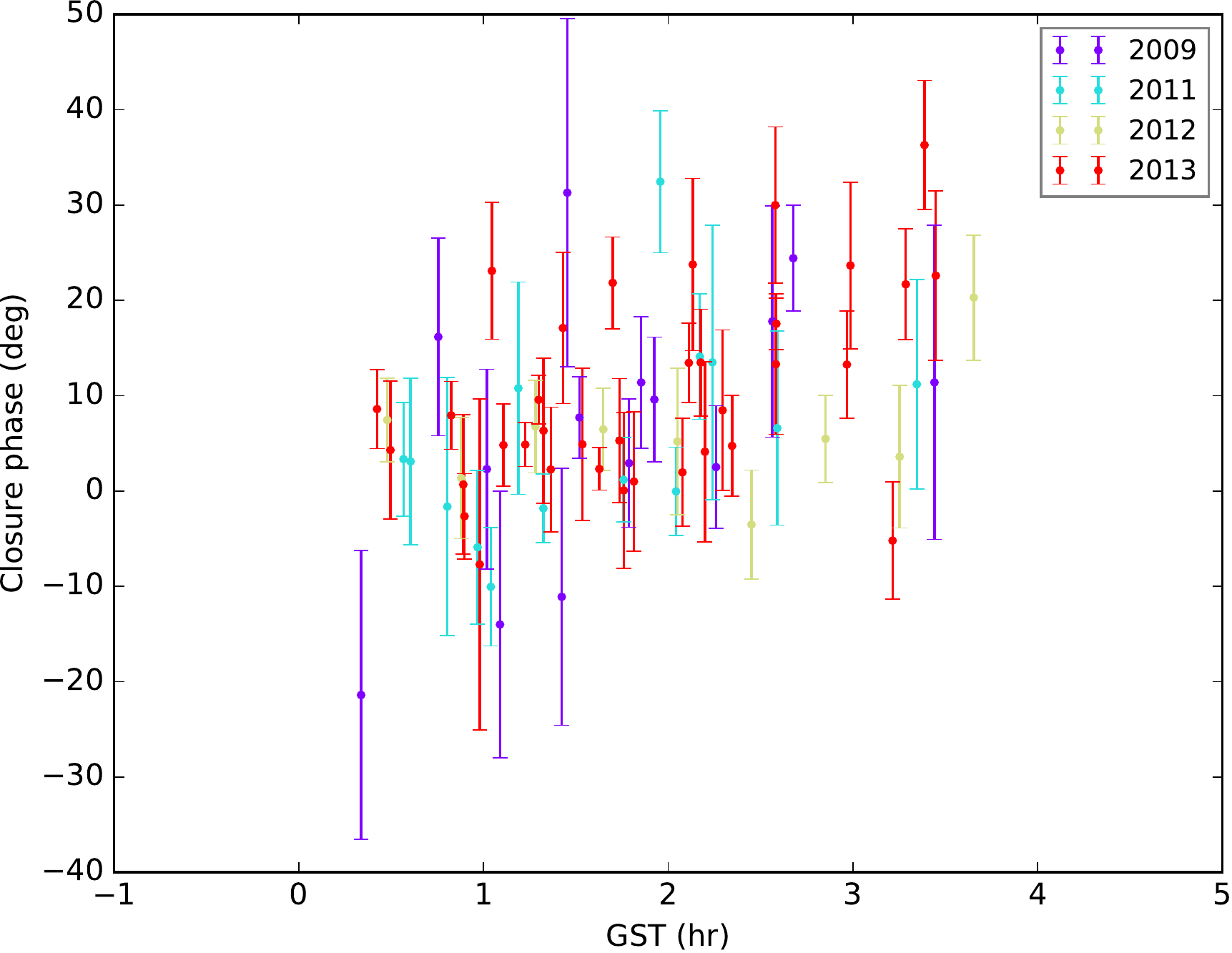}}
\resizebox{\hsize}{!}{\includegraphics{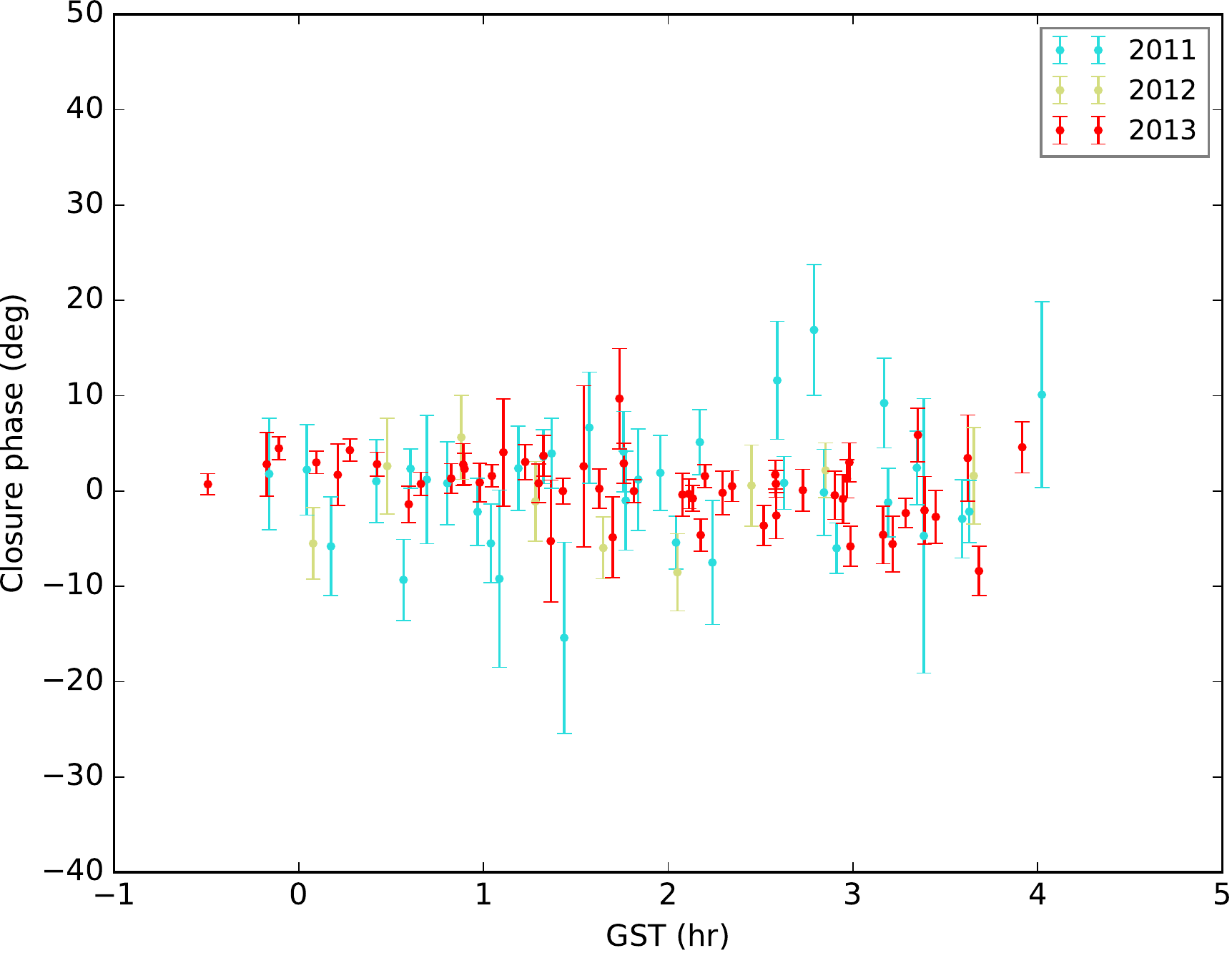}}
\end{center}
\caption{Measured Sgr A* closure phase as a function of sidereal time at the California-Hawaii-Arizona triangle (upper plot) and the trivial triangles (lower plot) in 2009, 2011, 2012, and 2013. Data are from \citet{Fish2016}, where measurements taken at the same time, but different triangles and frequency bands were circularly averaged with squared bispectral SNR weights.}
\label{fig:gst_year}
\end{figure}

Figure \ref{fig:gst_year} shows the resulting closure phases as a function of Greenwich Sidereal Time (GST), color-coded by year. GST was chosen rather than UT, because evolution in GST maps uniquely to evolution in baseline regardless of the date of observation. The metric $\mathcal{Q}$ described in Section \ref{sec:Nmetric} may now be applied to examine whether there is variability in the measured closure phases that is not due to thermal noise or Earth rotation.

After processing the data as described above an average of five and maximum of only nine data points occur within a single epoch. This amount of data points is considered too small for the statistical quantities that go into $\mathcal{Q}$ to contain reliable information, and for the detrending to work properly. $\mathcal{Q}$ has therefore only been applied to the combined multi-epoch data, and not to single-epoch observations. As a result, the $\mathcal{Q}$-value will reflect variability due to changing intrinsic source structure and interstellar scattering. As the scattering is expected to introduce significant variability on timescales of about a day, studying the $\mathcal{Q}$-metric in single-epoch data would reflect changes in intrinsic source structure only. 

\subsection{Generating simulations directly comparable to observations}
\label{sec:simproc}
Before comparing the metric value calculated from the observed closure phases to those calculated from different simulations, the simulated closure phase tracks should be made directly comparable to the measured closure phases. Closure phases due solely to source structure were calculated from the GRMHD movies starting at $8300t_G$ (after the MRI saturated at $8000t_G$ and almost all photons emitted then had reached the camera) for the CARMA-SMA-SMT triangle. Data points were taken at the same GST time stamps as the EHT data. In order to ensure that both the simulated and observed data sets have the same noise properties, the standard deviation of the Gaussian noise added to each simulated closure phase was set equal to the error on the EHT data point at the same time stamp. The resulting data set was detrended for baseline evolution, and the metric $\mathcal{Q}$ was calculated. As the EHT data do not have uniform sampling in time, the lag used for the differencing to correct for baseline evolution was set to a fixed number of 4 data points, corresponding to an average lag of 11 minutes, removing variations on time scales larger than $\sim$ a few ISCO periods (Fig. \ref{fig:pspec}). This process was repeated $10^4$ times in order to obtain different realizations of the thermal noise, after which an average $\bar{\mathcal{Q}}$ was calculated.

\subsection{Adding interstellar scattering to the simulated closure phases}
\label{sec:scat}
Refractive scattering by the ionized interstellar medium between the Earth and the Galactic Center introduces spurious substructure in the resulting image \citep{Johnson2015}. The scattering can be modeled as a phase-changing screen characterized by a phase structure function
\begin{equation}
D_{\phi}(\mathbf{r})\equiv \langle[\phi(\mathbf{r}'+\mathbf{r})-\phi(\mathbf{r}')]^2\rangle.
\end{equation}	 
Here, $\phi(\mathbf{r})$ is a statistically homogeneous Gaussian random phase field, with $\mathbf{r}$ a two-dimensional vector on the screen. $\langle\cdots\rangle$ denotes an ensemble average over different screen realizations. $D_{\phi}(\mathbf{r})$ is usually written as a power law $D_{\phi}(\mathbf{r})\propto |\mathbf{r}|^{\alpha}$, with $\alpha=5/3$ for Kolmogorov turbulence, which is thought to be the appropriate description for interstellar scattering \citep{Armstrong1995}. The screen is assumed to be frozen, so that the time evolution of the scattering only depends on the relative motions of the source, screen, and observer, characterized by a velocity $\mathbf{V}_{\bot}$. For Sgr A*, this relative motion causes the observed phase screen to change significantly on a timescale of about a day. The effect of variability due to scattering is thus not very large for a single observation, but it should be present in the \cite{Fish2016} data as they were taken during multiple epochs spanning four years. 

\begin{figure*}
\begin{center}
\resizebox{\hsize}{!}{\includegraphics{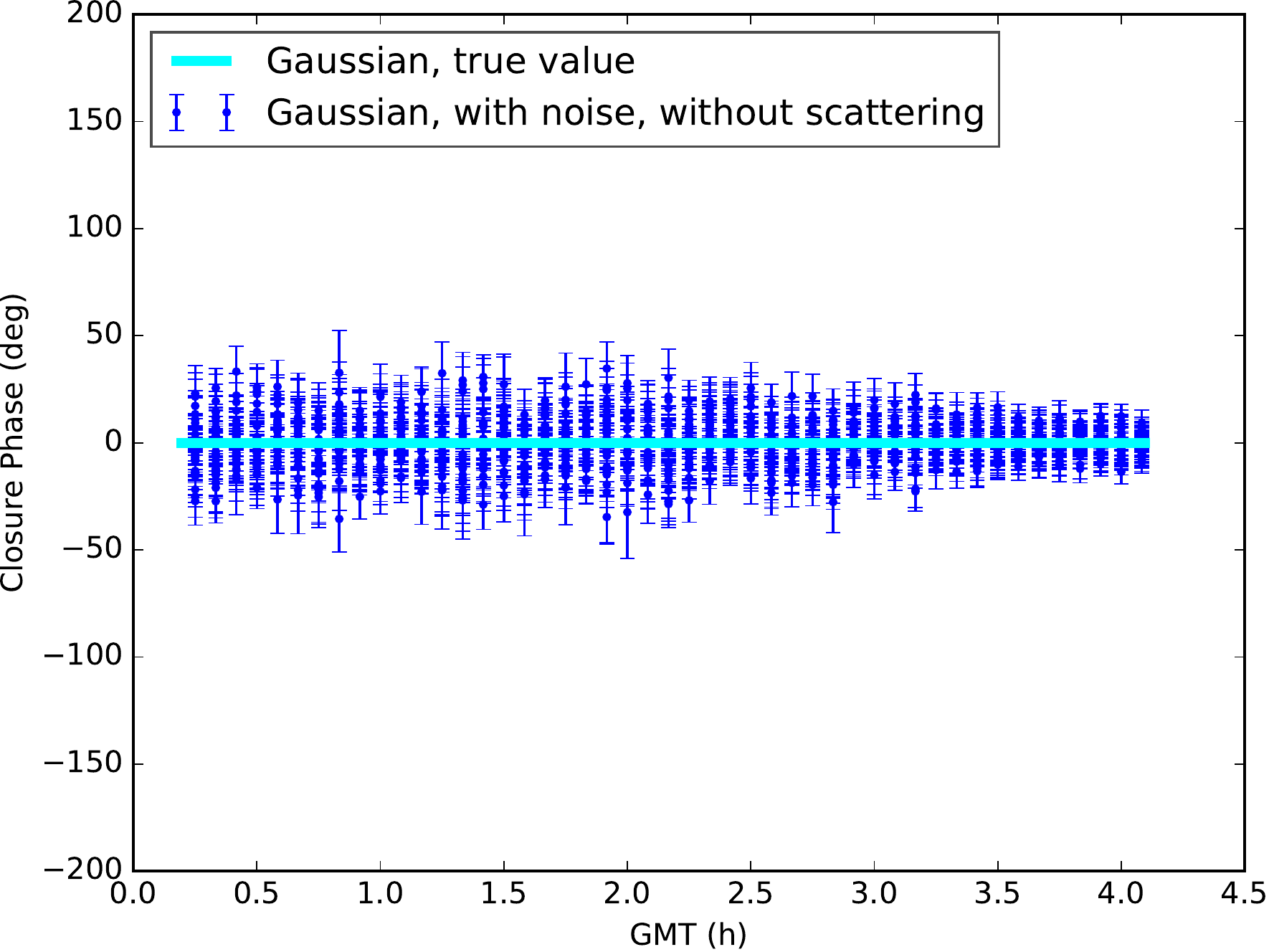}\includegraphics{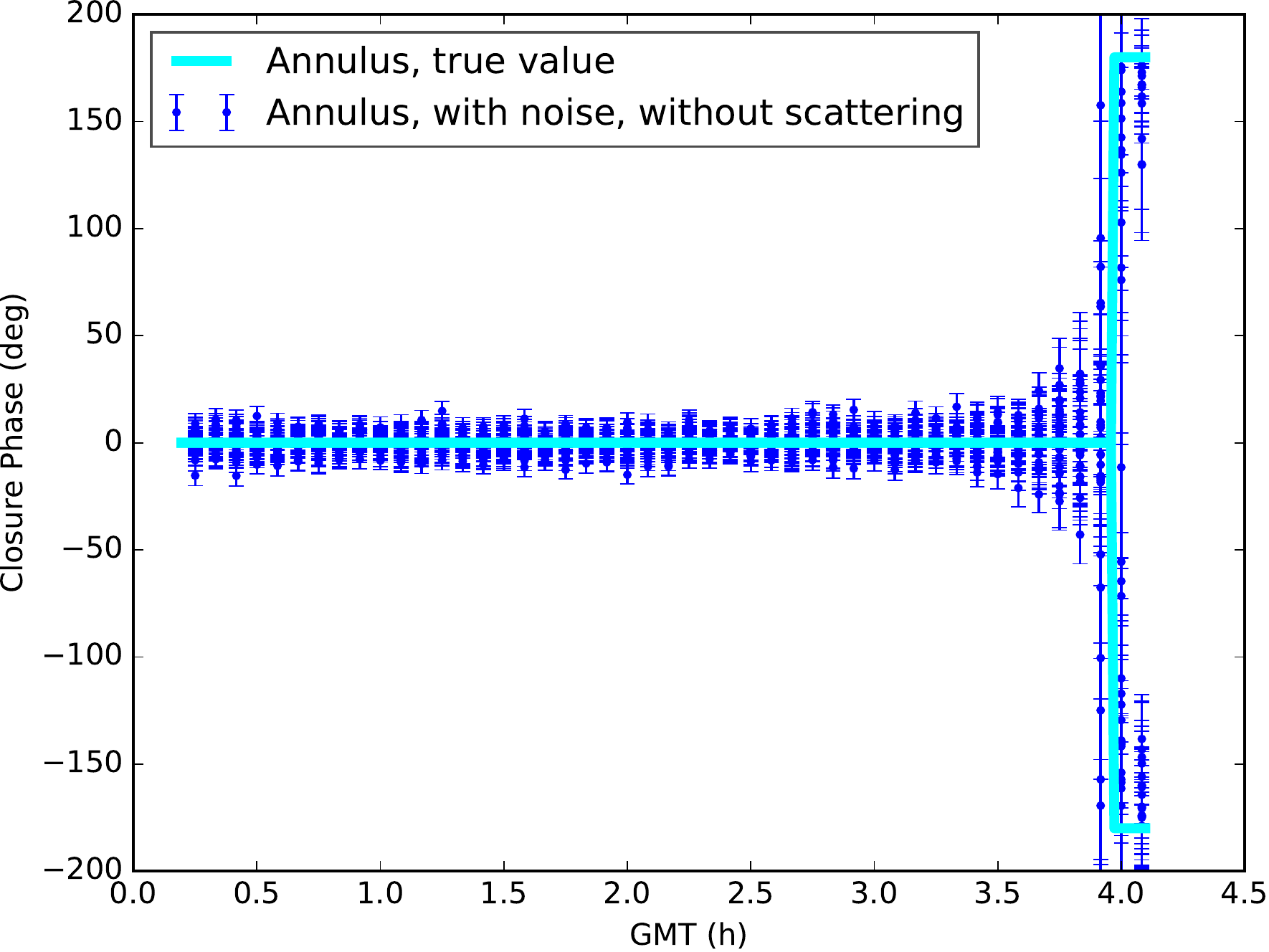}}
\resizebox{\hsize}{!}{\includegraphics{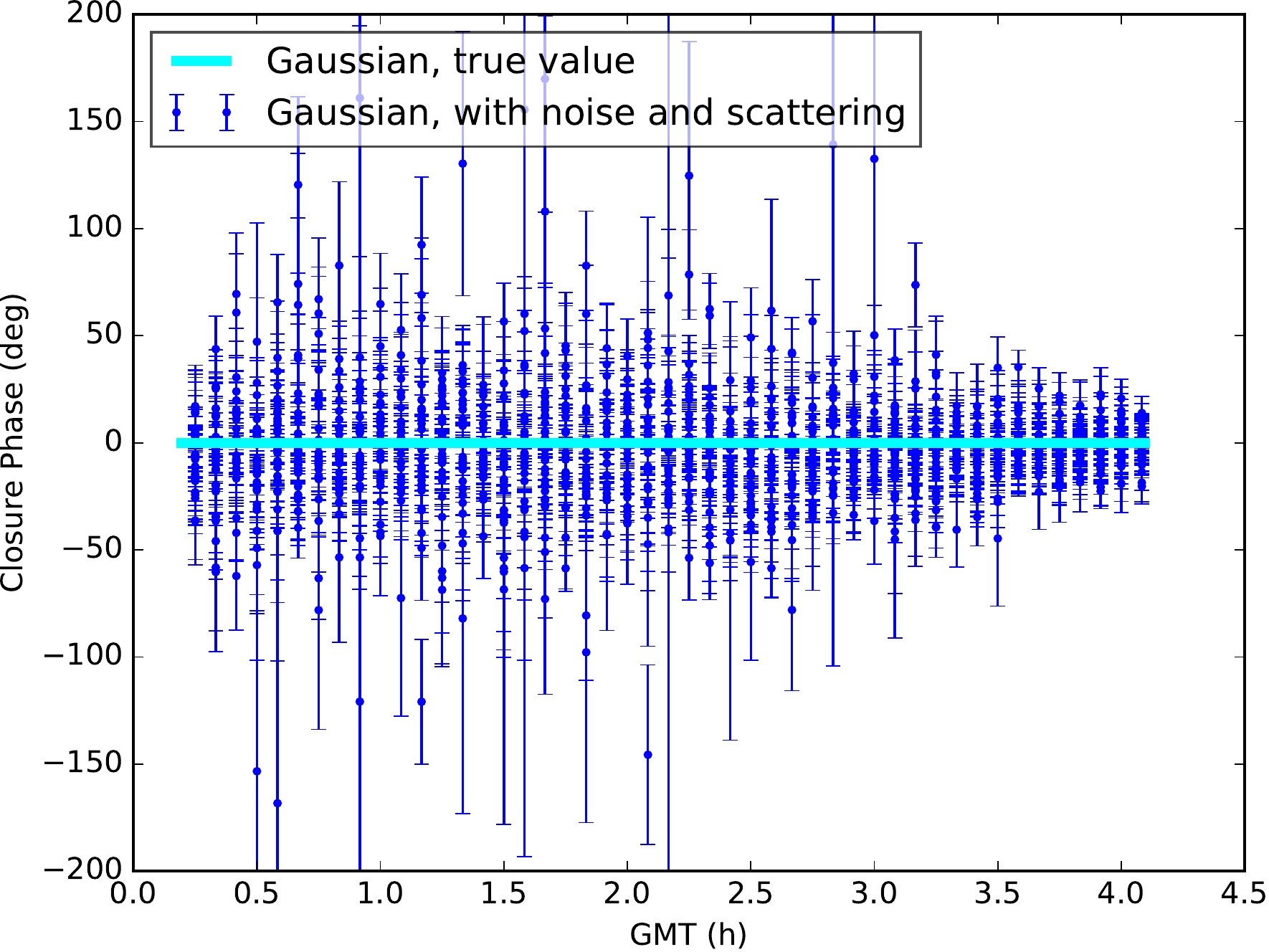}\includegraphics{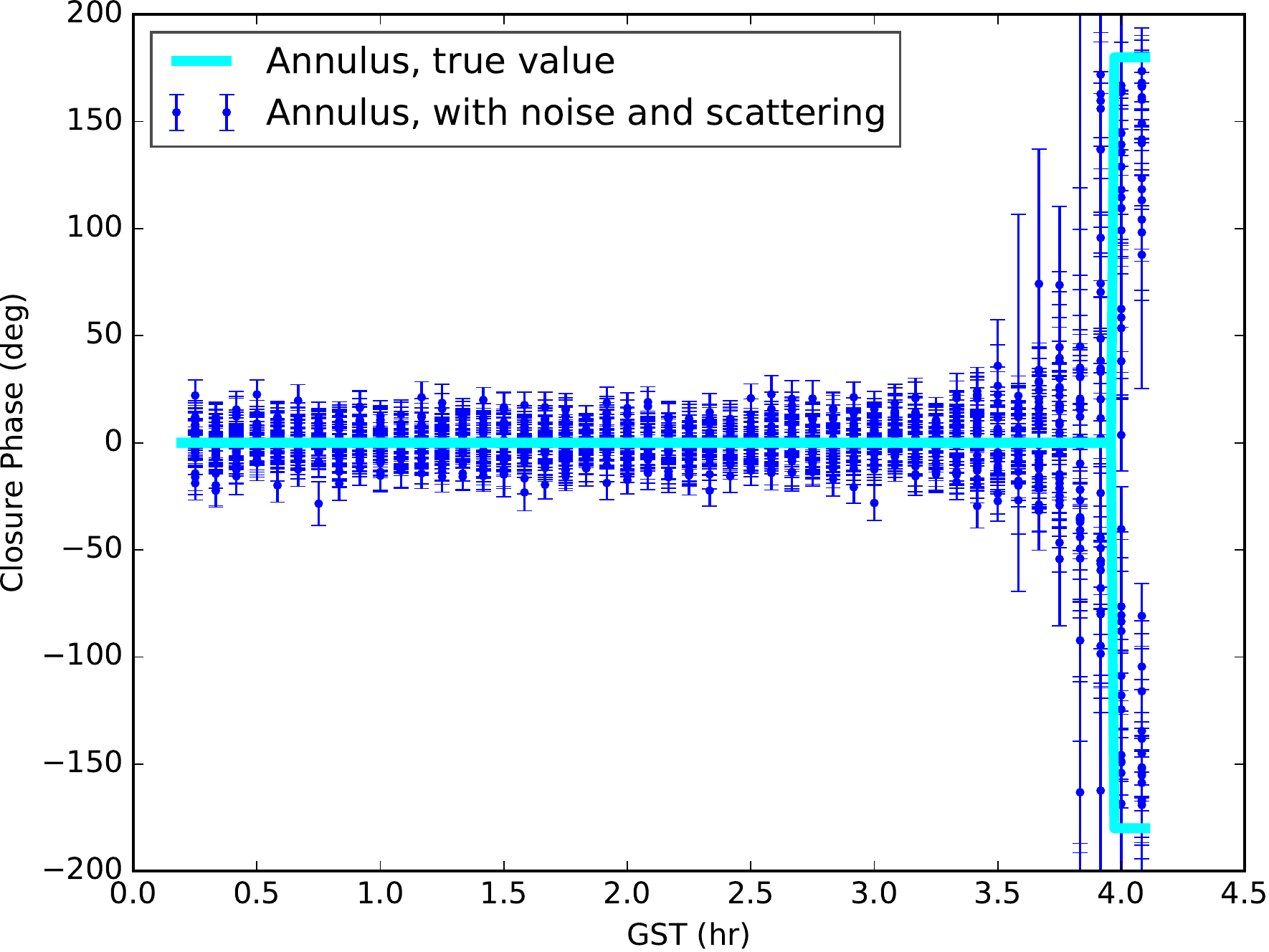}}
\end{center}
\caption{Upper plots: simulated closure phase tracks for a static source on the CARMA-SMA-SMT triangle, with (dark blue) and without (light blue) 30 realizations of the thermal noise ($t_{\mathrm{int}}=11$ s, $\Delta\nu=4$ GHz). The source models are the Gaussian model (left plots, FWHM: 52 $\mu$as) and annulus model (right plots, inner diameter: 21 $\mu$as, outer diameter: 97 $\mu$as) that provided the best fit to the visibility amplitudes measured by the EHT in 2013 \citep{Johnson2015pol}. The true value of the closure phase is always zero for these models, except when the annulus flips phase at GST $\sim3.9$ h. Lower plots: same as upper plots, but for 30 realizations of the thermal noise and scattering screen generated with the stochastic optics module in the $\texttt{eht-imaging}$ software \cite{Johnson2016so}. Interstellar scattering increases the jitter in closure phase. The phase flip in the annulus model corresponds to a null in the visibility amplitude.}
\label{fig:scatter}
\end{figure*}

The effect of a moving random phase screen on the closure phases is Gaussian jitter that is dependent on baseline and visibility amplitude \citep{Johnson2016}. The lower the visibility amplitude, the larger the root-mean-square of the closure phase jitter (see also the discussion in Section \ref{sec:behavior}). Here, we have considered the expected jitter for the Gaussian (FWHM: 52 $\mu$as) and annulus (inner diameter: 21 $\mu$as, outer diameter: 97 $\mu$as) source models that provided the best fits (within the Gaussian and annulus model classes) to the visibility amplitudes of Sgr A* measured by \cite{Johnson2015pol}. Only the annulus model fits the amplitude data well. The effect of different realizations of the scattering screen on the simulated closure phases of these models is shown in Figure \ref{fig:scatter}. The increase of the expected closure phase fluctuations of the annulus model towards the end of the track is due to the null in the model visibility amplitudes sampled at GST $\sim$3.9 h. The Gaussian model generally introduces larger closure phase jitter than the annulus model due to the lower visibility amplitudes at the $uv$-points sampled by the CARMA-SMA-SMT triangle. 

\citet{Johnson2016} developed a framework for calculating the expected RMS closure phase fluctuations due to interstellar scattering for different source models. The effect of interstellar scattering may thus be simulated by using the expected RMS fluctuations as the standard deviation of additional Gaussian jitter added to all simulated closure phases. However, the 2009-2013 EHT data contain on average 5 closure phases for each day of observations, and the scattering is not expected to introduce variability on time scales shorter than $\sim$a day. Adding the expected RMS to all data points may thus overestimate the variability introduced by interstellar scattering. 

A more realistic way to add the scattering is to scatter the model image or movie frames themselves with a different scattering screeen for each day of EHT observations. Using the stochastic optics module in the $\texttt{eht-imaging}$ software \citep{Johnson2016so}, a number of independent scattering screens equal to the number of days with EHT closure phase measurements can be generated, while assigning a particular scattering screen to each day. For each EHT data point, the model image or movie frame corresponding to the GST of the measurement can then be scattered with the scattering screen corresponding to the day of the measurement. Closure phases can then be simulated for multiple realizations of the set of scattering screens, and an average $\bar{\mathcal{Q}}$ may again be calculated.

\begin{figure*}
\begin{center}
\resizebox{\hsize}{!}{\includegraphics{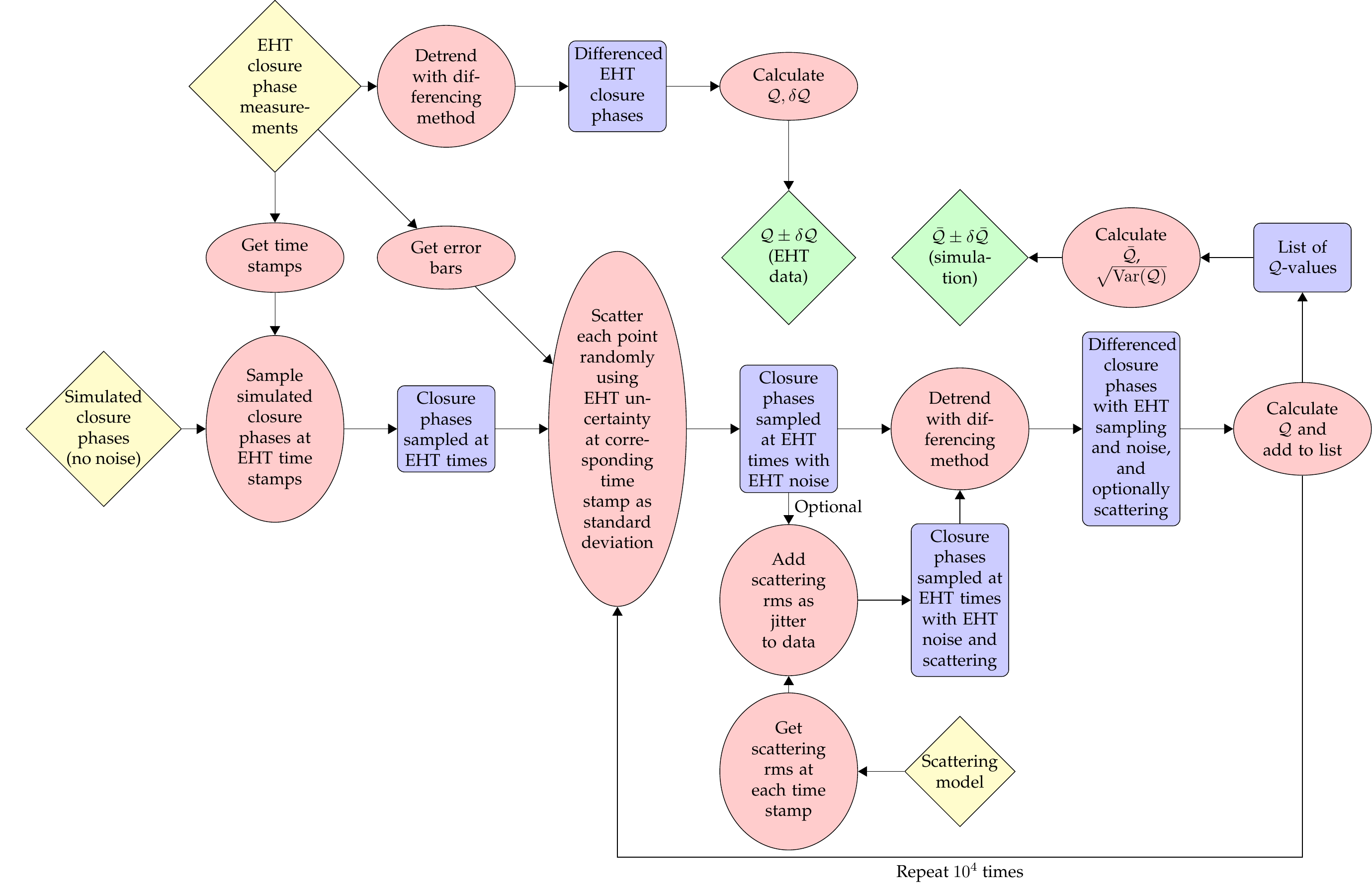}}
\end{center}
\caption{Flowchart describing the process of calculating $\mathcal{Q}$ from the EHT data and simulations. The yelow squares indicate the data products used as starting points, the red ovals are actions performed on data products, the blue rectangles indicate intermediate data products, and the green squares indicate the end products.}
\label{fig:flowchart}
\end{figure*}

The full process of calculating $\mathcal{Q}$ from the simulated and observed closure phases is shown schematically in Figure \ref{fig:flowchart}.

\section{$\mathcal{Q}$-metric results}
\label{sec:results}
We now calculate the $\mathcal{Q}$-metric for previously reported EHT data and simulations, and the results are compared in the case of a static (Sec. \ref{sec:resulttypes}) and variable (Sec. \ref{sec:comp}) source.
\subsection{Static source}
\label{sec:resulttypes}

\begin{figure}
\begin{center}
\resizebox{\hsize}{!}{\includegraphics{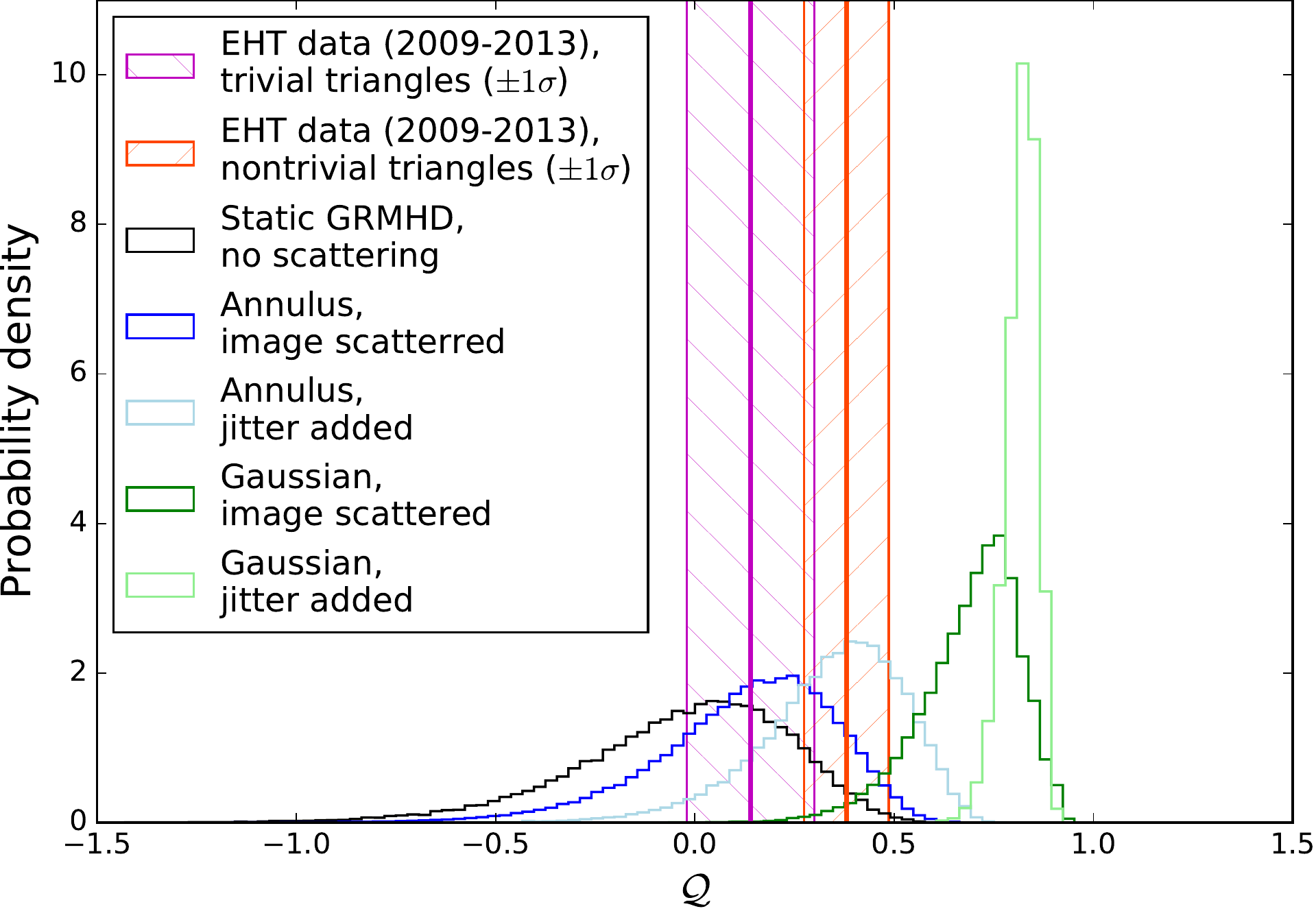}}
\end{center}
\caption{1$\sigma$ confidence intervals for the $\mathcal{Q}$ metric values calculated from the EHT data on the trivial (magenta) and non-trivial (orange) triangles, and simulations on the CARMA-SMA-SMT triangle. The distributions show different thermal noise and scattering realizations of the middle frame of the $a_*=0.94$ disk movie without scattering at all inclinations (black), the annulus (dark blue) and Gaussian (dark green) model scattered with a different scattering screen for each day of EHT observations, and the expected RMS closure phase jitter due to scattering calculated by \citet{Johnson2016} for the annulus (light blue) and Gaussian (light green) models added to the static source closure phases. The probability for a static source to give a $\mathcal{Q}$-value that is equal to or higher than the measured value of 0.38 is equal to the area beneath the static source distribution (black) at $\mathcal{Q}$-values higher than 0.38, which gives $3.0\%$. The probability that interstellar scattering by a Gaussian source gives a $\mathcal{Q}$-value that is equal to or lower than 0.38 is $2.2\%$.}
\label{fig:scatmodels_dist}
\end{figure}

Figure \ref{fig:scatmodels_dist} shows the $\mathcal{Q}$-values for the 2009-2013 EHT data and different simulations of static and scattered sources calculated with the procedure described above. The $\mathcal{Q}$-value of $0.38 \pm 0.11$ for the EHT data of the non-trivial triangle CARMA-SMA-SMT (orange band) is well above zero, which suggests that variability from sources other than thermal noise and Earth rotation is present in the data. The $\mathcal{Q}$-value for the EHT data on the trivial triangles of $0.14 \pm 0.16$ (magenta band) is consistent with zero, as one would expect since the closure phases should always be close to zero on trivial triangles. The $\mathcal{Q}$-values for the EHT data have also been estimated with bootstrapping \citep[Sec. \ref{sec:error}, ][]{Efron1979}. The mean and standard deviation of the bootstrapped distribution containing the $\mathcal{Q}$-values of $10^4$ resamplings of the EHT data are $0.37 \pm 0.11$ and $0.15 \pm 0.21$ for the non-trivial and trivial triangles, respectively. These values are in good agreement with the values calculated using standard error analysis (Sec. \ref{sec:error}) quoted above.

The combined distribution of $\mathcal{Q}$-values for different realizations of the thermal noise and all inclinations of the middle frame of the $a_*=0.94$ disk model without scattering (black) is consistent with zero, indicating that no intrinsic variability is present in these simulations. The distribution contains an equal amount of thermal noise realizations for all inclinations, as the basic behavior in $\mathcal{Q}$ should be the same for any static source after detrending for baseline evolution. The variability observed in the EHT data is inconsistent with these simulations at a level of $\sim2\sigma$: the probability that a static source gives a $\mathcal{Q}$-value that is equal to or higher than the $\mathcal{Q}$-value of 0.38 measured for the 2009-2013 EHT data is $3.0\%$. 

The distribution of the $\mathcal{Q}$-values for different realizations of the thermal noise and set of scattering screens for the annulus model (dark blue) is centered at a slightly higher $\mathcal{Q}$-value than the distribution for a static source without scattering, as expected from the difference in closure phase fluctuations shown in Figure \ref{fig:scatter}. As the Gaussian model gives stronger scattering fluctuations, the $\mathcal{Q}$-value distribution for the Gaussian source including interstellar scattering (dark green) is at significantly higher values than the distribution for the annulus model. It also overestimates the variability observed in the EHT data: the probability that interstellar scattering by a Gaussian source gives a $\mathcal{Q}$-value that is equal to or lower than the $\mathcal{Q}$-value of 0.38 measured for the EHT data is $2.2\%$. A Gaussian morphology of Sgr A* is thus not only inconsistent with the visibility amplitudes, but also with the observed closure phase variability (at a level of $\gtrsim2\sigma$).

The light blue and light green curves respectively show the $\mathcal{Q}$-value distributions for the simulations with the RMS jitter of the annulus and Gaussian model, as calculated by \citet{Johnson2016}, added to the static source closure phases, for different realizations of the thermal noise and scattering jitter. These are at higher values than the simulations where the models themselves were scattered with a different scattering screen for each day of EHT observations (dark blue and dark green, respectively). This illustrates that the scattering fluctuations on a closure phase track are correlated for a single realization of the scattering screen, and the additional variability arises mostly from the evolution of the scattering screen rather than from the scattering screen itself. 

\begin{figure}
\begin{center}
\resizebox{\hsize}{!}{\includegraphics{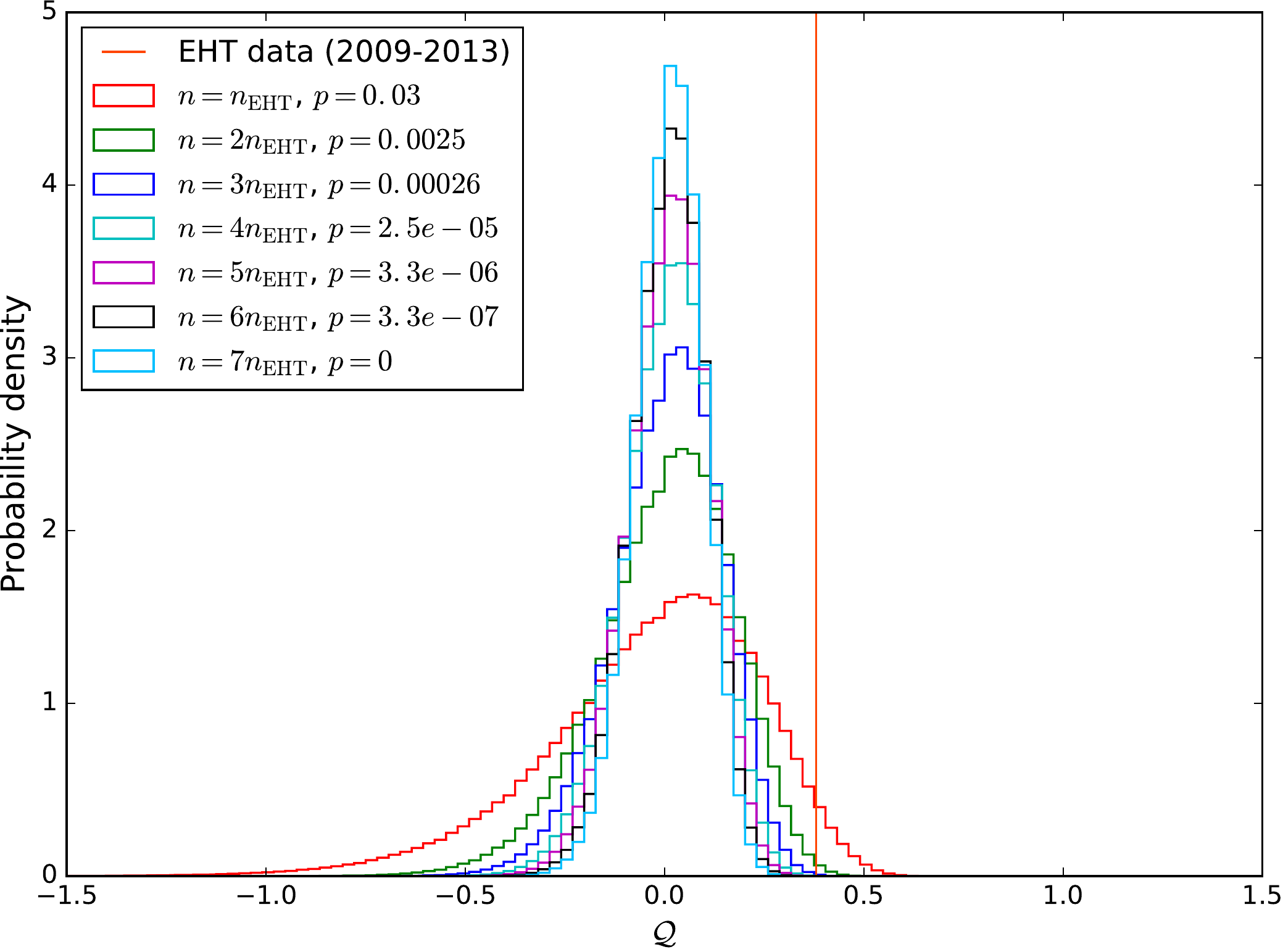}}
\end{center}
\caption{Improvement of the ability to detect image variability with increasing number of data points. The $\mathcal{Q}$-value distributions for a static source without scattering (the middle frame of the $a_*=0.94$ disk model at different inclinations with different realizations of the thermal noise) are shown for different numbers of data points expressed as multiples of the number of closure phases measured by the EHT in 2009-2013 ($n_{\mathrm{EHT}}=72$). The $p$-value indicates the probability that the model gives a $\mathcal{Q}$-value that is equal to or higher than the $\mathcal{Q}$-value of 0.38 measured for the EHT data. This value would have been a $5\sigma$ detection of image variability if the number of data points would have been $\sim6$ times as high.}
\label{fig:nextra}
\end{figure}

\begin{figure}
\begin{center}
\resizebox{\hsize}{!}{\includegraphics{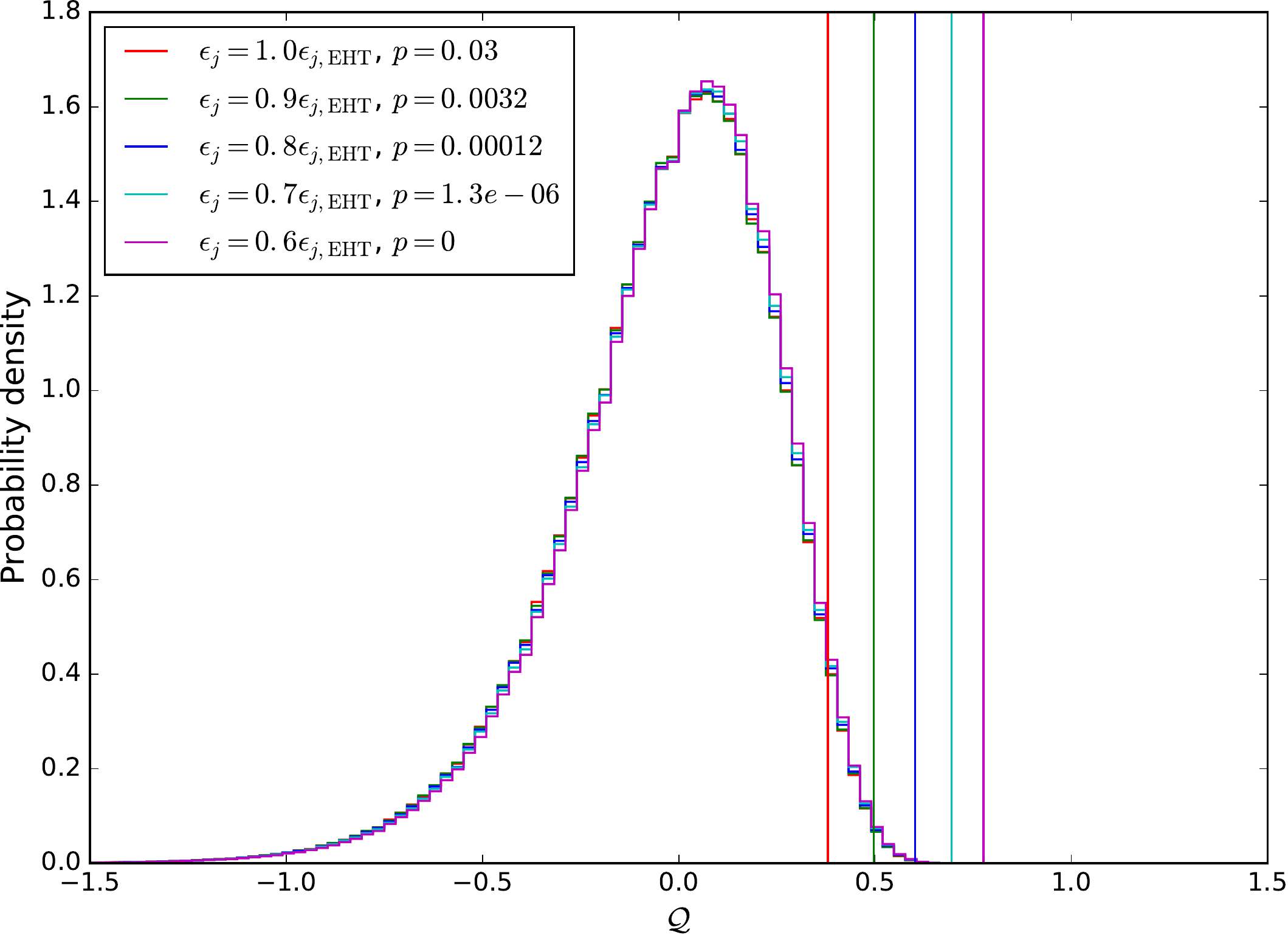}}
\end{center}
\caption{Improvement of the ability to detect image variability with decreasing error bars. The $\mathcal{Q}$-values for the EHT data and the $\mathcal{Q}$-value distributions for a static source without scattering (the middle frame of the $a_*=0.94$ disk model at different inclinations with different realizations of the thermal noise) are shown for different error bar sizes expressed as fractions of the error bar sizes of the 2009-2013 EHT data (Fig. \ref{fig:gst_year}). The $p$-value indicates the probability that the model gives a $\mathcal{Q}$-value that is equal to or higher than the $\mathcal{Q}$-value for the EHT data with corresponding error bar sizes. The current EHT data would give a $5\sigma$ detection of variability if the error bars had been $\sim0.6$ times as large.}
\label{fig:snr}
\end{figure}

The confidence of $97\%$ with which we detect image variability (due to structural and/or scattering variations) is expected to increase as we obain more data with better quality. As shown in Figure \ref{fig:nextra}, the distribution of $\mathcal{Q}$-values for a static source becomes more narrow and better centered at zero as the number of data points increases. The $\mathcal{Q}$-value of 0.38 measured for the EHT data would have been a $5\sigma$ detection of image variability if the number of data points would have been $\sim6$ times as high. Figure \ref{fig:snr} shows that decreasing the error bar size hardly affects the $\mathcal{Q}$-value distribution for a static source. This is not surprising as a higher SNR will not make the detrending do a better job. In fact, it should make detrending more difficult since closure phase variations due to baseline evolution relatively contribute more to the total variability when there is less thermal noise. On the other hand, the $\mathcal{Q}$-value for a variable source will be affected as the variations become more dominated by image variability. The current EHT data would give a $5\sigma$ detection of variability if the error bars had been $\sim0.6$ times as large.

\subsection{GRMHD movies}
\label{sec:comp}
Apart from the inclination, the orientation of the source in the plane of the sky also influences the closure phases measured. This sensitivity to source orientation occurs because a single baseline only measures the image structure parallel to the direction in which it points due to the projection-slice theorem \citep{Thompson2001}. Because the features in the movies move in certain directions, rotating the image will influence the phase variability measured by the individual baselines, and hence the closure phase variability. 

\begin{figure*}
\begin{center}
\resizebox{0.3\hsize}{!}{\includegraphics{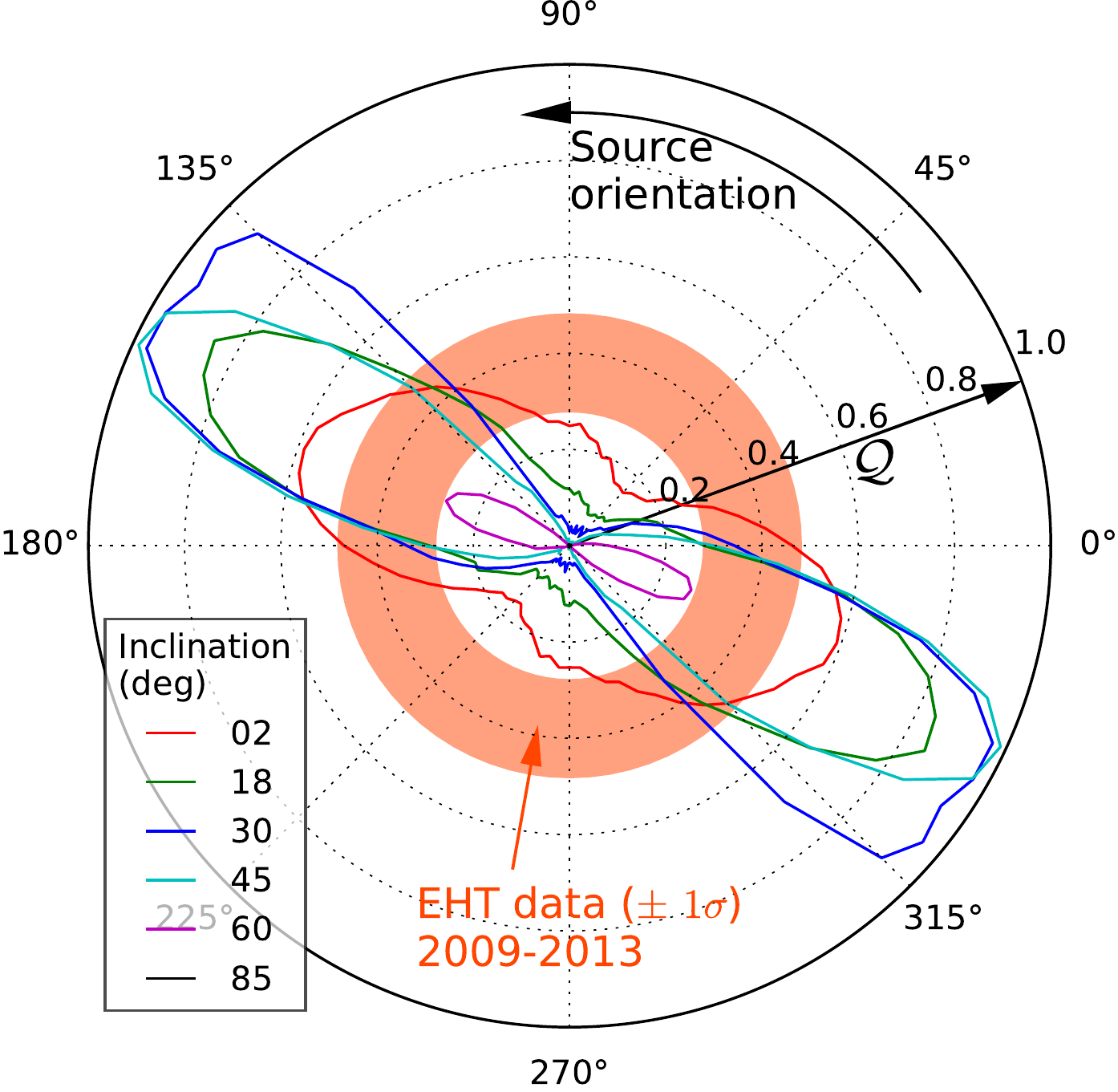}}
\resizebox{0.3\hsize}{!}{\includegraphics{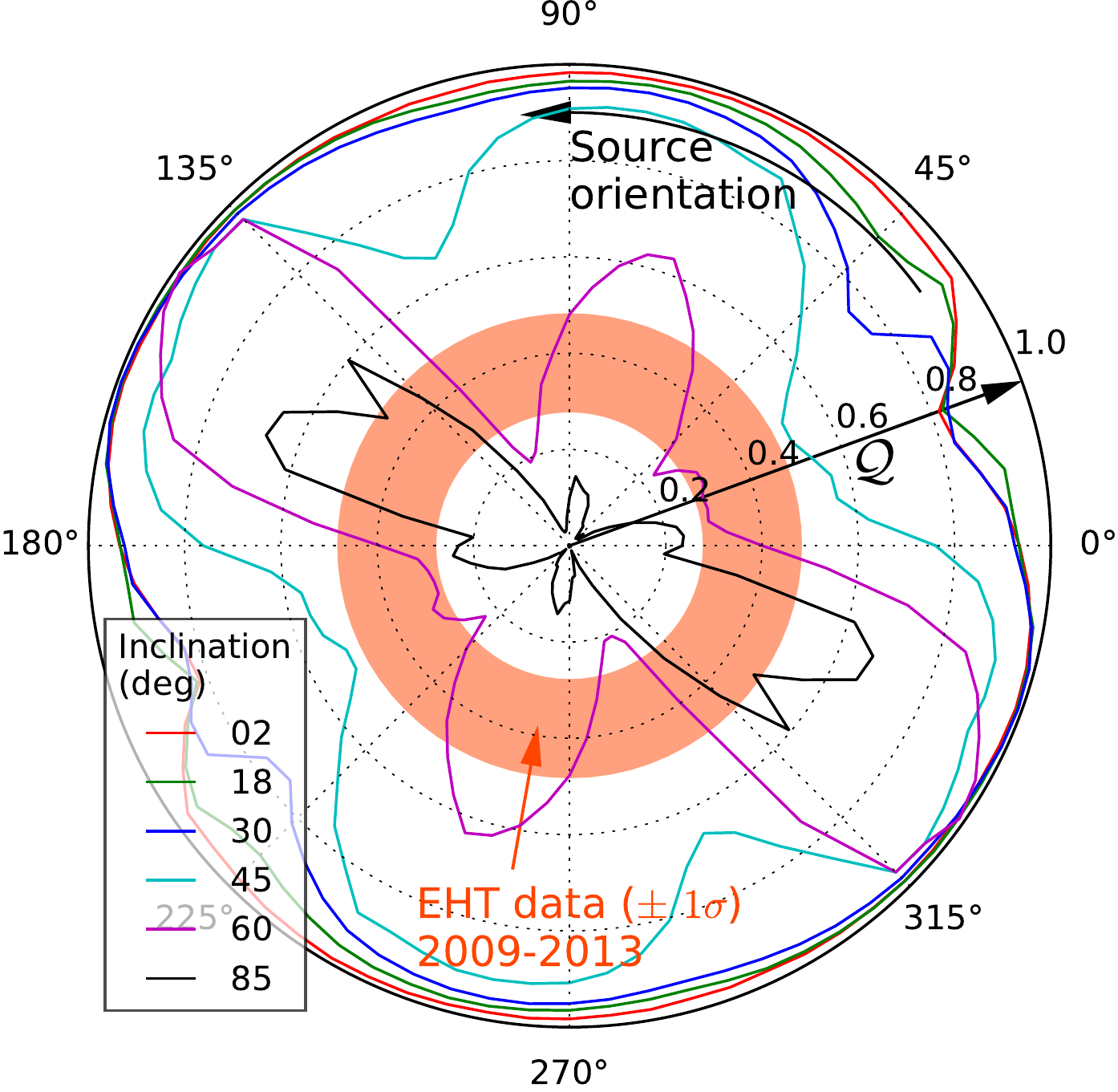}}
\resizebox{0.3\hsize}{!}{\includegraphics{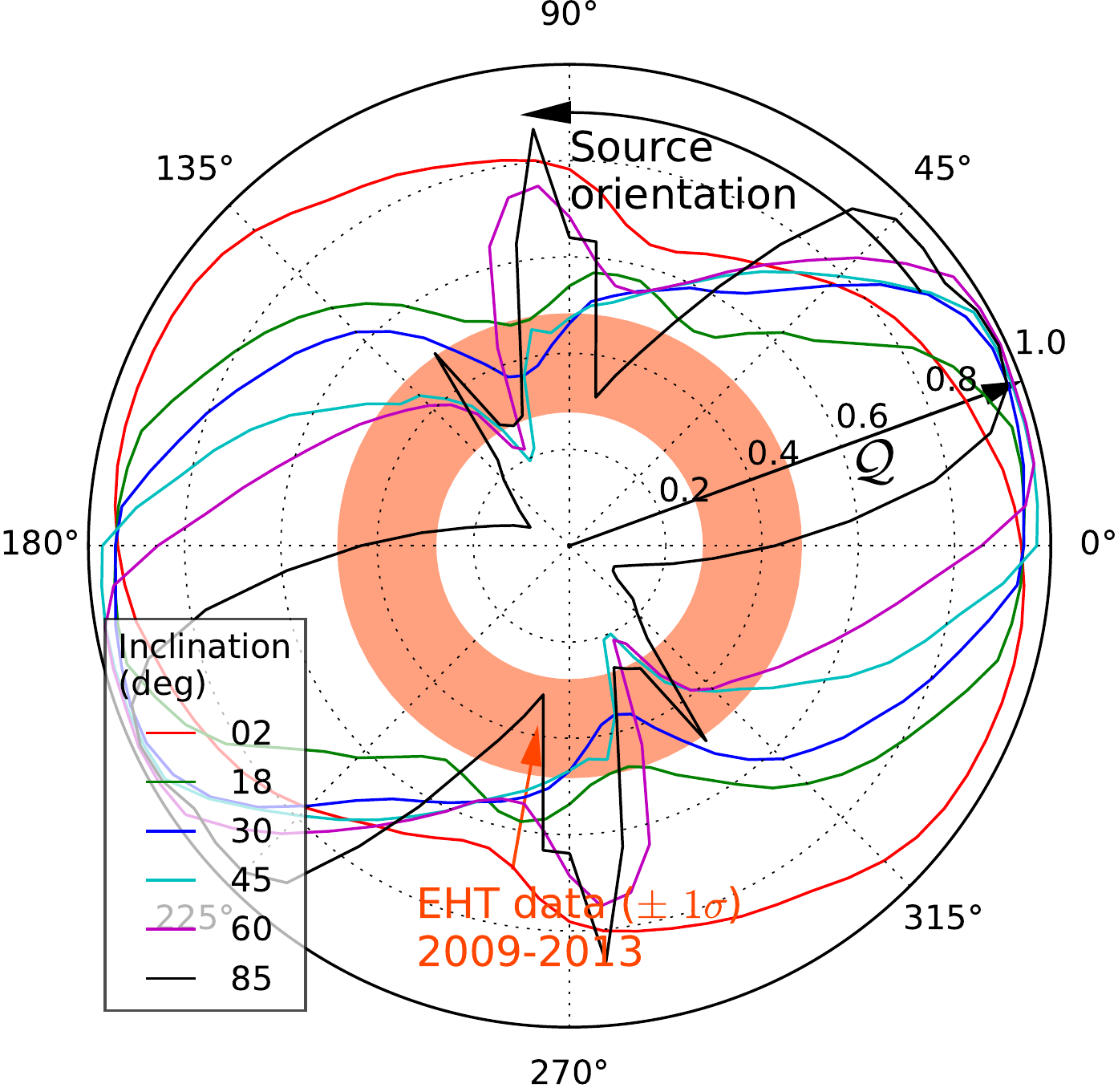}}
\resizebox{0.3\hsize}{!}{\includegraphics{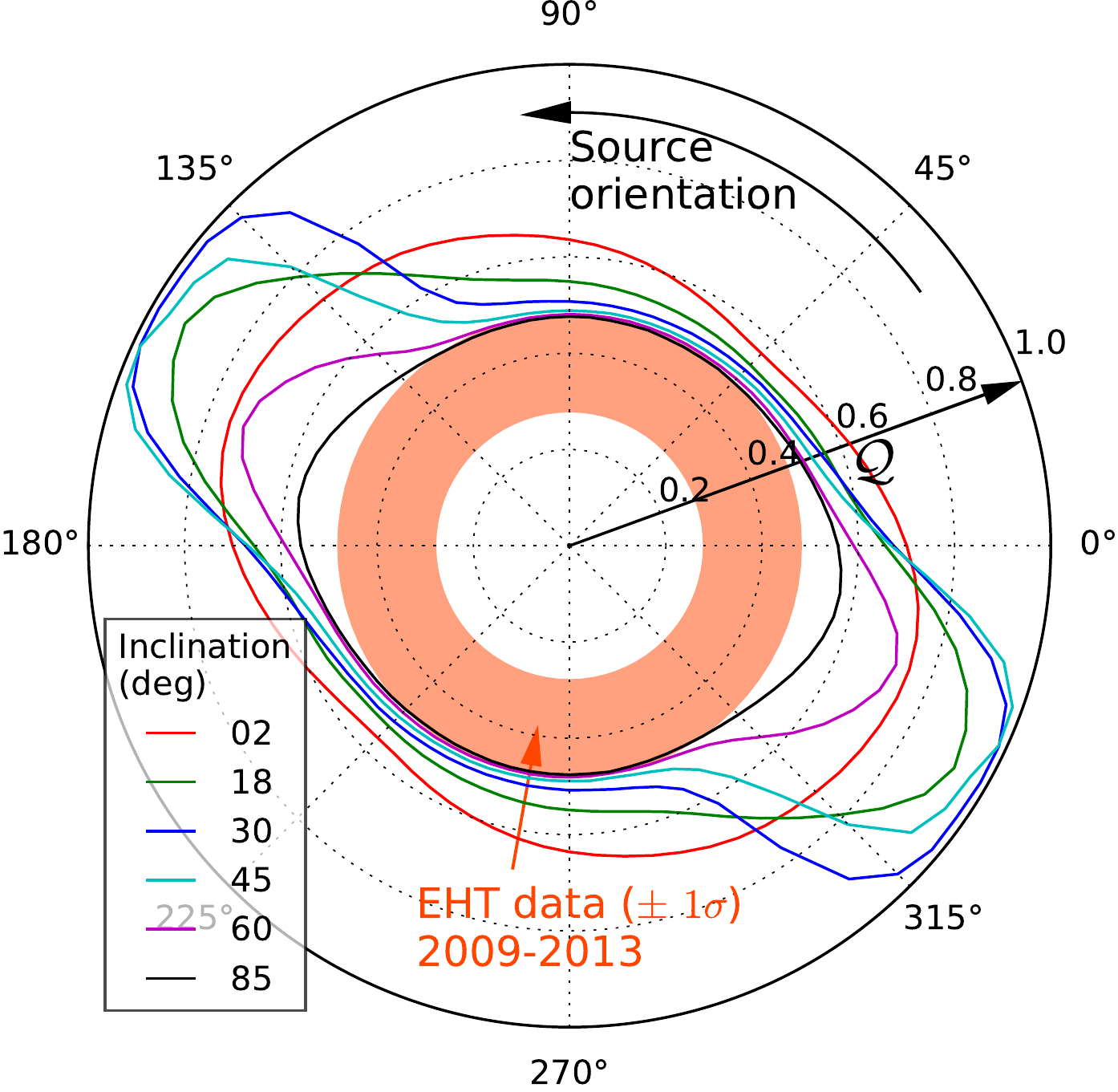}}
\resizebox{0.3\hsize}{!}{\includegraphics{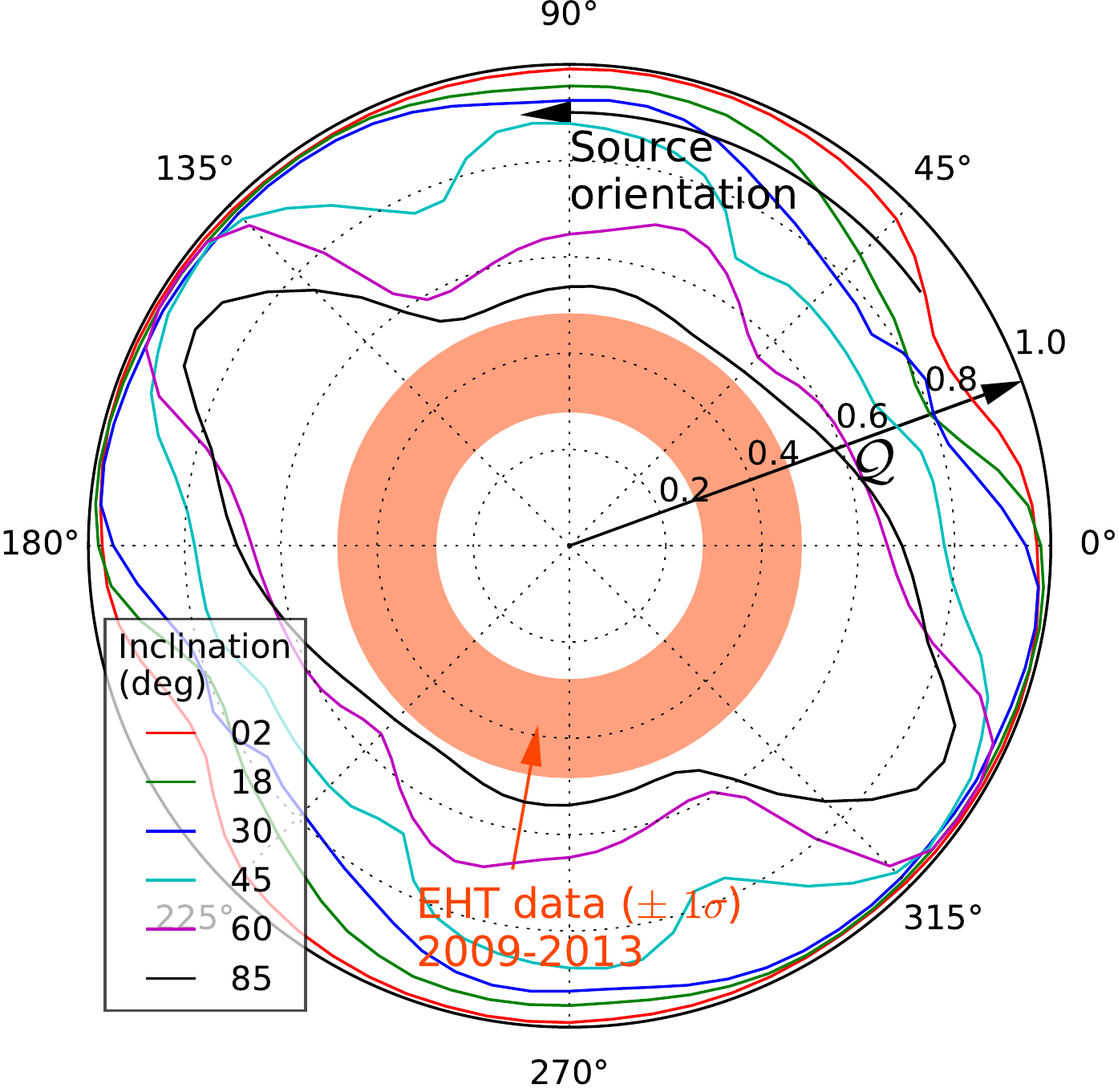}}
\resizebox{0.3\hsize}{!}{\includegraphics{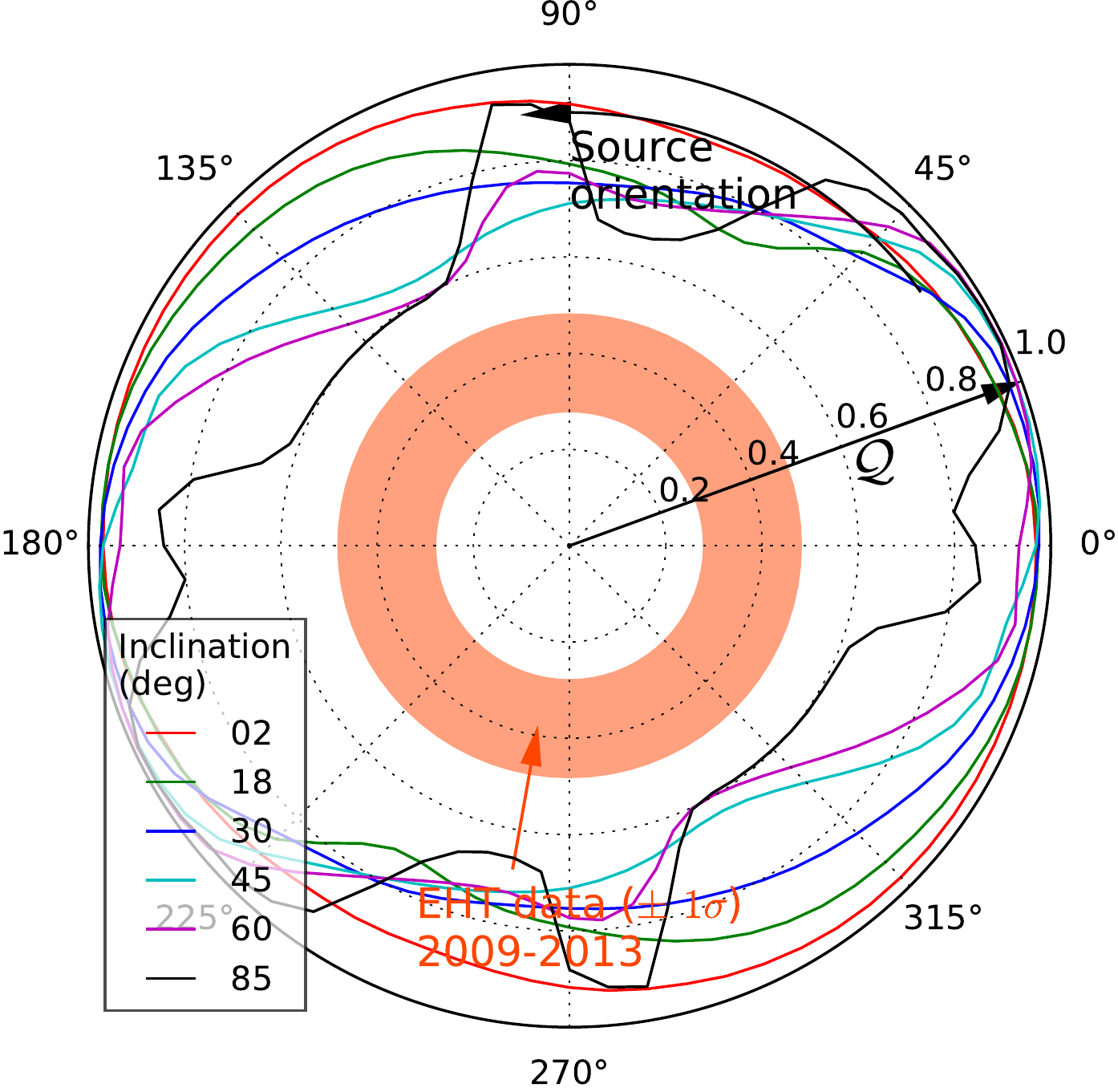}}
\end{center}
\caption{$\mathcal{Q}$-values for the GRMHD movies with (lower plots) and without (upper plots) interstellar scattering at different spin axis orientations and inclinations. Interstellar scattering was added by scattering the movie frames with a different realization of the scattering screen for each day of the 2009-2013 EHT observations. The $\mathcal{Q}$-value corresponds to the radius in the polar plot, while the spin axis orientation is represented by the polar angle. Different inclinations are shown in different colors. The orange band represents the 1$\sigma$ confidence interval for the $\mathcal{Q}$-value of the EHT data. The plots from left to right show the $\mathcal{Q}$-values for the $a_*=0.94$ disk model, the $a_*=0$ disk model, and the $a_*=0.94$ jet model, respectively.}
\label{fig:metric_orient}
\end{figure*}

Figure \ref{fig:metric_orient} shows the $\mathcal{Q}$-values for the movies with (lower plots) and without (upper plots) interstellar scattering at different projected spin axis orientations and inclinations. For the lower plots, interstellar scattering was added by scattering the movie frames with a different scattering screen for each day of the 2009-2013 EHT observations.

In most models and at most source orientations, the variability is highest for the lowest inclinations. This is caused by the decreasing apparent source size towards higher inclinations (Figure \ref{fig:movies}). When the apparent source size is large (low inclinations), the variable structures are well resolved by the baselines, leading to low visibility amplitudes and large (closure) phase fluctuations. For a small apparent source size (high inclinations), the baselines do not resolve the variable structures well enough to for the $\mathcal{Q}$-metric to detect strong intrinsic variability above the noise level. 

For the $a_*=0.94$ disk model (left plots), the $\mathcal{Q}$-metric does not detect any variability at the highest inclinations if no scattering is added (top left). The simulated closure phase variability including scattering for this model is generally slightly higher, but still compatible with the 2009-2013 EHT data. Comparing with Figure \ref{fig:scatmodels_dist}, scattering the movie itself generally introduces more variability than what one would get for the annulus model, but less than for the Gaussian model. For orientations $\sim$130-160 and $\sim$310-340 degrees, the simulated closure phases are much more variable than for the other orientations and the EHT data, especially for the intermediate inclinations. The sensitivity to orientation is probably strongest for the intermediate inclinations because of two effects. For low inclinations, the source variability looks symmetric as the features appear to follow circular orbits around the black hole (although they are sheared out by differential rotation before they complete a full orbit), so $\mathcal{Q}$ is not expected to be very sensitive to source orientation here. On the other hand, at high inclinations the apparent source size is small, so that closure phase fluctuations are suppressed. The intermediate inclinations show the most sensitivity to source orientation because the source dynamics appear to be asymmetric, while the source size is still relatively large and the variable structures are resolved by the baselines.

The same effect is visible for the $a_*=0$ simulation (middle plots), though the higher inclinations show more sensitivity to orientation here. Also, only the highest inclinations show variability comparable to what was observed in the EHT data, as the lower inclinations overpredict the observed variability. These differences with the $a_*=0.94$ model may again be explained from the apparent source size, which is larger for the $a_*=0$ simulation (see Figure \ref{fig:movies}), causing the closure phase to fluctuate more for all inclinations.

For the jet model (right plots), different orientations show peaks in variability because the source morphology is different from the disk models. There is generally more variability than in the $a_*=0.94$ disk model, but less than in the $a_*=0$ model. This is in agreement with the apparent source size being larger than the $a_*=0.94$ disk model and smaller than the $a_*=0$ model, especially for low inclinations (see Figure \ref{fig:movies}). High inclinations show the strongest dependence on orientation because the source appears less symmetric here, while the apparent source size is not very different from lower inclinations. Another effect visible for the jet model is that the peak variability direction changes with inclination, which is not the case for the disk model. This could be caused by the fact that while the disk and jet models qualitatively appear to evolve similarly at low inclinations, for the jet model an elongated structure becomes more apparent towards higher inclinations (Fig. \ref{fig:movies}, Sec. \ref{sec:grmhd}). This introduces a direction of variability that is not strongly present in the disk models, causing the directionality of the closure phase variability to change.

Considering the above, the $\mathcal{Q}$-metric is suitable for distinguishing between source models with different GRMHD parameters based on their effect on the closure phase variability. However, a given $\mathcal{Q}$-value on a certain triangle will not directly point towards, e.g., a range of inclinations, source orientations, black hole spins, or electron temperature distributions, as there is a lot of degeneracy between them. For example, a change in inclination may have the same effect on the measured $\mathcal{Q}$-value as a change of orientation on the sky. Other data, such as the visibility amplitudes and closure phase values, will be needed in addition to the $\mathcal{Q}$-metric to tightly constrain these parameters. As the analysis presented here suggests, the $\mathcal{Q}$-metric has the potential to be useful in ruling out parts of the GRMHD and scattering parameter space.

\section{Summary and outlook}
\label{sec:summary}
In this work, a framework for characterizing the amount of intrinsic variability in closure phase measurements has been developed. The $\mathcal{Q}$-metric is a normalized quantity indicating the amount of variability in a set of measured closure phase data that cannot be explained by the (known) thermal noise. In the limit of an infinite number of measurements, the value of $\mathcal{Q}$ is zero if all variability is due to thermal noise, and it will approach 1 as all observed variability becomes dominated by intrinsic variations in the data. This is independent of whether the noise is Gaussian: the $\mathcal{Q}$-value can be calculated for any error distribution.

This metric has a number of useful and novel applications in the context of the analysis of EHT closure phases.
\begin{itemize}
\item $\mathcal{Q}$ can be used as a test of intrinsic variability: if it is measured to be larger than 0 with high confidence, the variations in the data cannot be explained as purely due to noise.
\item As Earth rotation introduces a variation of the closure phase, it will cause the $\mathcal{Q}$-value to be nonzero even if no variability from a changing source structure or interstellar scattering (image variability) is present. However, as the variation is slow (Figures \ref{fig:cphase_time_1} and \ref{fig:cphase_time_2}), its effect on the closure phase variability can be mitigated by detrending it. We have demonstrated, using simulated observations of GRMHD simulations, that $\mathcal{Q}$ will indeed go to zero if an observation of a static source is detrended for baseline evolution using the method of differencing (Fig. \ref{fig:metric_seg}).
\item As the detrending does not work perfectly and noise fluctuations may cause $\mathcal{Q}$ to deviate slightly from zero for a static source, a measured $\mathcal{Q}$-value should be compared to the distribution of $\mathcal{Q}$-values of synthetic observations of a static source with different thermal noise realizations. These synthetic observations should have the same properties as the EHT observations for which $\mathcal{Q}$ is calculated. We have developed a framework for generating these synthetic data and comparing the observed variability.
\item If variability has been detected with high confidence, $\sqrt{\tilde{\mathcal{Q}}}$ (Eq. \ref{eq:qtilde}) gives the excess closure phase variability in degrees that is due to variability of the observed image.
\item Different GRMHD parameters such as the black hole spin, electron temperature distribution, inclination angle, and orientation of the source on the sky result in different closure phase behavior and $\mathcal{Q}$-values. Thus, the $\mathcal{Q}$-metric can distinguish between these models based purely on their effect on the closure phase variability. Regardless of whether image variability is detected with high confidence, the measured $\mathcal{Q}$-value may be compared to the $\mathcal{Q}$-values of synthetic observations of different GRMHD models in order to rule out (combinations of) GRMHD parameters. The $\mathcal{Q}$-values of different GRMHD simulations may also be compared to each other to investigate the effect of different GRMHD parameters on the closure phase variability.
\item When applied to data from multiple epochs, the $\mathcal{Q}$-metric may be used to distinguish between different source models based on the expected variability due to interstellar scattering.
\item For data from a single epoch, variability from interstellar scattering is expected to be small and the $\mathcal{Q}$-value will indicate variability dominated by changes of the intrinsic source structure.  
\end{itemize}

Using this framework, comparison of GRMHD simulations with 2009-2013 EHT observations is consistent with models exhibiting variability due to sources other than Earth rotation and thermal noise. This variability could be the result of a variable source, variable refractive substructure due to interstellar scattering, or a combination of the two. The probability that a static source without scattering substructure produces the same amount of variability as observed in the 2009-2013 EHT data was found to be 3.0\%.

Interstellar scattering by a Gaussian source model fitted to the visibility amplitudes measured by the EHT in 2013 would result in larger closure phase fluctuations than those present in the data for $97.8\%$ of the realizations of the scattering screen and thermal noise. Closure phase variability caused by interstellar scattering from an annulus model is consistent with the EHT data. This independently reinforces the existing evidence for a non-Gaussian source morphology based on the goodness-of-fit of the measured visibility amplitudes. 

An analysis comparing different GRMHD model parameters has shown that the amount of closure phase variability is strongly dependent on inclination and source orientation. The sensitivity to orientation is strongest when the source morphology is asymmetric and its size is not too small, which is the case for intermediate inclinations in the high ($a_* = 0.94$) spin disk model of Sgr A*, and higher inclinations for the $a_*=0$ and jet models. Closure phases from the zero spin black hole model were found to exhibit significantly more variability than closure phases from the high spin model. The apparent source size is an important determining factor for closure phase variability: the larger the source, the more closure phase variability. This is due to the lower visibility amplitudes for larger sources \citep[see also][]{Medeiros2016}.

The $\mathcal{Q}$-metric will be applied to future EHT data sets that include closure phases from many more station combinations, enabling a firmer detection and characterization of closure phase variability, which may again be compared to different models. In this work, the $\mathcal{Q}$-metric has been applied to multi-epoch EHT data only as there were not sufficient data points measured on a single day and triangle for the statistical quantities that $\mathcal{Q}$ consists of to contain reliable information. A large number of measurements on a single day, as are expected for future EHT observations, would allow for direct detection of source variability as the scattering screen is not expected to change significantly over a few hours. Such measurements would also allow for comparisons of the observed variability between different GST segments or epochs. The closure phases from \citet{Fish2016} were measured with a total bandwidth of 1 GHz, but the EHT observations carried out in April 2017 had a bandwidth of 4 GHz, and an upgrade to 8 GHz is expected in the near future. The increased sensitivity will allow for more high SNR data points on a single closure phase track.

The behavior of power spectra and structure functions of the closure phases and other data products may teach us about the time scales involved in accretion processes \citep{Roelofs2016, Shiokawa2017}. Relating the variability of simulated interferometric quantities directly to physical processes in the (GRMHD) simulations used to generate them is a challenging but important task for future studies of time variability.\\

This work is supported by the ERC Synergy Grant ``BlackHoleCam'' (Grant 610058). The EHT is supported by multiple grants from the National Science Foundation (AST-1440254, AST-1310896) and a grant from the Gordon \& Betty Moore Foundation (GBMF-3561) to S.S.D. The SMA is a joint project between the Smithsonian Astrophysical Observatory and the Academia Sinica Institute of Astronomy and Astrophysics. The ARO is partially supported through the NSF University Radio Observatories program. The JCMT was operated by the Joint Astronomy Centre on behalf of the Science and Technology Facilities Council of the UK, the Netherlands Organisation for Scientific Research, and the National Research Council of Canada. Funding for CARMA development and operations was supported by the NSF and CARMA partner universities. We would like to thank Scott Noble for providing GRMHD code HARM3D and Charles F. Gammie for providing computational resources to run the simulation on XSEDE supported by NSF grant ACI-1053575. We thank Andrew Chael for letting us utilize his eht-imaging library, and Lindy Blackburn and Jason Dexter for useful discussions and feedback. F.R. is grateful to the Harvard-Smithsonian Center for Astrophysics for hosting him during the period when much of the work presented here was carried out.


 \end{document}